\RequirePackage{ifpdf}
\documentclass[letterpaper]{article}
\usepackage{jheppub}
\usepackage{xcolor}
\usepackage{amsmath}
\usepackage{epsfig}
\usepackage{caption}
\usepackage{subcaption}
\usepackage{soul}

\usepackage{dsfont}
\usepackage{mathtools}
\usepackage{amssymb}
\usepackage{stmaryrd}
\usepackage{slashed}
\usepackage{bbold}
\usepackage{braket}
\usepackage{bm}
\usepackage{tikz}
\usepackage{color}
\usepackage{multirow}
\usetikzlibrary{decorations.markings}
\usepackage{scalerel}

\pdfoutput=1

\makeatletter
\gdef\@fpheader{}
\g@addto@macro\bfseries{\boldmath}
\makeatother

\newcommand{\dd}{\mathrm{d}}
\newcommand{\Rea}{\Re \mathrm{e}\,}
\newcommand{\Ima}{\Im \mathrm{m}\,}
\newcommand{\Eq}[1]{Eq.~(\ref{#1})}
\newcommand{\Fig}[1]{Fig.~{\ref{#1}}}

\newcommand{\roughly}[1]{\mathrel{\raise.3ex\hbox{$#1$\kern-0.85em
\lower1ex\hbox{$\sim$}}}}
\newcommand{\lsim}{\roughly<}
\newcommand{\gsim}{\roughly>}

\def\nn{\nonumber}

\newcommand{\be}{\begin{equation}}
\newcommand{\bee}{\begin{equation}}
\newcommand{\ee}{\end{equation}}
\newcommand{\beea}{\begin{eqnarray}}
\newcommand{\eea}{\end{eqnarray}}
\newcommand{\bea}{\begin{eqnarray}}

\def\nott#1{\setbox0=\hbox{$#1$}                
   \dimen0=\wd0                                 
   \setbox1=\hbox{/} \dimen1=\wd1               
   \ifdim\dimen0>\dimen1                        
      \rlap{\hbox to \dimen0{\hfil/\hfil}}      
      #1                                        
   \else                                        
      \rlap{\hbox to \dimen1{\hfil$#1$\hfil}}   
      /                                         
   \fi}                                         %

\def\uxsl{\hbox{/\kern-.4000em$u$}}
\def\uxslsm{\hbox{\smaller/\kern-.5600em$u$}}
\def\pxpsl{\hbox{/\kern-.5000em$p$}}
\def\epssl{\hbox{/\kern-.5600em$\epsilon$}}
\def\delsl{\hbox{/\kern-.7000em$\nabla$}}
\def\lxpsl{\hbox{/\kern-.5600em$l$}}
\def\kxpsl{\hbox{/\kern-.5600em$k$}}
\def\qxpsl{\hbox{/\kern-.3900em$q$}}

\def\pref#1{(\ref{#1})}
\def\exd{{\rm d}}
\def\ol#1{{\overline{#1}}}

\def\cA{{\cal A}}
\def\cB{{\cal B}}

\def\cF{{\cal F}}

\def\cI{{\cal I}}

\def\cL{{\cal L}}
\def\cM{{\cal M}}
\def\cN{{\cal N}}
\def\cO{{\cal O}}
\def\cP{{\cal P}}

\def\cR{{\cal R}}

\def\cV{{\cal V}}
\def\cW{{\cal W}}
\def\cX{{\cal X}}

\def\cZ{{\cal Z}}

\def\bfk{{\bf k}}
\def\bfl{{\bf l}}

\def\bfp{{\bf p}}
\def\bfq{{\bf q}}

\def\bfx{{\bf x}}
\def\bfy{{\bf y}}
\def\bfz{{\bf z}}

\def\mfa{{\mathfrak a}}

\def\mfc{{\mathfrak c}}

\def\mfg{{\mathfrak g}}
\def\mfh{{\mathfrak h}}

\def\mfC{{\mathfrak C}}

\def\intn{{\rm int}}
\def\vac{{\rm vac}}

\def\llangle{\langle\hspace{-2mm}\langle}
\def\rrangle{\rangle\hspace{-2mm}\rangle}
\def\trB{{\hbox{Tr}_{\rm env}}}

\def\Tr{{\rm Tr}}

\def\UV{{\scriptscriptstyle U\hbox{\kern-0.1em}V}}
\def\PPN{{\scriptscriptstyle P\hbox{\kern-0.1em}P\hbox{\kern-0.1em}N}}
\def\MN{{\scriptscriptstyle M\hbox{\kern-0.1em}N}}
\def\MNP{{\scriptscriptstyle M\hbox{\kern-0.1em}N\hbox{\kern-0.1em}P}}
\def\KK{{\scriptscriptstyle K\hbox{\kern-0.1em}K}}
\def\SM{{\scriptscriptstyle S\hbox{\kern-0.1em}M}}
\def\EH{{\scriptscriptstyle E\hbox{\kern-0.1em}H}}

\def\QCD{{\scriptscriptstyle Q\hbox{\kern-0.1em}C\hbox{\kern-0.1em}D}}
\def\IR{{\scriptscriptstyle I\hbox{\kern-0.1em}R}}
\def\TEV{{\scriptscriptstyle T\hbox{\kern-0.1em}E\hbox{\kern-0.1em}V}}
\def\aff{{a\hbox{\kern-0.1em}f\hbox{\kern-0.1em}f}}

\title{Cosmic Purity Lost: Perturbative and\\ Resummed Late-Time Inflationary Decoherence}

\author[a,b,c]{C.P.~Burgess,}
\author[d]{Thomas Colas,}
\author[e]{R.~Holman,}
\author[b,f]{Greg Kaplanek,}
\author[g]{Vincent Vennin}

\affiliation[a]{Department of Physics \& Astronomy, 
McMaster University, Hamilton, ON, Canada, L8S 4M1}

\affiliation[b]{Perimeter Institute for Theoretical Physics, 
Waterloo, ON, Canada, N2L 2Y5}

\affiliation[c]{School of Theoretical Physics, Dublin Institute for
 Advanced Studies, 10 Burlington Rd., 
 Dublin, Ireland}

\affiliation[d]{Department of Applied Mathematics and Theoretical Physics, University of Cambridge, Wilberforce Road,  Cambridge, CB3 0WA, UK}

\affiliation[e]{Minerva University, 14 Mint Plaza, San Francisco,
  CA 94103, USA}

\affiliation[f]{Theoretical Physics, Blackett Laboratory, Imperial
  College, London, SW7 2AZ, UK}

\affiliation[g]{Laboratoire de Physique de l'\'Ecole Normale
  Sup\'erieure, ENS, CNRS, Universit\'e PSL,
  Sorbonne Universit\'e, Universit\'e Paris Cit\'e, F-75005 Paris,
  France}

\emailAdd{cburgess@perimeterinstitute.ca}
\emailAdd{tc683@cam.ac.uk}
\emailAdd{rholman@minerva.edu}
\emailAdd{g.kaplanek@imperial.ac.uk}
\emailAdd{vincent.vennin@ens.fr}

\date{today}

\begin{document}

\sloppy

\abstract{We compute the rate with which unobserved fields decohere other fields to which they couple, both in flat space and in de Sitter space, for spectator scalar fields prepared in their standard adiabatic vacuum. The process is very efficient in de Sitter space once the modes in question pass outside the Hubble scale, displaying the tell-tale phenomenon of secular growth that indicates the breakdown of perturbative methods on a time scale parameterically long compared with the Hubble time. We show how to match the perturbative evolution valid at early times onto a late-time Lindblad evolution whose domain of validity extends to much later times, thereby allowing a reliable resummation of the perturbative result beyond the perturbative regime. Super-Hubble modes turn out to be dominantly decohered by unobserved modes that are themselves also super-Hubble. Although our calculation is done for spectator fields, if applied to curvature perturbations during inflation our observations here could close a potential loophole in recent calculations of the late-time purity of the observable primordial fluctuations.}

 
\maketitle

\section{Introduction}
\label{sec:intro}

The evolution of our understanding of large-scale structure formation from speculative roots to an observational science is one of the triumphs of science over the past 30 years. The cosmological consensus explains the observed distribution of matter as a cumulative consequence of gravitational instability acting on the small initial fluctuations also visible in the cosmic microwave background. Everything fits together -- assuming the existence of Dark Matter and Dark Energy -- provided there exists a spectrum of Gaussian adiabatic density fluctuations that are close to (but not exactly) scale invariant \cite{Blumenthal:1984bp, Planck:2018nkj} (for a theoretical review see \cite{Bernardeau:2001qr}).   

Even more remarkably, precisely this pattern of density fluctuations is predicted to emerge quite generically from quantum fluctuations during the much earlier universe provided these occur during an epoch of accelerated\footnote{Models differ about whether this expansion starts with an initially expanding or contracting universe. We restrict our attention here to inflationary models, due to the better control of approximations this allows.} universal expansion \cite{Mukhanov:1981xt, Guth:1982ec, Hawking:1982cz, Starobinsky:1982ee, Bardeen:1983qw, Mukhanov:1988jd}. Although current observations cannot distinguish whether primordial fluctuations are quantum or classical, one can imagine trying to do so \cite{Campo:2005sv, Martin:2015qta, Maldacena:2015bha, Martin:2016tbd, Choudhury:2016cso, Martin:2017zxs, Ando:2020kdz, Green:2020whw, Martin:2021znx, Espinosa-Portales:2022yok} and thereby test the central assumption underlying this framework. 

But everyday experience suggests that quantum coherence can be fragile so a key part of any testing of the quantum nature of primordial fluctuations is whether the theories in question predict that quantum-generated density fluctuations decohere or not after their production and before their observation. Preliminary calculations indicate that even the very weak gravitational interactions at the core of most models suffice to decohere primordial fluctuations for spectator fields \cite{Brandenberger:1990bx, Burgess:2006jn, Martineau:2006ki, Campo:2008ij,  Burgess:2014eoa, Boyanovsky:2015tba, Boyanovsky:2015jen, Hollowood:2017bil, Shandera:2017qkg,  Martin:2018zbe, Colas:2022hlq}
and for the metric itself \cite{Lombardo:2004fr, Lombardo:2005iz, Prokopec:2006fc, Nelson:2016kjm, Burgess:2022nwu, Colas:2022kfu, DaddiHammou:2022itk, Ning:2023ybc, Colas:2024xjy} though reliable determination is hindered by the presence of `secular growth': measures of decoherence grow with time and leave the domain of validity of the perturbative framework in which they are computed \cite{Tsamis:2005hd, Burgess:2009bs, Burgess:2010dd, Giddings:2011zd, Burgess:2015ajz, Gorbenko:2019rza, Kaplanek:2019dqu, Kaplanek:2019vzj, Green:2020txs, Colas:2022hlq}. This is an important practical problem since this breakdown happens well before the late-time universe in which observations are made. 

More recent calculations \cite{Burgess:2022nwu} seek to evade this obstacle by adapting Open EFT arguments \cite{Burgess:2015ajz} (for a review see \cite{Burgess:2020tbq, Burgess:2022rdo}) to resum the perturbative behaviour and obtain reliable predictions for the late-time regime where earlier methods fail. More specifically, ref.~\cite{Burgess:2022nwu} computes how gravitational self-interactions of General Relativity allow the short-wavelength `environment' of unmeasured scalar and tensor metric fluctuations to decohere the longer-wavelength `system' of observed scalar modes of the metric during inflation. A universal result was found for how the purity of primordial perturbations evolve in this regime, independent of the details of how the environment and system are defined and of how the initial state is prepared. Crucially, for technical reasons, the calculation of \cite{Burgess:2022nwu} works in a regime where {\it both} short- and long-wavelength modes are super-Hubble, leaving open the size of the contribution of Hubble-sized or sub-Hubble environmental modes to the decoherence of super-Hubble modes.  

One aim of this paper is to provide the missing argument that shows why the contribution of Hubble-size or sub-Hubble environmental modes is subdominant to the contribution of super-Hubble environmental modes in the decoherence of super-Hubble system modes. We show this by evaluating how a spectator\footnote{By `spectator' we mean a field whose contribution to the universal energy density (and so also whose mixing with the metric) is negligible.} scalar field decoheres the state of a second spectator field, doing so in both flat space and de Sitter space with the fields in question initially prepared in their respective adiabatic vacuum.\footnote{We regard the flat-space calculation to be a benchmark against which to compare the de Sitter result, but given the likely existence of unmeasured dark sector fields it might be of interest in its own right.} The calculation is also interesting in its own right, inasmuch as it can be done very explicitly and allows the determination of how the decoherence rate depends on the properties of the fields and of their interactions. It also allows an explicit matching of late-time Open EFT methods to early-time perturbative calculations and so shows in detail how the one evolves into the other. 

To this end we trace out the `environment' field and compute the evolution of the reduced density matrix of the `system' field. To identify how the system field is decohered we in particular compute the evolution of the purity of this reduced density matrix as a function of the masses and couplings of the fields and of the curvature of the background spacetime. We contrast two types of interactions between the fields: a purely Gaussian evolution in which the two fields simply mix through an off-diagonal mass term and an honest-to-God interaction between the fields (which we take to be given by a cubic term in the scalar potential).  

Making reliable late-time predictions despite the breakdown of perturbative methods is obviously a broader issue than just for the evolution of decoherence, and part of our motivation in this paper is to use the decoherence calculation as a vehicle for exploring tools for late-time resummation that have broader applicability. 

Our findings are organized as follows. First, in \S\ref{sec:System}, we define the model to be studied and define in particular both the mixing and cubic interactions used to couple the two fields to one another. The precise definition of the system's purity and its evolution equation are both also given in this section, together with some useful formulae for their evaluation. 

Then \S\ref{sec:MinkDec} computes the purity evolution for both types of interactions for fields on flat space to leading order in perturbation theory. Not surprisingly, we find that decoherence occurs but also find no secular-growth effects and so perturbative methods need not break down at late times. We use this calculation to explore how the decoherence depends on the properties of the environmental field, in particular as its mass $m$ becomes large. This limit is subtle because it involves taking two limits that do not commute, but in all cases the decoherence rate falls to zero when $m$ is taken to infinity.

\S\ref{sec:dSdec} repeats this purely perturbative exercise for fields in de Sitter space. The main difference relative to flat space arises for super-Hubble modes for which decoherence does experience secular growth, implying the eventual breakdown of the perturbative prediction. With this motivation \S\ref{sec:OpenEFT} briefly reviews the Open EFT late-time resummation tools and applies them to the two-scalar system of interest.\footnote{The evolution of the reduced system of interest turns out for our examples to be Gaussian and this provides an independent reason why the late-time resummation remains under control.} An evolution equation of the Lindblad form \cite{Burgess:2020tbq, Gorini:1976cm} valid at late times is derived and its solutions are found, showing in principle how late-time predictions can be made beyond leading order. 

\S\ref{sec:DecoApp} applies this solution to the decoherence calculation in particular and shows how the early-time perturbative evolution can be matched to the late-time Lindblad evolution because their domains of validity overlap in the super-Hubble regime. This allows the late-time purity solution to be matched onto the Bunch-Davies initial conditions despite these initial conditions not applying within the domain of validity of Lindblad evolution. Our conclusions are briefly summarized in \S\ref{sec:Conclusions}.

\section{The systems of interest}
\label{sec:System}

We start by summarizing well-understood general features of decoherence when system and environment interact weakly enough to justify perturbative methods. The resulting useful formulae are applied to systems in flat and de Sitter spaces in later sections.

\subsection{Field content and interactions}
\label{SystemDef}

Consider the action 
\be \label{SystemAction}
 S = - \int \exd^4 x \, \sqrt{-g} \left[ \frac{M_p^2}2 \, \cR +L_m + \frac12 (\partial \sigma)^2 + \frac12 (\partial \phi)^2 + V(\sigma, \phi) \right]
\ee
where $L_m$ is a `matter' sector chosen to dominate in the stress energy and so to produce the assumed background geometry. This could be as simple as a cosmological constant if we focus on de Sitter geometries (as we sometimes will) or it could consist of thermal matter if we instead focus on radiation or matter dominated geometries. Our main focus is on the fields $\sigma$ and $\phi$ and we ask how interactions with the environment $\phi$ affect the evolution of the state of the system $\sigma$. 

For spacetimes with $\exd s^2 = - \exd t^2 + \gamma_{ij} \, \exd x^i \, \exd x^j$ the action \pref{SystemAction} implies the Hamiltonian for the fields $\phi$ and $\sigma$ is
\be
   H = \int \exd^3x \sqrt{\gamma}  \left[  \frac12 \pi_\sigma^2 +  \frac12 \gamma^{ij} \partial_i \sigma \, \partial_j \sigma + \frac12 \pi_\phi^2 + \frac12 \gamma^{ij} \partial_i \phi \, \partial_j \phi  + \frac12 (m_{\rm env}^2 \phi^2 + m_{\rm sys}^2 \sigma^2) + V_{\rm int}(\sigma, \phi) \right]
\ee
in which $\pi_\sigma := \dot \sigma$ and $\pi_\phi := \dot \phi$ and over-dots denote differentiation with respect to $t$. 

At low energies the dominant interactions involve the fewest derivatives and so come from the scalar potential (if this is present). We assume decoherence due to these is faster than decoherence due to gravity-mediated interactions. Using the fairly general potential 
\be
  V = \tfrac12 \, m_{\rm env}^2 \phi^2 + \tfrac12 \, m_{\rm sys}^2 \sigma^2 + V_{\rm int}(\sigma,\phi) \,,
\ee
we describe decoherence below for various choices for the interaction potential $V_{\rm int}$. 

The simplest case just takes it bilinear in the `system' and `environment' fields, 
\be \label{int=mix}
  V_{\rm int} =  V_{\rm mix} (\sigma, \phi) :=    \mu^2 \phi \, \sigma   \,,
\ee
since then the spectator system is Gaussian at heart and can be solved exactly for specific background geometries. Once diagonalized the spectrum consists of two massive fields $\phi_\pm$ obtained by rotating the pair $\phi, \sigma$ through an angle $\vartheta$ with 
\be 
    \tan 2\vartheta = \frac{2\mu^2}{m_{\rm env}^2 - m_{\rm sys}^2} \,,
\ee
whose squared masses 
\be\label{masseigenvalues}
   M^2_\pm = \frac12 \Bigl[ (m_{\rm sys}^2+m_{\rm env}^2) \pm \sqrt{(m_{\rm sys}^2-m_{\rm env}^2)^2 + 4\mu^4} \Bigr] \,, 
\ee
are both positive when $0 \leq \mu^2 \leq m_{\rm env} m_{\rm sys}$ (as we henceforth assume).

The disadvantage of the Gaussian case is that it leaves all Fourier modes of the theory uncoupled, hence it may not be representative of honest-to-God interactions. The next simplest case we consider takes the simplest bona fide interaction: a cubic potential that we assume to be linear in the system field. Linearity in the system field simplifies later discussions because it implies the effects of the environment on the reduced system remains effectively Gaussian in a way made more explicit below. This leads to the choice
\be \label{int=c}
   V_{\rm int} = V_c(\sigma,\phi) := g\, \phi^2 \sigma \,,
\ee 
where the coupling $g$ has dimension mass in 4D. When using cubic interactions we work within the semiclassical approximation and expand to quadratic order in the fluctuation fields, working perturbively in powers of $g$ (more about this later). 

Both of the above choices are captured by the parameterization $V_{\rm int} = \sigma\, \cO(\phi)$ where
\be
   \cO_{\rm mix}(\phi) = \mu^2 \phi  \qquad \hbox{and} \qquad
   \cO_c(\phi) = g \phi^2   \,.
\ee 
When describing perturbative evolution we write $H = H_0 + H_\intn$ with $H_\intn = \int \exd^3x \, V_{\rm int}(\sigma,\phi)$ and consider the two cases $V_{\rm int} = V_{\rm mix}$ and $V_{\rm int} = V_c$ in turn.

\subsection{Time evolution and decoherence}

In the interaction picture the full density matrix $\rho(t)$ for the scalar system satisfies the Liouville equation
\be \label{Liouville}
   \partial_t \rho = -i \Bigl[ V_{\rm int}(t) \,, \rho \Bigr] \,,
\ee
where $V_{\rm int}(t)$ is the interaction-picture Hamiltonian. So far as measurements of the system are concerned all we really require is the evolution of the reduced density matrix for $\sigma$, defined by tracing out the $\phi$ sector:
\be
   \varrho(t) := \Tr_\phi [ \rho(t) ] \,. 
\ee
In perturbation theory this satisfies the evolution equation
\bea \label{rhoevoF}
  \partial_t \varrho(t) &\simeq&  -i \left[ \ol V_{\rm int}(t), \, \varrho \right] + \int_{t_0}^t \exd s \int \exd^3x \int \exd^3y \bigl\{ \left[ \sigma(s,\bfy) \, \varrho_0 \,, \sigma(t,\bfx) \right] \cW(t,\bfx; s,\bfy)  \\
  && \qquad\qquad \qquad \qquad \qquad \qquad \qquad\qquad \qquad + \left[ \sigma(t,\bfx) \,, \varrho_0 \sigma(s,\bfy) \right] \cW^*(t,\bfx; s,\bfy) \bigr\}+ \cdots \,,\nn
\eea 
where $\ol V_{\rm int} = \llangle \, \cO(\phi) \, \rrangle \, \sigma = \Tr_\phi[\cO(\phi) \Xi_{\rm env}] \, \sigma$ introduces the notation $\llangle \,\cdots\, \rrangle = \trB[(\cdots) \,\rho_{\rm env}]$ for averages only over the environment, and 
\bea
\label{eq:Wightmann:def}
  \cW(t,\bfx; s,\bfy)  &:=& \llangle \, \cO(t,\bfx) \, \cO(s,\bfy) \,\rrangle - \llangle \, \cO(t,\bfx) \, \rrangle \, \llangle \, \cO(s,\bfy) \, \rrangle  \\
  &=& \Tr_\phi \Bigl[ \cO(t,\bfx) \, \cO(s,\bfy) \, \Xi_{\rm env}\Bigr] - \Tr_\phi \Bigl[ \cO(t,\bfx)  \, \Xi_{\rm env}\Bigr]\Tr_\phi \Bigl[ \cO(s,\bfy) \, \Xi_{\rm env}\Bigr] \,. \nn  
\eea
These expressions assume the system and environment are uncorrelated at $t = t_0$, with
\be
   \rho(t_0) = \varrho_0 \otimes \Xi_{\rm env}
\ee
where the initial reduced density matrices for the $\sigma$ and $\phi$ sectors are respectively $\varrho_0$ and $\Xi_{\rm env}$, and so $\Tr_\phi[ \Xi_{\rm env}] = 1$ implies $\varrho(t_0) = \varrho_0$.

Our diagnostic for decoherence is the purity, $\gamma(t)$, defined as
\be 
   \gamma(t) := \Tr_\sigma [ \varrho^2(t) ]
\ee
and so $0 \leq \gamma \leq 1$, with $\gamma=1$ if and only if the reduced state $\varrho$ is pure. Its rate of change is\footnote{This result assumes $\Tr_\sigma[\varrho_0 \sigma(t,\bfx) \varrho_0 \sigma(s,\bfy)] = 0$ as would be the case if $\varrho_0 = | \Psi \rangle \, \langle \Psi |$ and $\langle \Psi | \sigma(t,\bfx) | \Psi \rangle = 0$, and so is true in particular if $|\Psi \rangle = | \hbox{vac} \rangle$.}
\be \label{PurityChangeMasterEq}
  \partial_t \gamma =2 \, \Tr_\sigma \left( \varrho \, \partial_t \varrho \right)  =  - 4 \int_{t_0}^t \exd s \int \exd^3x \int \exd^3y \; \hbox{Re}\Bigl[ W(t,\bfx; s,\bfy) \, \cW(t,\bfx; s,\bfy) \Bigr] \,,
\ee
where the Wightman function for $\sigma$ is 
\be
  W(t,\bfx;s,\bfy) := \langle \sigma(t,\bfx) \, \sigma(s,\bfy) \rangle = \Tr_\sigma\Bigl[ \sigma(t,\bfx) \, \sigma(s,\bfy) \, \varrho_0 \Bigr] \,. 
\ee
Much of the rest of this paper is devoted to computing this expression for various examples.

\subsubsection{Implications of translation invariance}

The above expressions can be written more conveniently for translation-invariant systems, for which the Wightman function satisfies $W(t,\bfx;s,\bfy)  = W(t,0;s,\bfy-\bfx) =: W(t,s;\bfy-\bfx)$ and similarly for $\cW(t,\bfx;s,\bfy) = \cW(t,s;\bfy-\bfx)$, so expression \pref{PurityChangeMasterEq} becomes
\be \label{PurityChangeMasterEqF}
  \partial_t \gamma     =  - 4 \cV \int_{t_0}^t \exd s\int \exd^3z \; \hbox{Re}\Bigl[ W(t,s;\bfz) \, \cW(t,s;\bfz) \Bigr]  \,,
\ee
where $\cV = \int \exd^3x$ is the volume of space. As usual, it is the rate per unit volume that has the sensible large-volume limit in a translationally invariant system. Equivalently, using the Fourier representation
\be \label{cWtocWk}
  \cW(t,\bfx; s,\bfy)   =   \int \frac{ \exd^3k}{(2\pi)^3} \; \cW_{\bf{k}}(t,s) \, e^{ i \bf{k} \cdot (\bf{x} - \bf{y})}
\ee
and the equivalent expression for $W(t,\bfx; s,\bfy)$ in terms of $W_{\bf{k}}(t,s)$, allows \pref{PurityChangeMasterEqF} to be written
\be \label{PurityChangeMasterEqFk}
  \partial_t \gamma   =  - 4 \cV \int_{t_0}^t \exd s\int \frac{\exd^3k}{(2\pi)^3} \; \hbox{Re}\Bigl[ W_\bfk(t,s) \, \cW_{-\bfk}(t,s) \Bigr] \,.
\ee

This expression is particularly easy to interpret in the special case that the interaction Hamiltonian is linear in the system field $\sigma$ (as we assume here). In this case we shall see that the system evolution remains Gaussian (to leading approximation) even once environmental effects are included and so each of the system's momentum modes evolves independently. These therefore remain uncorrelated if they were initially uncorrelated in the vacuum, so we can write
\be
   \varrho(t) = \prod_\bfk \otimes \varrho_\bfk(t) \qquad \hbox{and} \qquad \varrho^2(t) = \prod_\bfk \otimes \varrho^2_\bfk(t) 
\ee
where translation invariance allows the single-particle states to be labelled by their momentum and $\varrho_\bfk$ is the density matrix for the mode with momentum $\bfk$. When writing the product we switch to momentum states with discrete normalization (such as by putting the system into a huge box). The purity's rate of change can then be written
\be
   \partial_t \varrho^2 = \sum_\bfk \Bigl(\prod_{\bfq < \bfk} \otimes \varrho^2_\bfq \Bigr) \partial_t \varrho^2_\bfk \Bigl( \prod_{\bfq > \bfk} \otimes \varrho^2_\bfq \Bigr) 
\ee
and so comparing the trace of this with $\partial_t \gamma  = \sum_\bfk \partial_t \gamma_\bfk$ in eq.~\pref{PurityChangeMasterEqFk} (converted to discrete normalization using $\cV \int \exd^3 k = (2\pi)^3 \sum_\bfk$) gives
\be \label{PurityChangeMasterEqFSum}
  \Tr \Bigl( \partial_t \varrho^2_\bfk \Bigr) = \partial_t \gamma_\bfk =- 4 \int_{t_0}^t \exd s  \; \hbox{Re}\Bigl[ W_\bfk(t,s) \, \cW_{-\bfk}(t,s) \Bigr] \,,
\ee
showing how $\gamma_\bfk(t)$ tracks the decoherence mode-by-mode.

Eq.~\pref{PurityChangeMasterEqFSum} integrates to give a useful expression for the purity as a function of time
\be \label{PurityMasterEqFNested}
   \gamma_\bfk(t)       =  \gamma_\bfk(t_0) - 2  \int_{t_0}^t \exd s' \int_{t_0}^t \exd s  \; \hbox{Re}\Bigl[ W_\bfk(s',s) \, \cW_{-\bfk}(s',s) \Bigr] \,,
\ee 
where the limits of integration are disentangled using the conjugation property
\be \label{conjugationproperty}
   W^*_\bfk(t,s) = W_{-\bfk}(s,t)  \qquad \hbox{and} \qquad \cW^*_\bfk(t,s) = \cW_{-\bfk}(s,t)  \, ,
\ee
which is a consequence of the relations $W^*(t,\bfx; s,\bfy) = W(s,\bfy; t, \bfx)$ and $\cW^*(t,\bfx; s,\bfy) = \cW(s,\bfy; t, \bfx)$ that follow from $\sigma$ and $\cO$ being hermitian.

Notice that the above expressions superficially seem to imply that $\partial_t \gamma$ must vanish when evaluated at $t=t_0$. If true, its initial evolution would start out quadratically: $\gamma_\bfk(t) \simeq 1 - \frac12 \, \gamma_\bfk^{(2)} (t-t_0)^2  + \cdots$ (for an initially pure state), with
\be
  \gamma_\bfk^{(2)}(t=t_0) =   4 \hbox{Re}\Bigl[ W_\bfk(t_0,t_0) \, \cW_{-\bfk}(t_0,t_0) \Bigr] \,.
\ee
Unfortunately, the inference that $\partial_t \gamma$ vanishes at $t=t_0$ assumes the integrand is well-behaved there, which is generically {\it not} true for interacting systems (this is one of the cases where Gaussian systems are not representative). As we see below, for the interacting case the Hadamard singularity \cite{Hadamard:1923, DeWitt:1960fc, Fulling:1978ht} of $\cW_\bfk(t,s)$ in the coincident limit leads to a nonzero (but finite) expression for $\partial_t \gamma$ at $t=t_0$.

\subsubsection{Mode sums}

Translation invariance also leads to the field expansion
\be  \label{FieldExpansion}
\phi(t,{\bf x})   =   \int \frac{\exd^3 k}{(2\pi)^{3/2}} \Bigl[ v_{k}(t) \mfc_{{\bf k}} + v^{\ast}_{k}(t) \mfc^{\ast}_{-{\bf k}} \Bigr] \, e^{i \bfk \cdot \bfx} \,,
\ee
and a similarly for $\sigma(t,\bfx)$ in terms of $\mfa_\bfk$ and $\mfa^*_{-\bfk}$ and mode functions $u_k(t)$. The ladder operators satisfy $[\mfa_\bfp \,, \mfa^*_\bfq] = \delta^3(\bfp-\bfq)$ and $[\mfc_\bfp \,, \mfc^*_\bfq] = \delta^3(\bfp-\bfq)$ as usual. Using this expansion, the Wightman function $W_\bfk(t,s)$ for a single field can be written in terms of its modes $u_k(s)$ as
\be \label{ModeWk}
  W_{\bf{k}}(t,s) = \int \exd^3 q \; u_{q}(t)  u^{\ast}_{q}(s) \, \delta^{3}(\bfq - \bfk) =    u_{k}(t)  u^{\ast}_{k}(s)  \,.
\ee
The Wightman function for the composite operator $\cO = g \phi^2$ is similarly given by (see eq.~(\ref{AppWkCompo}))
\be \label{WkCompo}
\cW_{\bf{k}}(t,s)  =  2g^2 \int \frac{\exd^3p \, \exd^3 q}{(2\pi)^3}\; v_{q}(t) v_{p}(t) v^{\ast}_{q}(s) v^{\ast}_{p}(s) \; \delta^{3}(\bf{q} + \bf{p} - \bf{k}) \,.
\ee
Convergence of these integrals for large momentum is ensured by taking the time difference $t-s$ to have a small negative imaginary part.

Using these expressions in \pref{PurityMasterEqFNested} allows the integrated purity to be written in a simple way. In the case when the system and environment interact through $V_{\rm mix}$ then $\cO = \mu^2 \phi$ and so $\cW_\bfk(t,s) = \mu^4 W_\bfk(t,s)$ and  
\be \label{PurityMasterEqFNestedMix}
   \gamma_\bfk(t)   =  \gamma_\bfk(t_0) - 2\mu^4 \Bigl| \cM_k(t,t_0)  \Bigr|^2 \qquad \hbox{with} \qquad
   \cM_k(t,t_0) := \int_{t_0}^t \exd s \;  u_k(s) \, v_k(s) \,,
\ee 
where $u_k(s)$ denotes the system mode function while $v_k(s)$ is the environment mode function. The case where the system-environment interaction comes from $V_{c}$ similarly gives $\cO =g \phi^2$ and this, together with \pref{WkCompo}, implies 
\bea \label{PurityMasterEqFNestedCubic}
   \gamma_\bfk(t)   &=&  \gamma_\bfk(t_0) - 4g^2 \int \frac{\exd^3 p \, \exd^3q}{(2\pi)^3} \; \Bigl| \cN_{pqk}(t,t_0)  \Bigr|^2 \delta^3(\bfp+\bfq-\bfk) \nn\\  \hbox{with} \qquad
   \cN_{pqk}(t,t_0) &:=& \int_{t_0}^t \exd s \;   u_k(s) \, v_p(s) \, v_q(s) \,.
\eea 
Both cases predict $\gamma_\bfk(t) \leq \gamma_\bfk(t_0)$ for initially uncorrelated states.

\section{Decoherence in Minkowski space}
\label{sec:MinkDec}

We start by establishing a benchmark; evaluating the decoherence rate (for both mixing and cubic interactions) for flat-space examples with the initial state chosen to be the Minkowski vacuum.  In this case $\exd s^2 = - \exd t^2 + \exd \bfx^2$ and the field expansions \pref{FieldExpansion} involve
\begin{eqnarray} \label{AppMinkModes}
u_{k}(t) = \frac{e^{- i  \varepsilon_{\rm sys} \, t}}{\sqrt{ 2 \varepsilon_{\rm sys} }}  \qquad \hbox{and} \qquad 
v_{k}(t) = \frac{e^{- i  \varepsilon_{\rm env}\, t}}{\sqrt{ 2 \varepsilon_{\rm env}}}  \qquad \qquad (\mathrm{Minkowski}) 
\end{eqnarray}
where $\varepsilon_i(k) = \sqrt{k^2 + m_i^2}$ and $m_i$ is the mass of the relevant particle. 

The correlation function $W_\bfk(t,s)$ \pref{ModeWk} then is
\be \label{FlatWk}
 W^{(i)}_{\bfk}(t,s) =  \frac{1}{2 \varepsilon_i(k)} \, e^{-i\varepsilon_i(k)(t-s)}
\,,
\ee
where $t - s = x^0 - y^0$ has a small negative imaginary part to ensure convergence of the $k$-integration, leading to the position-space expression
\be \label{FinalDeltaComplex}
   W(x-y) =  \frac{im^2}{8\pi w} \;  H^{(2)}_1(w)  \qquad \hbox{where } \qquad w := m \sqrt{-(x-y)^2} \, ,
\ee
$H^{(2)}$ being the Hankel function of the second kind. This is a function only of the invariant interval $(x-y)^2 = \eta_{\mu\nu} (x-y)^\mu (x-y)^\nu$, and the phases are chosen such that $W(x-y)$ is real for spacelike separations. Also useful is the massless limit of this expression:
\be \label{WmasslessFlat}
  W(x-y)   = \frac{1}{4\pi^2 (x-y )^2} \,.
\ee

\subsection{Decoherence through mixing}

Consider first $V_{\rm int} = V_{\rm mix} = \mu^2 \phi \,\sigma$. In this case $\cO(\phi) = \mu^2 \phi$ and so $\cW_\bfk(t,s) = \mu^4 W_\bfk(t,s)$ and the purity evolution equation \pref{PurityChangeMasterEqF} becomes
\be  \label{PurityChangeMasterEqFflat}
  \partial_t \gamma_\bfk  
   =   - \frac{ \mu^4}{\varepsilon_{\rm sys} \varepsilon_{\rm env}}  \int_{t_0}^t \exd s\; \hbox{Re}\Bigl[ e^{-i(\varepsilon_{\rm sys}+\varepsilon_{\rm env})(t-s)} \Bigr]   
   =     \frac{\mu^4}{\varepsilon_{\rm sys} \varepsilon_{\rm env}(\varepsilon_{\rm sys} +\varepsilon_{\rm env})} \, \hbox{Im}\Bigl[  e^{-i(\varepsilon_{\rm sys}+\varepsilon_{\rm env})(t-t_0)} \Bigr] \,.
\ee 
Note that this equation solves for the purity {\it rate} (as opposed to the purity itself) as a simple check that the rate has the correct sign, so that the purity decreases away from unity once time evolution begins. It is tempting to go one step further and write the imaginary part as
 \be \label{PurityChangeMasterMixingWrong}
  \partial_t \gamma_\bfk  =   -  \frac{\mu^4}{\varepsilon_{\rm sys} \varepsilon_{\rm env}(\varepsilon_{\rm sys} +\varepsilon_{\rm env})}  \, \sin \Bigl[(\varepsilon_{\rm sys}+\varepsilon_{\rm env})(t-t_0) \Bigr]  \,,
\ee
though this treats the coordinates as if they are real (and so ignores the mandatory small negative imaginary part of $(t-t_0)$). Taking the small imaginary part of time to zero does not commute with the limit $m_{\rm env} \to \infty$, however, should we wish to explore $m_{\rm env} \gg k, \varepsilon_{\rm sys}$. For instance, the large-$m_{\rm env}$ limit of \pref{PurityChangeMasterEqFflat} is
\be \label{puritydecoupling}
    \partial_t \gamma_\bfk  \simeq     \frac{\mu^4}{m_{\rm env}^2\varepsilon_{\rm sys}} \, \hbox{Im}\Bigl[  e^{-im_{\rm env}(t-t_0)} \Bigr] \,,
\ee
and so vanishes exponentially quickly as $m_{\rm env} \to \infty$ with the imaginary part of $t-t_0$ held fixed, as opposed to the result $\partial_t \gamma_\bfk  \propto m_{\rm env}^{-2} \sin \Bigl[m_{\rm env}(t-t_0) \Bigr]$ found from \pref{PurityChangeMasterMixingWrong} if the imaginary part of the time coordinate were first taken to zero. We briefly discuss some of the implications of this difference for the decoupling of heavy states in \S\ref{sec:Conclusions} below. 

Integrating \pref{PurityChangeMasterMixingWrong} then gives 
\be  \label{PurityMasterEqFflat}
  \gamma_\bfk(t)   =  \gamma_\bfk(t_0)
  -  \frac{2\mu^4}{\varepsilon_{\rm sys} \varepsilon_{\rm env}(\varepsilon_{\rm sys} +\varepsilon_{\rm env})^2} \, \sin^2 \left[  \frac{ (\varepsilon_{\rm sys} +\varepsilon_{\rm env})(t-t_0) }{2}  \right]    \,,
\ee
consistent with \pref{PurityMasterEqFNestedMix}. For an initially pure state -- for which $\gamma_\bfk(t_0) = 1$ -- 
the deviation from unity starts off quadratically in $t-t_0$ because the single-field correlation function \pref{FlatWk} is nonsingular as $t \to s$. Notice that the mass eigenvalues $M_\pm^2$ of \pref{masseigenvalues} are only non-negative when $m_{\rm sys} m_{\rm env} \geq \mu^2$ and this implies the right-hand side of \pref{PurityMasterEqFflat} is always between 0 and 1. Our use of perturbation theory -- for which $\gamma_\bfk(t)$ is computed linearly in $\mu^4$ -- is only justified if $\mu$ is very small and in this case $\gamma_\bfk(t)$ is never that far from unity. 

Eq.~\pref{PurityMasterEqFflat} implies the purity of a specific mode can return to unity after a finite timescale $(\varepsilon_{\rm env}+ \varepsilon_{\rm sys})^{-1} \leq (m_{\rm env} + m_{\rm sys})^{-1}$. This might be an artefact of this particular system really being a noninteracting theory in disguise. The time scale for $\gamma_\bfk$ to return to unity differs for each mode, however, and so the total purity does not similarly return to unity once summed over $\bfk$. To see this explicity in a specific example choose $m_{\rm sys} = m_{\rm env} = m$, in which case the total purity rate of change per unit volume becomes
\be \label{PurityChangeMasterEqFflat2}
  \frac{\partial_t \gamma}{\cV}    =     -  \frac{\mu^4}{(2\pi)^2} \int_0^\infty \exd k \; \frac{k^2}{\varepsilon_k^3} \, \sin \Bigl[2\varepsilon_k(t-t_0) \Bigr]  
   =    -  \frac{\mu^4}{(2\pi)^2} \int_1^\infty  \frac{\exd v }{v^2} \,\sqrt{v^2-1}\; \sin \Bigl[2v \, m(t-t_0) \Bigr] \,.
\ee 

\begin{figure}[h]
\begin{center}
\includegraphics[width=90mm]{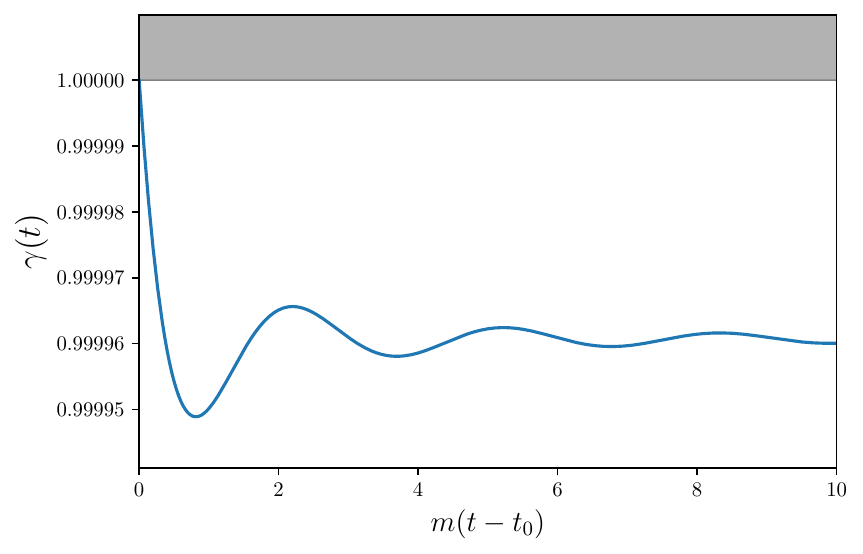}
\caption{\small Plot of the total purity $\gamma(t)$ as a function of $m(t-t_0)$ under the assumption that environment and system have the same mass $m_{\rm env}=m_{\rm sys}=m$ and that $\mu^4\cV/m = 10^{-4}$.} \label{Fig:PurityMixingPlot}
\end{center}
\end{figure}

Integrating \pref{PurityMasterEqFflat} -- assuming an initially pure state -- gives the total purity plotted in Figure \ref{Fig:PurityMixingPlot}. The oscillatory behaviour in this figure reflects the transients caused by assuming the interactions `turn on' at $t=t_0$ (implicit in the choice of an initially pure state). These oscillations disappear in an adiabatic calculation for which $t_0 \to -\infty(1 - i \epsilon)$ for infinitesimal positive $\epsilon$ (which projects onto the adiabatic vacuum), leaving only the constant displacement in $\gamma(t)$ to which the plot in Fig.~\ref{Fig:PurityMixingPlot} asymptotes.

\subsection{Decoherence through cubic interactions}

When system and environment interact through the cubic coupling then $\cO = g \phi^2$ and the environmental correlator that is required is
\bea  \label{cWposFlat}
  \cW(t,\bfx, s,\bfy) &=&  g^2 \Bigl[ \llangle \, \phi^2(t,\bfx) \, \phi^2(s,\bfy) \, \rrangle - \llangle \, \phi^2(t,\bfx) \, \rrangle \, \llangle \, \phi^2(s,\bfy) \, \rrangle \Bigr] \nn\\  
  &=& 2g^2 \Bigl[ W_{\rm env}(x - y) \Bigr]^2 =   \frac{g^2m_{\rm env}^2}{32\pi^2 (x-y)^2} \Bigl[  H^{(2)}_1(w_{\rm env}) \Bigr]^2
\eea
where $W_{\rm env}(x,y)$ is given by \pref{FinalDeltaComplex} with $m = m_{\rm env}$ and $w_{\rm env}:= m_{\rm env} \sqrt{-(x-y)^2}$. As for the single-field correlator, this expression is real when $x-y$ is spacelike. In momentum space $\cW_{\bf{k}}(t,s)$ is given by \pref{WkCompo}, which evaluates to:
\be  \label{FlatMassivecWk}
\cW_{\bf{k}}(t,s)
 =   \frac{ig^2}{2(2\pi)^2k(t-s)} \int_0^\infty \frac{p \, \exd p}{\varepsilon_{\rm env}}\, e^{-i \varepsilon_{\rm env} (t-s)} \left[ e^{-i \varepsilon_+(t-s)} - e^{-i \varepsilon_-(t-s)} \right] \,,
\ee
where $\varepsilon_\pm := \sqrt{(p \pm k)^2 + m_{\rm env}^2}$ and convergence of the $p$ integration is ensured because $t-s$ has a small negative imaginary part. 

The rate of change of the total purity is found by inserting \pref{cWposFlat} and \pref{FinalDeltaComplex} into \pref{PurityChangeMasterEq}, leading to the position-space expression 
\be \label{PurityChangeMasterEq2}
  \frac{\partial_t \gamma}{\cV} =   -  \frac{g^2m_{\rm env}^2m_{\rm sys}}{32\pi^2}    \int_{t_0}^t \exd s  \int_0^\infty \exd y \; \hbox{Im}\left\{  \frac{ y^2}{(s^2-y^2)^{3/2}} \;  H^{(2)}_1(m_{\rm sys}\sqrt{s^2-y^2})   \Bigl[  H^{(2)}_1(m_{\rm env}\sqrt{s^2-y^2}) \Bigr]^2\right\} \,.
\ee
The result for a specific mode $\bfk$ -- found by inserting \pref{FlatMassivecWk} and \pref{FlatWk} into \pref{PurityChangeMasterEqFSum} -- then is
\be \label{PurityChangeMasterEqFSum2}
  \partial_t \gamma_\bfk = - \frac{2}{\varepsilon_{\rm sys}}  \int_{t_0}^t \exd s  \; \hbox{Re}\left[  e^{-i\varepsilon_{\rm sys}(t-s)} \cW_{-\bfk}(t,s) \right] \, .
\ee
We next consider several informative limits of these expressions.

\subsubsection*{Massless environment}

An informative yet simple case takes the environment to be massless, in which case the correlators are
\be \label{cWmasslessFlat}
  \cW(x,y) = g^2 \Bigl[ W_{\rm env}(x - y) \Bigr]^2 =   \frac{g^2}{16\pi^4 (x-y)^4}  \qquad \hbox{(massless environment)} 
\ee
and
\be \label{PhiSqMasslessWk}
\cW_{\bf{k}}(t,s)  =\cW_{-\bf{k}}(t,s)  = -  \frac{ig^2}{2(2\pi)^2(t-s)} \,  e^{-ik(t-s)} \qquad \hbox{(massless environment)} \,.
\ee
These give the purity evolution equation  
\be \label{PurityChangeMasterEqFflatCubic}
  \partial_t \gamma_\bfk   =  - 4  \int_{t_0}^t \exd s\; \hbox{Re}\Bigl[ W_\bfk(t-s) \, \cW_{-\bfk}(t-s) \Bigr] 
  =  \frac{g^2}{4\pi^2\varepsilon_{\rm sys}}   \int_{t_0}^t \exd s\; \hbox{Re}\left[ \frac{i}{(t-s)} e^{-i (\varepsilon_{\rm sys} + k)(t-s)} \right] \,.
\ee

When taking the real part we use the identity
\be \label{iepsExp}
  \frac{1}{x - i \epsilon} = \frac{x+i \epsilon}{x^2 + \epsilon^2} \to \cP \left( \frac{1}{x} \right) + i \pi \delta(x) \,,
\ee
as $\epsilon \to 0$, where $\cP$ denotes the principal part, where the delta function satisfies 
\be
\label{eq:Dirac:prop}
  \int_{-a}^a \exd x \; \delta(x) = 1 \qquad \hbox{and} \qquad \int_{-a}^0 \exd x \; \delta(x) = \frac12 \,,
\ee
leading to the purity evolution rate 
\be \label{PurityChangeMasterEqFflatCubicxxx}
  \partial_t \gamma_\bfk  
   =   - \frac{g^2}{(2\pi)^2 \varepsilon_{\rm sys}} \left\{ \frac{\pi}{2} - \text{Si}[(k+\varepsilon_{\rm sys}) (t-t_0)] \right\} \,.
\ee

\begin{figure}[h]
\begin{center}
\includegraphics[width=90mm]{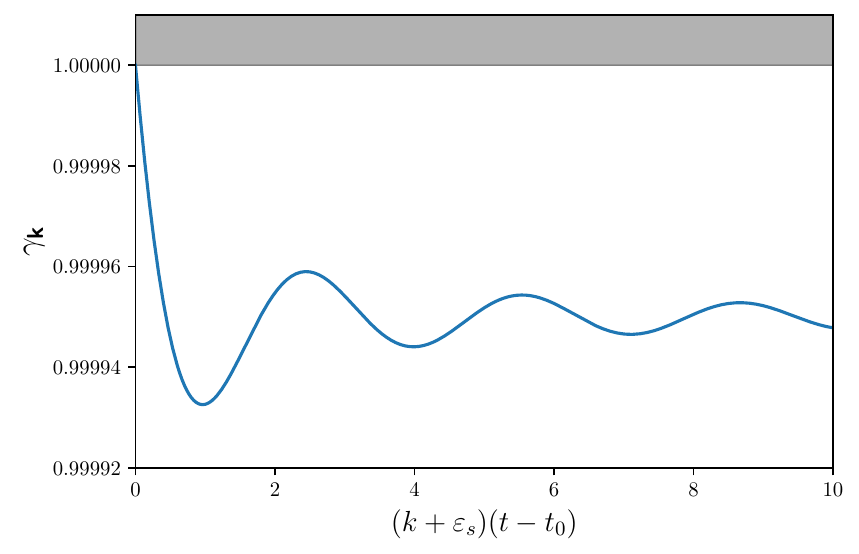}
\caption{\small Plot of the purity $\gamma_\bfk$ as a function of $(k + \varepsilon_{\rm sys})(t-t_0)$ with cubic system/environment coupling strength $g$, under the assumption that environment is massless and the system has mass $m_{\rm sys}$ and that the state at $t=t_0$ is pure. The numerics assume $g^2/[4\pi^2\varepsilon_{\rm sys}(\varepsilon_{\rm sys}+k)] = 10^{-4}$ where $\varepsilon_{\rm sys} = \sqrt{k^2+m_{\rm sys}^2}$. As discussed in the text, the initial rate does {\it not} vanish at $t = t_0$ and the instantaneous rate can be positive (although it does start off strictly negative).} \label{Fig:CubicFlatPlot}
\end{center}
\end{figure}

Asymptotic properties of the sine-integral function $\text{Si}(z)$ are listed in \S\ref{AppAsymptotic} for large and small arguments, and using these shows how the singular behaviour of the environmental correlator causes $\partial_t \gamma$ to approach a nonzero limit as $t \to t_0$, with
\be
   \partial_t \gamma_\bfk   \simeq  - \frac{g^2}{8\pi \varepsilon_{\rm sys}} \left[1 - \frac{2}{\pi}\left(\varepsilon_{\rm sys} + k\right)(t-t_0) + \cdots \right]  \qquad \hbox{if $k(t-t_0) \ll 1$}\,.  
\ee
Eq.~\pref{PurityChangeMasterEqFflatCubicxxx} integrates to give
\be 
 \gamma_\bfk (t) = 1 - \frac{g^2}{(2\pi)^2 \varepsilon_{\rm sys}} \left\{(t-t_0) \left[ \frac{\pi}{2} -   \text{Si}[2 (\varepsilon_{\rm sys} + k) (t-t_0 )] \right] +\frac{\sin ^2[(\varepsilon_{\rm sys} + k) (t-t_0 )]}{\varepsilon_{\rm sys} + k} \right\}
\ee
assuming $\gamma_\bfk(t_0) = 1$. This is plotted in Fig.~\ref{Fig:CubicFlatPlot} and shows an initial drop followed by transient oscillations about an asymptote displaced from unity by $g^2/[8\pi^2 \varepsilon_{\rm sys}(\varepsilon_{\rm sys}+k)]$. 

\subsubsection*{Massive environment}

Consider next the case $k,m_{\rm sys} \ll m_{\rm env}$ limit, for which \pref{FlatMassivecWk} becomes
\bea \label{FlatMassivecW0}
\cW_{\bf{k}}(t,s) &\simeq&  \frac{g^2 m_{\rm env}}{4\pi^2} \int_1^\infty \frac{\exd u}{u}\, \sqrt{u^2-1} \; e^{-2i u m_{\rm env} (t-s)} + \cO(k/m_{\rm env}) \qquad (\hbox{if $k\ll m_{\rm env}$})\,,
\eea
and so
\bea \label{PurityChangeMasterEqFSum2x}
  \partial_t \gamma_\bfk&\simeq& -    \frac{g^2 m_{\rm env}}{2\pi^2\varepsilon_{\rm sys}(k)} \int_{t_0}^t \exd s \; \hbox{Re}\left[  e^{-i(t-s)\varepsilon_{\rm sys}(k)} \int_1^\infty \frac{\exd u}{u}\, \sqrt{u^2-1}  \; e^{-2i u m_{\rm env} (t-s)} \right] \nn\\
  &\simeq& -  \frac{g^2}{4\pi^2\varepsilon_{\rm sys}} \int_1^\infty \frac{\exd u}{u^2} \sqrt{u^2-1} \; \hbox{Im}\left[ 1 -  e^{-2i u m_{\rm env} (t-t_0)} \right] \,.
\eea

This shares the same decoupling properties as the mixing case: it is exponentially small as $m_{\rm env} \to \infty$ with the imaginary part of $t-s$ fixed and negative. For the opposite limit -- where the imaginary part of time goes to zero with fixed $m_{\rm env}$ -- and assuming an initially pure state at $t= t_0$, the integrated purity instead is
\bea \label{PurityMassiveEnvEqFSum2x}
  \gamma_\bfk &\simeq& 
  1  -   \frac{g^2}{4\pi^2\varepsilon_{\rm sys} m_{\rm env}} \int_1^\infty \frac{\exd u}{u^3} \sqrt{u^2-1} \; \sin^2[ u m_{\rm env} (t-t_0)]  \\
  &=& 1 - \frac{g^2(t-t_0)}{32\pi \varepsilon_{\rm sys}} \Bigl( \left\lbrace\left[4 \pi  m_{\rm env}^2(t-t_0)^2 +\pi \right] \pmb{H}_0[2 m_{\rm env}(t-t_0)] - 4 m_{\rm env} (t-t_0)\right\rbrace J_1[2 m_{\rm env}(t-t_0)] \nn\\
  && \quad + \left\lbrace - \left[4 \pi  m_{\rm env}^2(t-t_0)^2+\pi \right] \pmb{H}_1[2 m_{\rm env}(t-t_0)] + 8 m_{\rm env}^2(t-t_0)^2 +4\right\rbrace J_0[2 m_{\rm env}(t-t_0)] - 4 m_{\rm env} (t-t_0) \Bigr) \,,\nn
\eea
where $J_\nu(z)$ is the usual Bessel function and $\pmb{H}_\nu(z)$ is a Struve function (whose asymptotic forms are listed in Appendix \ref{AppAsymptotic}). As is shown in Fig.~\ref{Fig:CubicFlatPlotMassive} this oscillates at late times about a fixed displacement from unity whose size goes to zero as $m_{\rm env} \to \infty$ (with the same qualitative shape as for the massless environment).

\begin{figure}[h]
\begin{center}
\includegraphics[width=90mm]{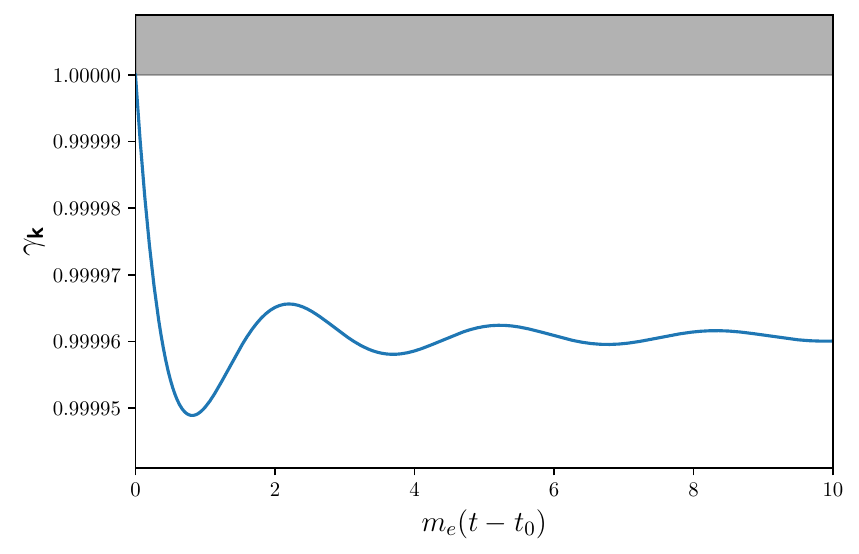}
\caption{\small Plot of the purity $\gamma_\bfk$ as a function of $m_{\rm env}(t-t_0)$ with cubic system/environment coupling strength $g$, under the assumption that environment is very massive compared to the system mass $m_{\rm sys}$ and $k$ and that the state at $t=t_0$ is pure. The numerics assume $g^2/[4\pi^2\varepsilon_{\rm sys} m_{\rm env}] = 10^{-4}$ where $\varepsilon_{\rm sys} = \sqrt{k^2+m_{\rm sys}^2}$. } \label{Fig:CubicFlatPlotMassive}
\end{center}
\end{figure}

With these expressions in hand for flat space as a benchmark we next turn to calculating the corresponding quantities for de Sitter geometries.

\section{Decoherence in de Sitter}
\label{sec:dSdec}

Consider next the de Sitter geometry with metric
\be
  \exd s^2 = - \exd t^2 + a^2(t) \, \exd \bfx^2 = a^2(\eta) \left( - \exd \eta^2 + \exd \bfx^2 \right) \,,
\ee
where $a(t) = e^{Ht}$ and the second equality introduces conformal time, which satisfies $\exd t = a \, \exd \eta$ and so (as usual) $\eta = -H^{-1} \, e^{-Ht} = - {1}/({aH})$ runs over $-\infty < \eta < 0$ when $-\infty < t < \infty$. In particular, the scale factor in conformal time is $a(\eta) = -1/(H\eta)$ and late times correspond to $\eta \to 0$.

In this geometry the basic evolution equation acquires extra contributions from the background metric. For instance \pref{rhoevoF} becomes
\bea \label{rhoevoC}
  \partial_t \varrho(t)  &=&  \int_{t_0}^t \exd s \int \exd^3x \, a^3(t) \int \exd^3y \, a^3(s) \Bigl\{ \Bigl[ \sigma(s,\bfy) \, \varrho_0 \,, \sigma(t,\bfx) \Bigr] \cW(t,\bfx; s,\bfy) \\
  && \qquad\qquad\qquad\qquad\qquad\qquad\qquad\qquad\qquad + \Bigl[ \sigma(t,\bfx) \,, \varrho_0 \,\sigma(s,\bfy) \Bigr] \cW^*(t,\bfx;s,\bfy) \Bigr\} \,,\nn
\eea 
where the spatial volume element $\sqrt{-g_{(3)}} = a^3$ appears when the evolution Hamiltonian generates translations in cosmic time $t$. Correlation functions like $W(x,y)$ transform as bi-scalars under coordinate transformations.

The purity evolution equation \pref{PurityChangeMasterEq} then becomes
\be \label{PurityChangeMasterEqC}
  \partial_t \gamma   = - 4 \int_{t_0}^t \exd s \int \exd^3x \, a^3(t) \int \exd^3y \, a^3(s) \; \hbox{Re}\Bigl[ W(t,\bfx,s,\bfy) \, \cW(t,\bfx,s,\bfy) \Bigr] \,,
\ee
which in momentum space and in conformal time modifies \pref{PurityChangeMasterEqFSum} to
\be \label{PurityChangeMasterEqFflatdS}
  \partial_\eta \gamma_\bfk  =   - 4 a^4(\eta)  \int_{\eta_0}^\eta  \exd \eta'\, a^4(\eta')  \; \hbox{Re}\Bigl[ W_\bfk(\eta,\eta') \, \cW_{-\bfk}(\eta,\eta') \Bigr] \,,
\ee
where $\bfk$ is the mode's co-moving momentum label. 

\subsection{de Sitter correlators}

To evaluate these expressions we require the correlation functions in de Sitter space~\cite{Birrell:1982ix}. We choose the initial state in the remote past to be the Bunch-Davies vacuum, corresponding to modes that start in the remote past in their adiabatic vacuum. In this case modes labelled by $\bfk$ are uncorrelated in the remote past and so use of Bunch-Davies correlators carries the implicit ancillary choice $t_0 \,, \eta_0 \to - \infty$. 

For free scalar fields in de Sitter space we use the Bunch-Davies mode functions $u_k(\eta) \, e^{i \bfk \cdot \bfx}$ which satisfy the Klein-Gordon equation with the boundary condition that they become conformal to flat-space mode functions in the remote past \cite{Bunch:1978yq} so that $a u_k \simeq e^{- i k \eta} / \sqrt{2k}$. This gives
\be \label{GendSMode}
 u_k(\eta) = \tfrac12  \sqrt\pi \, H e^{\tfrac{i\pi}{4}(2 \nu + 1) } ( - \eta)^{3/2}  \,  H_\nu^{(1)}(-k\eta)  \,, 
\ee
where\footnote{We here supplement the action \pref{SystemAction} with an addition nonminimal coupling $\delta S = \frac12 \int \exd^4x \,\cR (\xi_{\rm env}\phi^2 + \xi_{\rm sys} \sigma^2)$ because it is no more difficult to do so and allows us to include the case $\xi = \frac16$ corresponding to conformally coupled scalars.}
\be \label{nu_xi_def}
   \nu^2 =  \frac94 - \zeta^2 \qquad \hbox{where} \qquad \zeta^2 :=  \left( \frac{m}{H} \right)^2 + 12 \, \xi\,.
\ee

The de Sitter correlation function implied by this choice then becomes
\be \label{dSWk}
   W_\bfk(\eta,\eta') = u_k(\eta) \, u^*_k(\eta') =   \frac{\pi \, H^2 }{4} (\eta \eta')^{3/2}  H_\nu^{(1)}(-k\eta) \, \Bigl[  H_\nu^{(1)}(-k\eta') \Bigr]^*  \,,
\ee
where the times $\eta$ and $\eta'$ are required to have a small negative imaginary part for the same reasons as in flat space. The corresponding position-space result is given by
\cite{Bunch:1978yq}
\be \label{WkFTdefdSmass}
  W(\eta,\bfx; \eta',\bfy) = \frac{H^2}{16\pi} \left(\tfrac14 - \nu^2 \right) \sec(\pi \nu)\;  {}_2F_1\left[\tfrac32 + \nu, \tfrac32 - \nu; 2; z(\eta,\bfx; \eta', \bfy) \right]    \,, 
\ee
where
\be \label{zetadef}
   z(\eta,\bfx;\eta', \bfy) := 1 - \frac{1}{4\eta\eta'} \Bigl[-(\eta - \eta')^2 + (\bfx - \bfy)^2 \Bigr] \,,
\ee
and ${}_2F_1[a,b;c;z]$ is the hypergeometric function. 

\subsubsection*{Massless limits}

There are two important special massless cases. The first is a conformally coupled massless scalar (CCMS) $(m, \xi) = (0, \frac16)$ for which $\zeta^2 = 2$ and $\nu = \frac12$ and so \pref{GendSMode} becomes
\be \label{CCMSmodes}
   u_k(\eta) = \frac{iH\eta}{\sqrt{2 k}} \, e^{-ik\eta} \qquad\qquad \hbox{(CCMS)} \,,
\ee
and the correlators \pref{dSWk} and \pref{WkFTdefdSmass} are
\be \label{CCMScorrelatork}
  W_k(\eta,\eta') = \frac{H^2 \eta \eta'}{2k} \, e^{-ik(\eta - \eta')}   \qquad\qquad \hbox{(CCMS)} \,,
\ee
and
\be  \label{WkFTdefdSm0}
  W(\eta,\bfx; \eta',\bfy) = \frac{H^2 \eta\eta'}{4\pi^2 [-(\eta - \eta')^2 +(\bfx-\bfy)^2]} \qquad\qquad \hbox{(CCMS)}  \,.
\ee

The second special massless case is a minimially coupled massless scalar (MCMS) for which $(m,\xi) = (0,0)$ and so $\zeta^2 = 0$ and $\nu = \frac32$. In this case \pref{GendSMode} reduces to
\begin{eqnarray} \label{dSModes}
u_{k}(\eta) = ( - H \eta ) \;  \frac{e^{- i k \eta}}{\sqrt{ 2 k}} \left(1 - \frac{i}{k\eta} \right)  \qquad \qquad \hbox{(MCMS)}\,,
\end{eqnarray}
and so 
\be \label{dSWkMM}
  W_{\bf{k}}(\eta,\eta') =    u_{k}(\eta)  u^{\ast}_{k}(\eta')  =  \frac{H^2\eta\eta'}{2 k} \, e^{-ik(\eta-\eta')} \left(1 - \frac{i}{k\eta} \right)  \left(1 + \frac{i}{k\eta'} \right)  \qquad \hbox{(MCMS)}\,.
\ee

In this case the position-space correlator is not well-defined (though its derivatives are) because large IR fluctuations in de Sitter space contribute an additive IR-divergent constant. This divergence can be seen when the $k^{-3}$ dependence of \pref{dSWkMM} is integrated in the region $k \to 0$. Equivalently, it can be seen from the divergence of \pref{WkFTdefdSmass} as $\nu \to \frac32$ due to the $\sec(\pi \nu)$ prefactor. For $\zeta^2:=(m/H)^2 + 12 \xi$ small but nonzero we have $\nu  \simeq \frac32 - \frac13 \, \zeta^2 + \cO(\zeta^4)$ and \pref{WkFTdefdSmass} becomes 
\begin{eqnarray} \label{WkFTdefdSsmallmass}
  W(\eta,\bfx; \eta',\bfy) &\simeq&   -  \frac{H^2}{16\pi^2} \left\{  \frac{2}{ (\nu - \frac32)} + 3  +  \left[ \frac{z}{z-1}+2 \log (1-z) \right] _{z=z(\eta,\bfx;\eta',\bfy)} + \cO\left[ (\nu-\tfrac32 ) \right] \right\} \\
  &=&    \frac{H^2}{8\pi^2} \left\{  \frac{3}{ \zeta^2} - 2  +  \frac{2\eta\eta'}{-(\eta-\eta')^2+(\bfx-\bfy)^2} + \log \left[ \frac{4\eta\eta'}{-(\eta-\eta')^2+(\bfx-\bfy)^2} \right]  + \cO\left( \zeta^2 \right) \right\} \,. \nn
\end{eqnarray}
The leading term contains the divergence as $\zeta \to 0$ and corresponds to the well-known de Sitter result $\langle \phi^2(x) \rangle = 3H^4/(8\pi^2 m^2)$ for a light (but massive) minimally coupled scalar field prepared in its Bunch-Davies vacuum. These strong IR fluctuations are ultimately what frustrate the use of mean-field intuition for minimally coupled massless fields, leading to the breakdown of semiclassical perturbative methods at late times \cite{Tsamis:2005hd,Weinberg:2005vy,Boubekeur:2005fj,Sloth:2006az,Sloth:2006nu,Weinberg:2006ac,Seery:2007we,Seery:2007wf,Bartolo:2007ti,Burgess:2009bs, Burgess:2010dd, Giddings:2011zd, Burgess:2015ajz, Gorbenko:2019rza, Kaplanek:2019dqu, Kaplanek:2019vzj, Green:2020txs, Colas:2022hlq}. 

Another view of this divergence is found by asking how the correlation function \pref{WkFTdefdSmass} falls off as $|\bfx-\bfy| \to \infty$ (with $\eta$ and $\eta'$ held fixed). Appendix \ref{AppdSFreeCorr} shows this to be given by the asymptotic form 
\bea \label{WkFTdefdSmass5}
  W(\eta,\bfx; \eta',\bfy)  
  &\simeq&    \frac{H^2}{16\pi^2} \left[ \frac{\Gamma(\frac32+ \nu) \Gamma(-2\nu)}{ \Gamma(\frac12-\nu)} \left( \frac{4\eta\eta'}{\vert\bfx-\bfy\vert^2} \right)^{\frac32+\nu}\left(1 + \cdots \right)  \right. \\
  && \qquad \qquad\qquad \qquad \left. +  \frac{\Gamma(\frac32-\nu) \Gamma(2\nu)}{ \Gamma(\frac12+\nu)}  \left( \frac{4\eta\eta'}{\vert\bfx-\bfy\vert^2 }\right)^{\frac32-\nu} \left(1 + \cdots \right)  \right] \,, \nn 
\eea
as $|\bfx-\bfy| \to \infty$. Both terms fall like $|\bfx-\bfy|^{-3}$ when $\nu$ is imaginary ({\it e.g.}~when $m \gg H$). Eq.~\pref{WkFTdefdSmass5} predicts a falloff like $|\bfx-\bfy|^{-2}$ when $\nu = \frac12$, in agreement with expression \pref{WkFTdefdSm0} for a conformal scalar. But there is no falloff at all if $\nu = \frac32$, as appropriate for a massless minimally coupled scalar. For small nonzero $\zeta^2$ the falloff goes like $|\bfx-\bfy|^{-2\zeta^2/3}$, consistent with the logarithm of $|\bfx-\bfy|$ seen in \pref{WkFTdefdSsmallmass}.

\subsection{Decoherence through mixing}
\label{ssec:DecoMixingdS}

We now return to the main event, and compute the decoherence rate when the fields only mix through the potential $V_{\rm mix} = \mu^2 \phi \,\sigma$, but now do so prepared in the Bunch-Davies vacuum in de Sitter space. With this choice the environment correlator is $\cW_\bfk(\eta,\eta') = \mu^4 W_\bfk(\eta,\eta')$ with $W_\bfk(\eta,\eta')$ given in \pref{dSWk}. 

Consider first the simplest case where both system and environment fields are massless and minimally coupled. Then the correlator is given by \pref{dSWkMM}, and using this in \pref{PurityChangeMasterEqFflatdS} implies  
\bea  \label{dSpurityEvo}
 \partial_\eta \gamma_\bfk &=& -   \frac{ \mu^4 }{H^4 k^2 \eta^2} \int_{\eta_0}^\eta \frac{\exd \eta'}{{\eta'}^2} \; \hbox{Re}\left[ e^{-2ik(\eta-\eta')}  \left(1 - \frac{i}{k\eta} \right)^2  \left(1 + \frac{i}{k\eta'} \right)^2 \right]    \nn \\
    &=&  -    \frac{ \mu^4 }{3 H^4 k^6 \eta^7 } \hbox{Re}\left(  (k \eta-i)^2 \left\{-2 i (k\eta)^3  \left[ \text{Ei}(2 i k \eta) - \text{Ei}(2 i k \eta_0) \right] e^{-2 i k \eta} \phantom{\frac12} \right.\right.\\
    && \qquad\qquad \left.\left. + \Bigl( 1  -2 i k \eta + k^2 \eta^2  \Bigr) - \frac{\eta^3}{\eta_0^3} \Bigl(1  -2 i k  \eta_0 + k^2  \eta_0^2 \Bigr) e^{-2 i k (\eta - \eta_0)}   \right\} \right) \,, \nn
\eea
where the exponential integral function $\hbox{Ei}(z)$ is defined in \pref{AppEizDef}. 

As a check we take the $H\to 0$ limit of this expression and compare with our earlier flat-space result. Using $-k\eta = ({k}/{H}) \, e^{-Ht} \simeq ({k}/{H}) - k t + \cO(H)$ shows that the flat limit uses the asymptotic form when both $-k\eta$ and $- k \eta_0$ are large. The relevant asymptotic form for the exponential integral is given in \pref{AppEizImag} and leads to
\be  \label{dSpurityEvoFlat}
 \partial_t \gamma_\bfk   \simeq    -  \frac{ \mu^4 }{2 H^3 (k \eta)^6 } \hbox{Re}\left[ i (k \eta)^3 + (k \eta)^2 + i k \eta -1 -i (k^2 \eta^2-2 i k \eta -1)\frac{\eta^3}{\eta_0^3} \left(k\eta_0 + i \right)   e^{-2i k(\eta-\eta_0)} + \cdots \right]   \,.
\ee
Only the $(k\eta)^3$ and $(k \eta)^2 (k\eta_0)$ terms in the curly braces survive the $H\to 0$ limit, leaving
\bea  \label{dSpurityEvoFlatLimit}
 \partial_t \gamma_\bfk &=& -  \frac{ \mu^4 }{2 H^3 (k \eta)^3 } \hbox{Re}\left\{ i \left[1 - \frac{\eta^2}{\eta_0^2} \, e^{-2i k(\eta-\eta_0)}\right] + \hbox{($H$-suppressed)} \right\}  \nn\\
 &\to&-  \frac{ \mu^4 }{2 H^3 (k \eta)^3 } \hbox{Im} \, \Bigl[  e^{-2i k(\eta-\eta_0)} \Bigr] =    \frac{ \mu^4 }{2 k^3 }  \,\hbox{Im} \, \Bigl[  e^{-2i k(t-t_0)} \Bigr] \,,
\eea
in agreement with the massless limit of \pref{PurityChangeMasterEqFflat}.

Because we use Bunch-Davies states we must take $\eta_0 \to - \infty$. Using the asymptotic form \pref{AppEizImag} in this limit allows  \pref{dSpurityEvo} to be written 
\be  \label{dSpurityEvoBD}
 \partial_\eta \gamma_\bfk = -  \frac{ \mu^4 }{3 H^4 k^6 \eta^7 } \hbox{Re}\Bigl\{  (k \eta-i)^2 \Bigl[-2 i (k\eta)^3 \Bigl( \text{Ei}(2 i k \eta) + i \pi \Bigr) \, e^{-2ik\eta}  + \Bigl( 1  -2 i k \eta + k^2 \eta^2  \Bigr)    \Bigr] \Bigr\}   \,.
\ee
Switching to cosmic time and using $k\eta = - k/(aH)$ then leads to the result
\bea  \label{dSpurityEvoBDCT}
\frac{ \partial_t \gamma_\bfk}{H} &\simeq&   \frac{ \mu^4}{3 H^4 } \left( \frac{ aH}{ k } \right)^6\hbox{Re}\left\{  \left(\frac{k}{aH} +i \right)^2 \left[2 i \left( \frac{k}{aH} \right)^3 \Bigl( \text{Ei}[-2 i k /(aH)] + i \pi \Bigr) \, e^{2ik/(aH)}  \right. \right. \nn\\
 && \qquad\qquad\qquad\qquad\qquad\qquad\qquad\qquad\qquad \left. \left. + \left( 1  +  \frac{2 ik}{aH}  + \frac{k^2}{(aH)^2}  \right)    \right] \right\}   \,.
\eea

Eq.~\pref{dSpurityEvoBDCT} confirms that the decoherence rate is suppressed relative to the Hubble rate within the perturbative regime (for which $\mu \ll H$) and provided we focus on modes with $k \gsim aH$. But the rate grows dramatically once  $k \ll aH$ (super-Hubble modes). For instance, in the late-time super-Hubble limit $0 < - k\eta \ll 1$ the asymptotic form of \pref{dSpurityEvoBDCT} becomes
\be  \label{dSpurityEvoBDSH}
 \partial_\eta \gamma_\bfk \simeq  +  \frac{ \mu^4 }{3 H^4 k^6 \eta^7 } 
 \qquad \hbox{and so} \qquad
 \frac{\partial_t \gamma_\bfk }{H}   \simeq 
 - \frac{\mu^4 H^2}{3k^6}   \, e^{6Ht} \propto a^6(t)  \,.
\ee
This grows so strongly with time that it eventually invalidates the perturbative assumptions used in its derivation. Once this occurs a better method is required, such as is described in \S\ref{sec:OpenEFT} below.

\subsection{Decoherence through cubic interactions}

Consider next the case where system and environment interact with one another through the cubic interaction $V_c = g \phi^2 \sigma$. In this case the relevant environmental correlator is given by (compare with \pref{cWposFlat})
\be  \label{cWposdS}
  \cW(x,y)  =  2g^2 \Bigl[ W_{\rm env}(x , y) \Bigr]^2  
   =   \frac{g^2H^4}{128\pi^2} \left(\tfrac14 - \nu_{\rm env}^2 \right)^2 \sec^2(\pi \nu_{\rm env}) \left\lbrace  {}_2F_1\left[ \tfrac32 + \nu_{\rm env}, \tfrac32 - \nu_{\rm env}; 2; z(x,y) \right]   \right\rbrace^2
\ee 
where $W_{\rm env}(x,y)$ denotes the expression \pref{WkFTdefdSmass} evaluated with $m = m_{\rm env}$ and $\xi = \xi_{\rm env}$.  

\subsubsection{Evolution of total purity}

Using \pref{cWposdS} in the rate of evolution for the total purity \pref{PurityChangeMasterEqC} leads to 
\bea \label{PurityChangeMasterEqFPos}
  \partial_t \gamma 
   &=&  - 8g^2  \int_{t_0}^t \exd s \int a^3(t) \, \exd^3x \int a^3(s) \, \exd^3y \; \hbox{Re} \left\{W_{\rm sys}(t,\bfx;s,\bfy)  \Bigl[ W_{\rm env}(t,\bfx;s,\bfy)  \Bigr]^2\right\} \\
   &=& -8 g^2\, \cV(t) \int_{t_0}^t \exd s \; a^3(s) \int \exd^3y  \; \hbox{Re}  \left\{W_{\rm sys}(t,0;s,\bfy)  \Bigl[ W_{\rm env}(t,0;s,\bfy)  \Bigr]^2\right\}   \,,\nn
\eea
where `$s$' and `$e$' flag that the system and environment correlators can depend on different masses and nonminimal couplings. Translation invariance allows the $\exd^3 x$ integral to be performed trivially, giving the comoving volume $\cV_c$, in terms of which the physical volume is $\cV(t) = \cV_c \, a^3(t)$. Eq.~\pref{PurityChangeMasterEqFPos} underlines that it is the cosmic-time evolution rate per unit physical volume that has the sensible large-volume limit for translationally invariant systems. 

Performing the angular part of the $\exd^3 y$ integral gives
\bea \label{PurityChangeMasterEqPS}
  \partial_\eta \gamma   &=&   - \frac{ g^2 \cV_c}{128 \pi^2 H^2} \left(\tfrac14 - \nu_{\rm sys}^2 \right)  \left(\tfrac14 - \nu_{\rm env}^2 \right)^2 \sec(\pi \nu_{\rm sys})\sec^2(\pi \nu_{\rm env}) \\
  && \qquad \times  \int_{-\infty}^\eta \frac{\exd \eta' }{(\eta\eta')^4}  \int_0^\infty \exd y \; y^2  \; \hbox{Re} \left\{{}_2F_1\left[\tfrac32 + \nu_{\rm sys}, \tfrac32 - \nu_{\rm sys}; 2; z \right]  \Bigl[ {}_2F_1\left[\tfrac32 + \nu_{\rm env}, \tfrac32 - \nu_{\rm env}; 2; z \right] \Bigr]^2\right\} \,, \nn
\eea
where $z=z(\eta,\eta',y)$ is as given in \pref{zetadef} and we send $\eta_0 \to - \infty$ because we use Bunch-Davies correlators. It is useful to extract the hidden time-dependence in the argument of the hypergeometric functions by trading the $y$ integral for an integral over $z$, leading to
\bea \label{PurityChangeMasterEqPS2}
  \partial_\eta \gamma   &=&    - \frac{ g^2 \cV_c}{64 \pi^2 H^2} \left(\tfrac14 - \nu_{\rm sys}^2 \right)  \left(\tfrac14 - \nu_{\rm env}^2 \right)^2 \sec(\pi \nu_{\rm sys})\sec(\pi \nu_{\rm env})  \int_{-\infty}^\eta \frac{\exd \eta' }{(\eta\eta')^3}  \int_{-\infty}^{z_0} \exd z \; \sqrt{(\eta-\eta')^2+4\eta\eta'(1-z)} \nn \\
  && \qquad\qquad\qquad\qquad \times  \; \hbox{Re} \left\{{}_2F_1\left[\tfrac32 + \nu_{\rm sys}, \tfrac32 - \nu_{\rm sys}; 2; z \right]  \Bigl[ {}_2F_1\left[\tfrac32 + \nu_{\rm env}, \tfrac32 - \nu_{\rm env}; 2; z \right] \Bigr]^2\right\} \,, 
\eea
where the upper limit of the $z$ integration is
\be \label{z0def}
   z_0 := z(v=0) = 1 + \frac{(\eta-\eta')^2}{4\eta\eta'} \,.
\ee

The $y$ (or $z$) integral can diverge in the infrared {\it even when the particle masses are not zero}. To see why, recall that \pref{WkFTdefdSmass5} states the correlation function falls off for large $y$ like $(1/y)^{p}$ where $p = 3 - 2 \, \hbox{Re} \, \nu$. Convergence of the $y$ integral in \pref{PurityChangeMasterEqPS} -- or the $z$ integral in \pref{PurityChangeMasterEqPS2} -- therefore requires Re$(\nu_{\rm sys} + 2\nu_{\rm env}) < 3$. From here on we assume the environment to be a field with mass large enough to ensure convergence. For instance if the system field is massless and minimally coupled (so $\nu_{\rm sys} = \frac32$) then convergence requires Re $\nu_{\rm env}< \frac34$ and so $\zeta_{\rm env}^2 = (m_{\rm env}/H)^2 + 12 \, \xi_{\rm env}\geq \frac{27}{16}$.  

We see below that the decoherence per mode remains well-defined for nonzero wavenumber $\bfk$  for general $\nu_{\rm sys}$ and $\nu_{\rm env}$, but blows up as $k \to 0$ in a way that causes the same infrared divergence as seen above once all of the modes are summed. This IR divergence is an artefact of our use of non-derivative couplings and so is typically not shared by gravitational examples. The IR divergence also underlines how the long-distance properties of the quantum state on global de Sitter can become pathological  even when shorter-distance observables amenable to observation remain well-defined.

Once convergence is ensured the purity rate for small $\eta$ can be identified on very general grounds by dropping all subdominant powers of $\eta$ within the integrand, with $\sqrt{(\eta - \eta')^2 +4\eta\eta'(1-z)} \simeq |\eta'| = - \eta'$ in the small-$\eta$ limit. This leads to
\be \label{PurityChangeMasterEqPS3}
  \partial_\eta \gamma  \simeq  - \frac{C g^2 \cV_c}{64 \pi^3 H^2\eta^4} \left(\tfrac14 - \nu_{\rm sys}^2 \right)  \left(\tfrac14 - \nu_{\rm env}^2 \right)^2 \sec(\pi \nu_{\rm sys})\sec^2(\pi \nu_{\rm env})  \Bigl[ 1 + \cO(\eta) \Bigr] \,,
\ee
where
\be
 C(\nu_{\rm sys},\nu_{\rm env}) := \int_{-\infty}^{\infty} \exd z \; \hbox{Re} \left({}_2F_1\left[\tfrac32 + \nu_{\rm sys}, \tfrac32 - \nu_{\rm sys}; 2; z \right]  \left\{ {}_2F_1\left[\tfrac32 + \nu_{\rm env}, \tfrac32 - \nu_{\rm env}; 2; z \right] \right\}^2\right) \,.
\ee
The integrand of this expression has singularities at $z =1$ and $z \to \infty$ and the imaginary part of time together with \pref{zetadef} shows that the $z$ contour negotiates these by passing just below the real axis (with a small negative imaginary part). 

The cosmic time evolution rate per unit physical volume therefore becomes
\bea \label{PurityChangeMasterEqPS3t}
  \frac{\partial_t \gamma}{\cV(t)} &\simeq& -  \frac{C g^2 H^2}{64 \pi^3} \left(\tfrac14 - \nu_{\rm sys}^2 \right)  \left(\tfrac14 - \nu_{\rm env}^2 \right)^2 \sec(\pi \nu_{\rm sys})\sec^2(\pi \nu_{\rm env})  \Bigl[ 1 + \cO(\eta) \Bigr] \,,
\eea
and so for small $\eta$ becomes approximately time-independent -- and perturbatively small in Hubble units (Hubble rate per Hubble volume) provided $g \ll  H$.  

\subsubsection{Fourier-space result (conformal environment)}
\label{ssec:ModeByModedS}

Similar considerations give the decoherence as a function of mode number $\bfk$, and the $k$-dependence of the result helps explain the origins of the infrared divergence encountered above. The starting point of this calculation is \pref{PurityChangeMasterEqFflatdS},
\be  \label{PurityChangeMasterEqFflatdS2}
  \partial_\eta \gamma_\bfk  =  - 4  \int_{\eta_0}^\eta \frac{\exd \eta'}{ (H^2\eta\eta')^4} \hbox{Re}\Bigl[ W_k(\eta,\eta') \, \cW_{\bfk}(\eta,\eta') \Bigr]   \,,
\ee
with the environment correlator given by either of the equivalent expressions
\be   \label{WkCompodS}
\cW_{\bf{k}}(\eta,\eta')   =   2g^2 \int \exd^3y  \; e^{i  \bf{k} \cdot \bfy}  \Bigl[ W_{\rm env}(\eta, 0;\eta', \bfy) \Bigr]^2 
=    2g^2 \int \frac{\exd^3p}{(2\pi)^3}\; W_{|\bfk-\bfp|}(\eta,\eta') W_{\bfp}(\eta,\eta')   \,,
\ee 
where $W_{\rm env}(\eta,\bfx; \eta', \bfy)$ is given by \pref{WkFTdefdSmass} and $W_\bfp(\eta,\eta')$ given by \pref{dSWk}. 

These expressions can be made explicit in the special case where the environment is a conformally coupled massless scalar ($\nu_{\rm env}= \frac12$). In this case the integrals converge (unlike for a minimally coupled massless scalar) but the environmental correlator is quite simple (see Eq~(\ref{AppcWkdScubicc})):
\be \label{cWkdScubicc}
 \cW_{-\bfk}(\eta, \eta')  = -\frac{ig^2( H^2 \eta \eta')^2}{2(2\pi)^2(\eta - \eta')} \, e^{-ik(\eta - \eta')}  \,.
\ee
With this choice \pref{PurityChangeMasterEqFflatdS2} evaluates to
\bea  \label{PurityChangeMasterEqFflatdS2w}
  \partial_\eta \gamma_\bfk^{\rm conf} 
  &=& - \frac{g^2 }{2\pi^2}  \int_{\eta_0}^\eta \frac{\exd \eta'}{ (H^2\eta\eta')^2} \hbox{Im}\left[ W_\bfk(\eta,\eta') \, \frac{ e^{-ik(\eta - \eta')}}{\eta - \eta'} \right]  \\
  &=&- \frac{g^2}{8\pi H}   \int_{\eta_0}^\eta \frac{\exd \eta'}{ (H^2\eta\eta')^{1/2}} \hbox{Im}\left\{  H^{(1)}_{\nu_{\rm sys}}(-k\eta) \left[ H^{(1)}_{\nu_{\rm sys}}(-k\eta') \right]^* \frac{ e^{-ik(\eta - \eta')}}{\eta - \eta'} \right\}  \,.\nn
\eea
Further progress comes if we use \pref{iepsExp} to write
\be \label{0+Identity}
   \frac{1}{\eta - \eta' - i0^+} = \cP \left( \frac{1}{\eta - \eta'} \right) + i \pi \delta(\eta - \eta') \,,
\ee
where coordinates on the right-hand side can be regarded as real when taking the imaginary part. This leads to
\bea  \label{PurityChangeSingParty}
  \partial_\eta \gamma^{\rm conf}_\bfk &=& \frac{g^2}{16 H^2\eta} \left|  H_{\nu_{\rm sys}}^{(1)}(-k\eta) \right|^2  - \frac{g^2}{8\pi H}   \int_{-\infty}^\eta \frac{ \exd \eta' }{(H^2 \eta\eta')^{1/2}} \\
&& \qquad\qquad\qquad\qquad\times \cP \left(\frac{1}{\eta-\eta'} \right) \hbox{Im} \left\{ e^{-ik(\eta - \eta')} \;  H^{(1)}_{\nu_{\rm sys}}(-k\eta)  \left[ H^{(1)}_{\nu_{\rm sys}}(-k\eta') \right]^* \right\}   \,.  \nn
\eea
Notice that the principal part integral in the final line  converges as $\eta'\to\eta$ because the imaginary part must vanish as $\eta' \to \eta$ due to its antisymmetry under $\eta \leftrightarrow \eta'$. 

The principal part contribution is subdominant in the small $k\eta$ regime (as is verified explicitly in \pref{AppPurityChangeMasterEqFflatdS2w22} for the case when the system is a minimally coupled massless field). In this super-Hubble regime the small-$k \eta$ and small-$k\eta'$ limit of the Hankel functions can be used, allowing \pref{PurityChangeSingParty} to be written as
\bea  \label{PurityChangeSingPart2}
   \partial_\eta \gamma_\bfk^{\rm conf} 
  &\simeq& - \frac{g^2}{8\pi^3H^2} \Bigl| 2^{\nu_{\rm sys}} \Gamma(\nu_{\rm sys}) \Bigr|^2 \int_{-\infty}^\eta \frac{ \exd \eta'}{\sqrt{\eta \eta' }}  \;  \hbox{Im} \left[ \frac{e^{-ik(\eta - \eta')}}{\eta - \eta'} \;  (k^2 \eta \eta')^{ - \nu_{\rm sys}}\right]  \\
    &\simeq&   \frac{g^2}{16\pi^2H^2\eta} \Bigl| 2^{\nu_{\rm sys}} \Gamma(\nu_{\rm sys}) \Bigr|^2 (-k\eta)^{-2\nu_{\rm sys}} \Bigl[1 + \cO(\eta) \Bigr] \,, \nn
\eea
which assumes $\nu_{\rm sys}$ is real. This agrees with the leading part of \pref{AppPurityChangeMasterEqFflatdS2w22} in the limit $\nu_{\rm sys} \to \frac32$. Notice also that  the small-$\eta$ limit of \pref{PurityChangeSingPart2} is consistent with the $\eta^{-4}$ dependence of \pref{PurityChangeMasterEqPS3} for any $\nu_{\rm sys}$ as may be seen by rescaling the integration variable in any convergent integral of the form
\be \label{k_to_ell}
   \int \frac{\exd^3 k}{(2\pi)^3} f(k\eta) = \eta^{-3} \int \frac{\exd^3 l}{(2\pi)^3} f(l) \propto \eta^{-3}  \,.
\ee

\subsubsection{Fourier-space result (general environment)}
\label{sssec:FourierGeneral}

We next consider more general choices for the environmental field. In this case the explicit integrations required to get the purity evolution are more difficult to perform. We start by summarizing a numerical exploration of systems where the environment is not a conformal scalar and then provide an analytic understanding of this behaviour. 

\subsubsection*{Numerical evaluation}

\begin{figure}[h]
\begin{center}
\includegraphics[width=0.48 \linewidth]{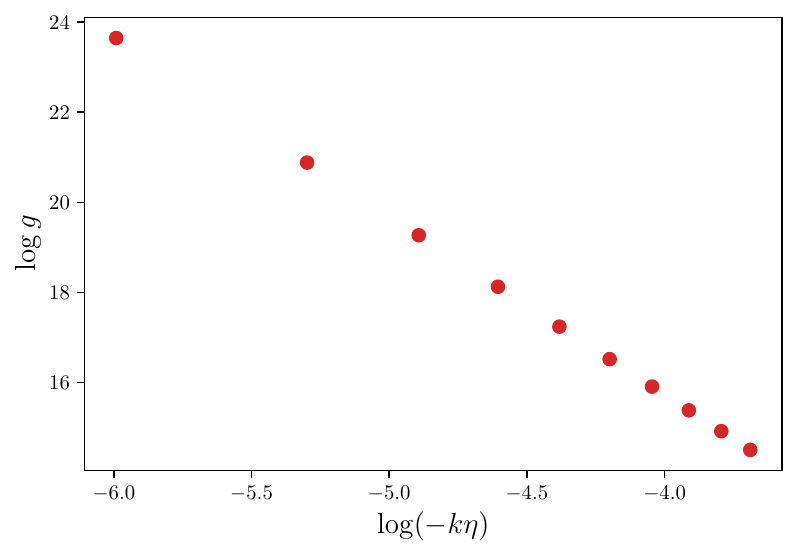}
\includegraphics[width=0.5 \linewidth]{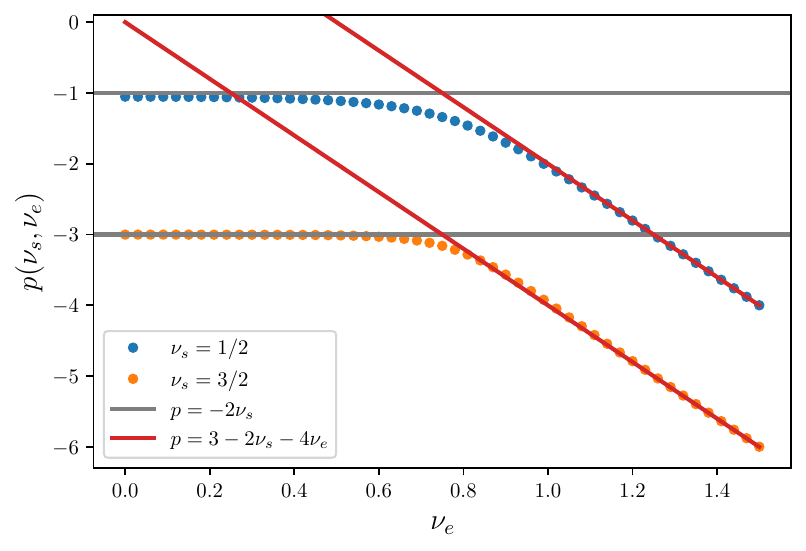}
\caption{\small Plots summarizing the numerical evidence for power-law behaviour of $1 - \gamma_\bfk(\eta)$. As argued in Appendix \ref{App:PT_mass}, the function $g(- k \eta, a )$ with parameter $0 < a  < 1$ defined in (\ref{gnu_def}) has the same late-time scaling as $1 - \gamma_\bfk(\eta)$, although is significantly less costly to numerically integrate which is why we use it as a proxy for the purity's scaling here such that $1 - \gamma_\bfk(\eta) \propto g(\eta,a) \propto (-k \eta)^p$. The left panel considers a minimally coupled massless scalar with $\nu_{\rm sys} = \frac32$ and plots the value of $g \propto 1 - \gamma_\bfk$ as a function of $-k\eta$, where a straight line indicates a power-law evolution. The right panel shows the power $p(\nu_{\rm sys},\nu_{\rm env})$ inferred from graphs like the left panel as a function of the environment particle's mass-dependent parameter $\nu_{\rm env}$, done for two choices $\nu_{\rm sys} = \frac{1}{2}, \frac{3}{2}$. Notice the crossover between $\gamma_\bfk \propto (-k\eta)^{-2\nu_{\rm sys}}$ for Re $\nu_{\rm env}< \frac34$ and $\gamma_\bfk \propto (-k\eta)^{3 - 2\nu_{\rm sys} -4\nu_{\rm env}}$ for Re $\nu_{\rm env}> \frac34$. The fact that the crossover is continuous and not sudden is a numerical artifact, since we numerically probed values of $-k\eta \sim \mathcal{O}(10^{-3})$ which are not that small.
} \label{Fig:NumericalEnvironment}
\end{center}
\end{figure}

Fig.~\ref{Fig:NumericalEnvironment} plots the results of a numerical evaluation for the scaling of $1-\gamma_\bfk(\eta)$ as a function of conformal time $\eta$. The figure assumes the system is either conformal ($\nu_{\rm sys} = \frac12$) or massless minimally coupled $\nu_{\rm sys} = \frac32$ while exploring different values of $\nu_{\rm env}$ for the environment. The figure's left-hand panel plots the scaling of the purity in $-k\eta$ in the super-Hubble regime $- k\eta \ll 1$ and shows evidence for power-law behaviour $1 - \gamma_\bfk \propto (-k\eta)^p$ (because the figure is a straight line for logarithmic axes). 

The power $p$ can be obtained from the slope of this line and the result of such a determination is shown as a function of Re $\nu_{\rm env}$ in the right panel of Fig.~\ref{Fig:NumericalEnvironment}, which is done for the two cases of $\nu_{\rm sys} = \frac{1}{2}, \frac{3}{2}$. This plot agrees with the prediction $p = -2 \nu_{\rm sys}$ seen in eqs.~\pref{PurityChangeMasterEqPS3} as well as \pref{PurityChangeSingPart2} in the special case of a conformal environment $\nu_{\rm env}= \frac12$ and minimally coupled massless system $\nu_{\rm sys} = \frac32$, but the figure shows that this prediction applies more broadly for Re$\,\nu_{\rm env}\lsim 0.6$ (including when $\nu_{\rm env}$ is pure imaginary). For Re $\nu_{\rm env}\gsim 0.8$ the figure shows a crossover to a new regime $p \simeq3 - 2 \nu_{\rm sys} -4\nu_{\rm env}$, with $p \to -6$ for the minimally coupled massless scalar ($\nu_{\rm env}\to \frac32$), in agreement with the power found in \S\ref{ssec:DecoMixingdS} for decoherence due to mixing between minimally coupled massless fields in de Sitter. 

The crossover in the power $p$ at $\nu_{\rm env}\simeq \frac34$ is related to the infrared convergence of the total purity $\gamma(\eta,-\infty)$, obtained from $\gamma_\bfk$ by integrating over $\bfk$. Although the Fourier component $\gamma_\bfk$ is infrared finite for all Re$\,\nu_{\rm env}< \frac32$, the integral $\exd^3 k$ of $\gamma_\bfk \propto k^{-4\nu_{\rm env}}$ is only infrared finite when Re$\,\nu_{\rm env}< \frac34$. This is related to the condition for infrared convergence discussed below eq.~\pref{z0def} when $\gamma$ was instead obtained as an integral over position. 

\subsubsection*{Analytical discussion}

The asymptotic small-$\eta$ behaviour of $1-\gamma_\bfk$ can also be understood analytically by referring back to expression \pref{PurityMasterEqFNestedCubic}, which in an expanding universe with an initially pure state becomes
\bea \label{PurityMasterEqFNestedMixCosmo}
   \gamma_\bfk(\eta)   &=&  1 - \frac{g^2}{2\pi^3} \int \exd^3 p \, \exd^3q \; \Bigl| \cN_{pqk}(\eta,\eta_0)  \Bigr|^2 \delta^3(\bfp+\bfq-\bfk) \nn\\  \hbox{with} \qquad
   \cN_{pqk}(\eta,\eta_0) &:=& \int_{\eta_0}^\eta \exd s \;  a^4(s) \, u_k(s) \, v_p(s) \, v_q(s) \,,
\eea 
and $u_k(s)$ and $v_p(s)$ respectively representing the system and environmental mode functions.  Specializing to the de Sitter case we take $\eta_0 \to -\infty$ and use the Bunch-Davies modes given by \pref{GendSMode}, in which case 
\be
  \gamma_\bfk = 1 - \frac{g^2}{128 H^2} \, \cF(-k\eta) \,,
\ee
with the dimensionless function $\cF(x)$ defined by
\bea \label{PurityMasterEqFNestedMixCosmo321}
   \cF(-k\eta) &:=& \int \exd^3 p \, \exd^3q \; \Bigl| \widehat \cN_{pqk}(\eta)  \Bigr|^2 \delta^3(\bfp+\bfq-\bfk) \\
   \hbox{with} \qquad 
  \widehat \cN_{pqk}(\eta)  &=& \int_{-\infty}^\eta \exd s \;  \sqrt{ - s } \; H_{\nu_{\rm sys}}^{(1)}(- k s) H_{\nu_{\rm env}}^{(1)}(- p s) H_{\nu_{\rm env}}^{(1)}(- q s)  \nn\, .
\eea
We see from this that $1-\gamma_\bfk$ remains bounded as $\eta \to 0$ if the $\exd s$ integral in \pref{PurityMasterEqFNestedMixCosmo321} converges in this limit, and this convergence is controlled by the small-argument limit of the Hankel functions: $H_\nu(z) \propto z^{- r}$ for $r = \hbox{Re}\,\nu$ (see \pref{AppHankelSmall}). Convergence as $\eta \to 0$ therefore requires Re$\,(2\nu_{\rm env}+\nu_{\rm sys}) <  \frac32$ (and so never happens when $\nu_{\rm sys} = \frac32$ but is possible for other choices of $\nu_{\rm env}$ and $\nu_{\rm sys}$).

When Re$\,(2\nu_{\rm env}+\nu_{\rm sys}) \geq  \frac32$ the $\exd s$ integral does not converge as $\eta \to 0$ and the small-$\eta$ form of $\widehat \cN_{pqk}(\eta)$ is  
\be \label{WidehatN}
   \Bigl|\widehat\cN_{pqk}(\eta) \Bigr|^2 \propto \frac{1}{k^3} \left( - p \eta \right)^{-2\nu_{\rm env}} \left( - q \eta \right)^{-2\nu_{\rm env}} \left( - k \eta \right)^{3 - 2\nu_{\rm sys}} \,.
\ee
This must be integrated over $\exd^3p$ and $\exd^3q$ as in \pref{PurityMasterEqFNestedMixCosmo321} and the final $\eta$-dependence of the result depends on whether or not this integration converges in the IR. Performing the $\exd^3q$ integral using the delta function implies $q = |\bfk-\bfp|$ and so for nonzero $k$ we see that the $\exd^3p$ integration converges near $p \to 0$ when Re $\nu_{\rm env}< \frac32$ ({\it i.e.}~always apart from the edge case of a massless minimally coupled environment). But for $k = 0$ the integral is more strongly IR divergent because $q \to 0$ as $p \to 0$ and convergence instead requires Re $\nu_{\rm env}< \frac34$.  

Now comes the main point. When Re $\nu_{\rm env}< \frac34$ the convergence of the momentum integrals implies the purity varies like $(-k \eta)^{-2\nu_{\rm sys}}$ as $- k\eta \to 0$ -- in agreement with \pref{PurityChangeSingPart2} and the numerics presented in Figure \ref{Fig:NumericalEnvironment}. As discussed in Appendix \ref{App:PT_mass} the scaling in this case is given by 
\be  \label{PurityMasterEqFNestedMixCosmo2}
   \cF(-k\eta)  \simeq  A(\nu_{\rm sys}, \nu_{\rm env}) \; (-k\eta)^{-2\nu_{\rm sys}} \qquad (\mathrm{when} \ \mathrm{Re}[\nu_{\rm env}] < \tfrac{3}{4})  \,,
\ee 
with the coefficient $A(\nu_{\rm sys}, \nu_{\rm env})$ given in Eq.~\pref{A_coeff_def}. 

\begin{figure}[h]
\begin{center}
\includegraphics[width=0.49 \linewidth]{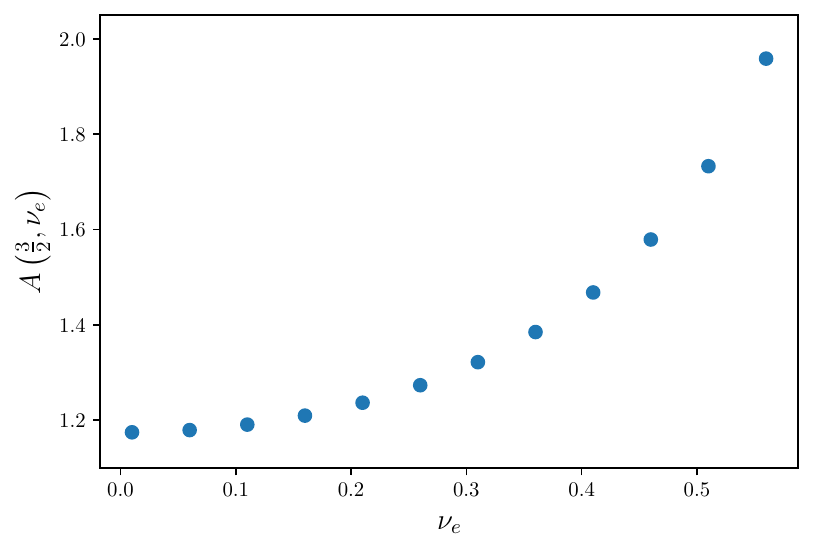}
\includegraphics[width=0.49 \linewidth]{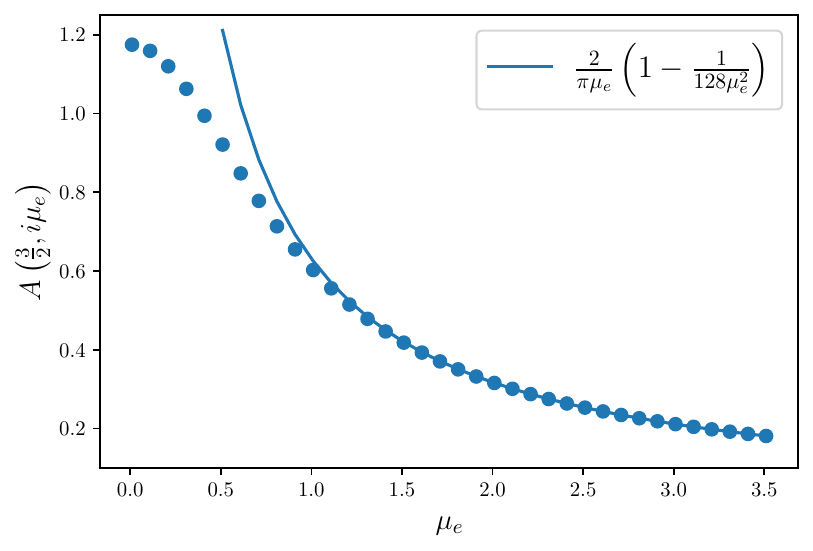}
\caption{\small Dots show the result of numerical evaluation of the coefficient $A(\nu_{\rm sys},\nu_{\rm env})$ as a function of Re$\,\nu_{\rm env}$ (left panel) and $\mu_{\rm env}= \hbox{Im}\,\nu_{\rm env}$ (right panel) for the specific case $\nu_{\rm sys} = \frac32$. The solid line shows the asymptotic large-mass form given in eq.~\pref{Aasymptotic} of the text.} \label{Fig:NumericalCoeff}
\end{center}
\end{figure}

For the special case of $\nu_{\rm sys} = \frac32$ the coefficient $A(\nu_{\rm sys},\nu_{\rm env})$ can be written as (see Eq.~\pref{A32_App})
\be \label{A32}
A(\tfrac{3}{2}, \nu_{\rm env}) = \int_{0}^\infty \exd x \; x^2 \; \left| \; \frac{x}{4 \nu_{\rm env}} \left[ H_{\nu_{\rm env}}^{(1)}\big(\tfrac{x}{2} \big) \frac{\partial H_{\nu_{\rm env}+1}^{(1)}\big(\tfrac{x}{2}\big) }{\partial \nu_{\rm env}} - \frac{\partial H_{\nu_{\rm env}}^{(1)}\big(\tfrac{x}{2} \big) }{\partial \nu_{\rm env}} H_{\nu_{\rm env}+1}^{(1)}\left(\tfrac{x}{2} \right) \right] - \frac{\big[ H_{\nu_{\rm env}}^{(1)}\big(\tfrac{x}{2} \big) \big]^2}{2\nu_{\rm env}} \; \right|^2\ .
\ee
In the limit where $\nu_{\rm env}= i \mu_{\rm env}$ with $\mu_{\rm env} \gg 1$ this leads to the asymptotic form (see Eq.~\pref{Aint_asympt_mu})
\be \label{Aasymptotic}
  A\left( \tfrac32, i \mu_{\rm env}\right) \simeq \frac{2}{\pi \mu_{\rm env}} \left[ 1 - \frac{1}{128 \mu_{\rm env}^2} + \cO(\mu_{\rm env}^{-4} ) \right] \,,
\ee
which vanishes only as a power of $1/m$ because this is equivalent to using the ordering of limits where $\epsilon \to 0$ before $\mu_{\rm env}$ is taken to be very large -- see the discussion below eq.~\pref{puritydecoupling}.  Figure \ref{Fig:NumericalCoeff} shows the results of a numerical evaluation of $A(\nu_{\rm sys} ,\nu_{\rm env})$ as a function of $\nu_{\rm env}$ when $\nu_{\rm sys} = \frac32$ is chosen (as appropriate for a massless minimally coupled system field). The left panel of the figure shows the result when $\nu_{\rm env}$ is real and the right panel shows the result as a function of $\mu_{\rm env}$ when $\nu_{\rm env}= i \mu_{\rm env}$ is pure imaginary. Also plotted is the approximate expression \pref{Aasymptotic} for large $\mu_{\rm env}$, which works well even for $\mu_{\rm env}$ not much larger than unity. 

On the other hand, for $\frac34 < \hbox{Re}\,\nu_{\rm env}< \frac32$ the integral converges for $k$ nonzero but diverges as $k \to 0$, showing that $k$ acts to regulate the would-be IR divergence. In fact, using Eq.~\pref{B_coeff_result} the leading order behaviour in this case can be written down explicitly, giving
\bea \label{B_coeff}
\cF(k\eta) &\simeq& B(\nu_{\rm sys},\nu_{\rm sys}) \; (-k\eta)^{3-4\nu_{\rm env}-2\nu_{\rm sys}} \qquad (\mathrm{when} \ \tfrac34 < \nu_{\rm env}< \tfrac32) \nn\\
\hbox{with} \quad 
B(\nu_{\rm sys},\nu_{\rm sys}) &=& \frac{ 4^{\nu _{\rm sys}+ 4 \nu _{\rm env}-1} \Gamma^2(\nu _{\rm sys}) \Gamma^4(\nu _{\rm env}) \tan \left(\pi  \nu _{\rm env}\right) \sec \left(2 \pi  \nu _{\rm env}\right)  }{\pi ^{7/2} (3 - 2 \nu_{\rm sys} - 4 \nu_{\rm env})^2 \left(\nu _{\rm env}-1\right) \Gamma \left(\frac{5}{2}-2 \nu _{\rm env}\right) \Gamma \left(2 \nu _{\rm env}-1\right) } \ .
\eea

\subsubsection*{Summary for general environment}

The upshot from the above examples is that the cosmic-time evolution rate for the per-mode purity can be written
\be
  \partial_t \gamma_\bfk = -  \frac{g^2}{8\pi H} \, G\left( \nu_{\rm env}, \nu_{\rm sys}, \frac{k}{aH}\right) 
\ee
where $G(\nu_{\rm env}, \nu_{\rm sys}, x)$ is a dimensionless function that is obtained in principle by evaluating the integrals in \pref{PurityChangeMasterEqFflatdS2} and \pref{WkCompodS}. $G$ behaves much as in flat space and so does nothing remarkable for sub-Hubble modes and as a result the decoherence rate remains much smaller than the Hubble expansion rate within the perturbative regime $g \ll H$. 

If  Re$\,(2\nu_{\rm env}+\nu_{\rm sys}) < \frac32$ that is pretty much the end of the story. But if Re$\,(2\nu_{\rm env}+\nu_{\rm sys}) \geq  \frac32$ then things change once the mode in question becomes super-Hubble because in this case the function $G$ begins to grow dramatically with time, with the asymptotic form 
\be \label{Glateform}
  G\left(\nu_{\rm env}, \nu_{\rm sys} , \frac{k}{aH} \right) \propto  \begin{cases} \left( {aH}/{k} \right)^{2\nu_{\rm sys}} & \hbox{if Re$\,\nu_{\rm env}\lsim \frac34$} \cr  \left( {aH}/{k} \right)^{2\nu_{\rm sys} + 4\nu_{\rm env}- 3} & \hbox{if Re$\,\nu_{\rm env}\gsim \frac34$}\,. \end{cases}
\ee
Because $a = e^{Ht}$ this shows exponential growth in time for light system fields. Once this growth begins to overwhelm the small prefactors of $g/H$ the perturbative methods used up to this point break down and other methods are required, as we discuss in \S\ref{sec:OpenEFT}.

\section{Open EFTs and late-time resummation}
\label{sec:OpenEFT}

This section describes how to make reliable late-time inferences about the purity evolution given that we largely only have access to perturbation theory. We do so using Open EFT techniques \cite{Koks:1996ga, Lombardo:2004fr, Lombardo:2005iz, Burgess:2006jn, Anastopoulos:2013zya, Fukuma:2013uxa, Burgess:2014eoa, Agon:2014uxa, Burgess:2015ajz, Boyanovsky:2015xoa, Boyanovsky:2015jen, Boyanovsky:2015tba, Braaten:2016nyq, Braaten:2016sja, Nelson:2016kjm, Hollowood:2017bil, Shandera:2017qkg, Boyanovsky:2018fxl, Boyanovsky:2018soy, Martin:2018zbe, Bohra:2019wxu, Akhtar:2019qdn, Kaplanek:2019dqu, Brahma:2020zpk, Kaplanek:2020iay, Rai:2020edx, Burgess:2021luo, Kaplanek:2021fnl, Brahma:2021mng, Banerjee:2021lqu, Oppenheim:2022xjr, Brahma:2022yxu, Kaplanek:2022xrr, Kaplanek:2022opa, Colas:2022hlq, Colas:2022kfu, DaddiHammou:2022itk, Burgess:2022nwu, Burgess:2022rdo, Cao:2022kjn, Prudhoe:2022pte, Colas:2023wxa, Brahma:2023hki, Sharifian:2023jem, Alicki:2023tfz, Alicki:2023rfv, Ning:2023ybc}
to resum the late-time behaviour of the purity evolution that allows the exploration of the late-time growth of super-Hubble modes  (see \cite{breuerTheoryOpenQuantum2002} for a review of open quantum systems and \cite{Burgess:2022rdo} for a recent review of these techniques in a gravitational context).  

\subsection{Lindblad evolution}

Open EFT techniques also start with the underlying Liouville evolution \pref{Liouville} of the full theory, whose consequences are again explored perturbatively. But the idea is to integrate them perturbatively without directly passing through the intermediate step of \pref{rhoevoF} on which we have hitherto relied. 

To this end one sets up and solves the Liouville equation for the reduced density matrix of the system as well as for its complement (the projection of the full density matrix onto the environmental degrees of freedom). Both of these evolutions in general depend on both the system and environmental variables. Because the Liouville equation is linear the equations of motion for the environment can be formally integrated (usually within perturbation theory) as a function of the system variables with the result re-substituted back into the evolution of the system
\cite{Nakajima:1958pnl, Zwanzig:1960gvu}. 

This procedure leads to an evolution equation for the system's density matrix $\varrho(t)$ that is nonlocal in time (see \cite{Burgess:2020tbq} and \cite{breuerTheoryOpenQuantum2002} for a review). The result can be given quite explicitly in perturbation theory \cite{Nakajima:1958pnl, Zwanzig:1960gvu}. When the system-environment interaction factorizes
\be \label{VvsAB}
   V_{\rm int}(t) = \sum_n A_n (t) \otimes B_n(t) \,,
\ee
with $A_n$ acting only within the system's Hilbert space and $B_n$ acting only on the environment -- as is the case for both \pref{int=mix} and \pref{int=c} -- and if system and environment are chosen to be uncorrelated at the initial time $t_0$ -- as we have done above -- then the resulting evolution equation for $\varrho(t)$ reduces (in interaction picture) to the following approximate expression:
\bea \label{NakaZwanExplicit}
  \partial_t \varrho(t) &=& - i \sum_n \Bigl[ A_n(t) , \varrho(t) \Bigr] \, \llangle \; B_n(t) \; \rrangle 
  + (-i )^2 \sum_{mn} \int_{t_0}^t \exd s \Biggl\{ \Bigl[ A_m(t) , A_n(s) \,\varrho(s) \Bigr] \, \llangle \, \delta B_m(t) \, \delta B_n(s) \, \rrangle \nn  \\
  && \qquad\qquad\qquad\qquad\qquad\qquad   -  \Bigl[ A_m(t) , \varrho(s)\,A_n(s)  \Bigr] \, \llangle \, \delta B_n(s) \,\delta B_m(t) \, \rrangle \,\Biggr\} + \cO(V_{\rm int}^3)\,.
\eea
This stops at second order in $V_{\rm int}$ and (as before) $\llangle \,\cdots\, \rrangle = \trB[(\cdots) \,\rho_{\rm env}]$ for averages over the environment, with $\delta B_n := B_n - \llangle \, B_n \, \rrangle\,$. Because the first-order term in \pref{NakaZwanExplicit} describes Hamiltonian evolution using $V_{\rm eff} = \llangle \, V_{\rm int} \, \rrangle$ it cannot contribute to decoherence.  Our focus is therefore on the second-order term that provides the leading decoherence effect. 

Notice that if $\varrho(s)$ on the right-hand side is regarded as differing from its initial value $\varrho_0$ by terms that are suppressed by $V_{\rm int}$ then we can replace $\varrho(s) \to \varrho_0$ and \pref{NakaZwanExplicit} reduces to \pref{rhoevoF}. It is tempting to therefore think that the same reasoning should therefore make it a good approximation to replace $\varrho(s)$ on the right-hand side with $\varrho$ evaluated at any other time -- such as $\varrho(t)$, for instance (which would considerably simplify finding solutions). The fallacy in this argument is that it assumes $\varrho(s)$ and $\varrho(t)$ must be perturbatively different for {\it all} times. What makes this reasoning false is that perturbative time evolution {\it always} fails at {\it late} times: no matter how small $V_{\rm int}$ is relative to $H_0$, there is always a time for which $e^{-i(H_0+V_{\rm int})t}$ is not well-approximated by $e^{-iH_0t}(1 - i V_{\rm int} t)$. 

A useful alternative version of \pref{NakaZwanExplicit} that differs from \pref{rhoevoF} requires something in addition to simply perturbing in $V_{\rm int}$. That something is often the existence of a hierarchy of scales, and in this case that hierarchy is provided by the ratio between the Hubble scale (which controls the size of environmental correlations) and the time-scale $\tau$ of decoherence (which in the examples above is controlled by $\mu\ll H$ or $g \ll H$). If the correlation function $\llangle \, \delta B_m(s) \,\delta B_n(t) \, \rrangle$ falls off sharply enough once $H|t-s| \gg 1$ then for evolution over times longer than $H^{-1}$ we can Taylor expand the rest of the integrand in \pref{NakaZwanExplicit} about $s = t$ with successive terms suppressed to the extent that $(H \partial_t)^n \ll 1$ when acting on the rest of the integrand. This becomes a suppression by powers of $(H\tau)^{-1}$ if $\tau$ is the time-scale that governs the long-time evolution. 

When this expansion is a good approximation the leading evolution is given by
\be \label{LindbladH}
  \partial_t \varrho  \simeq - i   \Bigl[ V_{\rm eff}  \, , \varrho   \Bigr] -  \sum_{mn} \kappa_{mn} \Bigl[  \Bigl\{ A_{m} A_{n}, \varrho \Bigr\} - 2\,A_n \varrho  A_m \Bigr] \,,
\ee 
where  the coefficients and operators on the right-hand side are all evaluated at the same time, $t$, as for the left-hand side and
\be \label{VeffDef}
  V_{\rm eff} :=  \sum_n A_n \, \llangle \, B_n \, \rrangle + \sum_{mn} h_{mn} A_{m} A_{n}   \,.
\ee 
The coefficients are $\kappa_{mn} =\tfrac12 (C_{mn} + C_{nm}^*)$ and $h_{mn} =- \tfrac{i}2 (C_{mn} - C_{nm}^*)$ where
\begin{equation}\label{CDefEq}
 C_{mn}(t) := \int_{t_0}^t \exd s \  \llangle \, \delta B_m(t)\, \delta B_n(s) \, \rrangle \,.
\end{equation}
These satisfy $\kappa_{nm}^* = \kappa_{mn}$ and $h_{nm}^* = h_{mn}$ (and $h_{mn}$ would vanish if the $C_{mn}$'s were hermitian). Eq.~\pref{LindbladH} is intuitive inasmuch as it is also what would have been obtained if the environmental correlations were local in time: $\llangle \, \delta B_m(t)\, \delta B_n(s) \, \rrangle \simeq C_{mn}(t) \, \delta(t-s)$. 

Eq.~\pref{LindbladH} has the {\em Lindblad} --- or GKSL (Gorini, Kossakowski, Sudarshan \& Lindblad) \cite{Burgess:2020tbq, Gorini:1976cm} --- form that is known to preserve the positivity and normalization of $\varrho$ provided $\kappa_{mn}$ is hermitian and positive definite (as it must therefore be if derived using a reliable approximation within a sensible theory). For the present purposes two things about it are useful. First, it is easier to integrate than is \pref{NakaZwanExplicit} because it is Markovian in the sense that $\partial_t\varrho (t)$ depends only on other variables at time $t$ and not on the entire history of evolution prior to this time. Second -- as reviewed in \cite{Burgess:2022rdo, Burgess:2020tbq} -- it can be used to resum late-time behaviour because its solutions can be trusted over much longer timescales than can those of \pref{NakaZwanExplicit} provided that the coefficients $\kappa_{mn}$ and $h_{mn}$ do not depend explicitly on $t_0$. 

All of these nice consequences flow from the existence of a hierarchy of scales, but how peaked must the correlation function  $\llangle \, \delta B_n(s) \,\delta B_m(t) \, \rrangle$ be in order for them to be true? Given a specific environment the correlation function can be computed; is there a simple criterion that when satisfied justifies the Lindblad limit? One might think that perturbation theory in $V_{\rm int}$ itself might justify this limit because within the interaction picture this is what controls the system's evolution rate $\partial_t \varrho$. The smaller this is the more slowly $\varrho$ varies and one might hope that the expansion in inverse powers of $H\tau$ could be justified solely as part of the expansion in powers of $V_{\rm int}$. 

What complicates this argument is the time-dependence of operators like $A_n(s)$ in expressions like \pref{NakaZwanExplicit}. The evolution of operators in the interaction picture is governed by the unperturbed Hamiltonian $H_0$ and so in general is not $V_{\rm int}$ suppressed. But it can be nevertheless small for other reasons -- for instance during inflation the generic freezing of modes in the super-Hubble regime can suppress the time evolution of operators by powers of $-k \eta = k/(aH)$ rather than $V_{\rm int}$. We know of no royal road and in specific cases the validity of the Lindblad approximation must be checked by verifying that the subdominant terms in the expansion in $t-s$ are suppressed once inserted into the $s$ integration. Given that the environmental correlators are sharply peaked in the example considered here, we assume that the aforementioned non-Markovian corrections depending on $t-s$ are suppressed by $k/aH$. This assumption is based on the results found in \cite{Burgess:2022nwu}, where an explicit check of this suppression was performed, although admittedly for a different choice of system and environment.

\subsubsection{Specialization to spectator scalars}

The systems described in \S\ref{SystemDef} have interactions of the form assumed in \pref{VvsAB} where $A_n(t) \to \sigma(\bfx,t)$ and $B_n(t) \to \cO(\bfx,t)$ with $\cO_{\rm mix} = \mu^2 \phi$ for mixing and $\cO_c = g\phi^2$ for cubic interactions. The sum over $n$ corresponds to the integration over position (with the scale factor $a(t)$ entering for cosmological applications as dictated by the metric-dependence required by general covariance). For this specific system the arguments leading to eq.~\pref{NakaZwanExplicit} at second-order perturbation theory then go through as above and lead to the result
\bea \label{NakaZwanExplicitF}
  \partial_t \varrho(t) &=& - i \int \exd^3x \; a^3(t) \Bigl[ \sigma(t,\bfx) , \varrho(t) \Bigr] \, \llangle \; \cO(t,\bfx) \; \rrangle\\
  && \quad  
  + (-i )^2 \int \exd^3x \; a^3(t) \int \exd^3y \int_{t_0}^t \exd s \; a^3(s) \Biggl\{ \Bigl[ \sigma(t,\bfx) , \sigma(s,\bfy) \,\varrho(s) \Bigr] \, \llangle \, \delta \cO(t,\bfx) \, \delta \cO(s,\bfy) \, \rrangle  \nn \\
  && \qquad\qquad\qquad\qquad\qquad\qquad   -  \Bigl[ \sigma(t,\bfx) , \varrho(s) \,\sigma(s,\bfy)  \Bigr] \, \llangle \, \delta \cO(s,\bfy) \,\delta \cO(t,\bfx) \, \rrangle \,\Biggr\} + \cO(V_{\rm int}^3)\,.\nn
\eea

Whether this expression simplifies into a Lindblad form depends on how sharply peaked the relevant environmental correlator 
\begin{equation}
    \llangle \, \delta \cO(t,\bfx)\, \delta \cO(s,\bfy) \, \rrangle = \cW(t,\bfx ; s,\bfy)  = \begin{cases}  \mu^4 W_{\rm env}(t,\bfx; s, \bfy) & \hbox{(mixing)} \cr  2 g^2 \Bigl[W_{\rm env}(t,\bfx; s, \bfy) \Bigr]^2 & \hbox{(cubic)}   \end{cases}  \, ,
\end{equation}
is in time. If it is sufficiently peaked Taylor expanding the rest of the integrand and dropping subdominant terms leads to the analog of \pref{LindbladH}, which in this case has the time-local form
\bea \label{LindbladHF}
  &&\partial_t \varrho  (t) \simeq - i   \Bigl[ V_{\rm eff}(t)  \, , \varrho (t)  \Bigr] -  \int \exd^3x \int \exd^3y \; a^6(t) \; \kappa(t,\bfx,\bfy) \Bigl[  \Bigl\{ \sigma(t,\bfx) \sigma(t,\bfy), \varrho(t) \Bigr\} - 2\, \sigma(t,\bfy) \, \varrho (t) \sigma(t,\bfx) \Bigr] \nn\\
&&\qquad\hbox{with} \quad  V_{\rm eff}(t) :=  \int \exd^3x \; a^3(t) \sigma(t,\bfx) \, \llangle \, \cO(t,\bfx) \, \rrangle + \int \exd^3x \int \exd^3y \; a^6(t) h(t, \bfx,\bfy)\,  \sigma(t,\bfx) \sigma(t,\bfy)   \,,
\eea
where
\be \label{gammahdef}
  \kappa(t,\bfx,\bfy) = \frac12 \Bigl[ C(t, \bfx , \bfy) + C^*(t, \bfy , \bfx) \Bigr] \quad \hbox{and} \quad 
  h(t,\bfx,\bfy) = - \frac{i}2 \Bigl[ C(t, \bfx,\bfy) - C^*(t, \bfy,\bfx) \Bigr]
\ee
for
\begin{equation}\label{CDefEqF}
 C(t,\bfx,\bfy) := \int_{t_0}^t \exd s \  \llangle \, \delta \cO(t,\bfx)\, \delta \cO(s,\bfy) \, \rrangle = \int_{t_0}^t \exd s \; \cW(t,\bfx ; s,\bfy)  \,.
\end{equation}

\subsubsection{Emergent spatial locality}

Further simplification occurs if the correlation function also falls off sufficiently quickly as a function of position, which we note here for completeness despite this step not directly applying to the cases of interest in the rest of the paper (and despite not in itself being crucial from the point of view of resumming late-time evolution).

If the falloff is sufficiently steep the spatial integrals are well-approximated by expanding any fields evaluated at position $\bfy$ in powers of $|\bfy-\bfx|$ and the leading order evolution equation becomes local in space as well as time:
\be \label{LindbladHFloc}
  \partial_t \varrho  (t) \simeq - i   \Bigl[ V_{\rm eff}(t)  \, , \varrho (t)  \Bigr] -  a^3(t) \int \exd^3x\, \mfg(t, \bfx)  \Bigl[  \Bigl\{ \sigma^2(t,\bfx) , \varrho(t) \Bigr\} - 2\, \sigma(t,\bfx) \, \varrho (t) \sigma(t,\bfx) \Bigr] \,,
\ee 
with
\be \label{VeffDefFloc}
  V_{\rm eff}(t) := a^3(t) \int \exd^3x \Bigl[ \llangle \, \cO(t,\bfx) \, \rrangle  \,  \sigma(t,\bfx) +  \mfh(t, \bfx)\,  \sigma^2(t,\bfx) \Bigr]    \,,
\ee 
where
\be
   \mfg(t, \bfx) := a^3(t) \int \exd^3y \; \kappa(t, \bfx, \bfy) \quad
   \hbox{and} \quad \mfh(t,\bfx) := a^3(t) \int \exd^3y \; h(t, \bfx, \bfy) \,.
\ee
This result is again {\it as if} we'd assumed
\be \label{localityconditions}
  \kappa(t,\bfx,\bfy) \simeq \frac{\mfg(t,\bfx) }{a^3(t)} \, \delta^3(\bfx-\bfy) \quad \hbox{and} \quad 
  h(t,\bfx,\bfy) \simeq \frac{\mfh(t,\bfx) }{a^3(t)} \, \delta^3(\bfx-\bfy) \,,
\ee
with corrections depending on $|\mathbf{x} - \mathbf{y}|$ which in principle need to be checked are subdominant, and where the $a^3(t)$ factors are required on general grounds because of general covariance. The proportionality of \pref{LindbladHFloc} to $a^3$ in the spatially local limit when \pref{localityconditions} applies agrees with earlier work \cite{Burgess:2014eoa}. 

\subsubsection{Translation invariant systems}

Returning to the general case where the correlations need not be local in space, we next record expressions for later use that apply when the system is translationally invariant. In this case eq.~\pref{LindbladHF} governing the evolution of $\varrho$ can be written
\bea \label{LindbladHFmom}
  \partial_t \varrho  (t) &=& - i   \Bigl[ V_{\rm eff}(t)  \, , \varrho (t)  \Bigr] -  \int \exd^3x \int \exd^3y \; a^6(t) \; \kappa(t,\bfx-\bfy) \Bigl[  \Bigl\{ \sigma(t,\bfx) \sigma(t,\bfy), \varrho(t) \Bigr\} - 2\, \sigma(t,\bfy) \, \varrho (t) \sigma(t,\bfx) \Bigr]\nn\\
  &=&  - i   \Bigl[ V_{\rm eff}(t)  \, , \varrho (t)  \Bigr] - a^6(t) \int \exd^3k  \; \kappa_\bfk(t) \left[  \Bigl\{ \sigma_\bfk(t) \sigma_{-\bfk}(t), \varrho(t) \Bigr\} \phantom{\frac12} \right. \\
  && \qquad\qquad\qquad\qquad\qquad\qquad\qquad\qquad\qquad\qquad \left. \phantom{\frac12} -  \sigma_\bfk(t) \, \varrho (t) \,\sigma_{-\bfk}(t) - \sigma_{-\bfk}(t) \, \varrho (t) \,\sigma_{\bfk}(t) \right]  \,, \nn
\eea 
where we use \pref{FieldExpansion} with
\be
   \sigma_\bfk(t) := u_k(t) \, \mfa_\bfk + u^*_k(t) \, \mfa^\star_{-\bfk} = \sigma_{-\bfk}^*(t) \,.
\ee

Eqs.~\pref{gammahdef} and \pref{CDefEqF} imply $\kappa_\bfk(t)$ is given by
\be \label{kappakdef}
    \kappa_\bfk(t) =  \int_{t_0}^t \exd s \; \hbox{Re}\, \Bigl[\cW_\bfk(t,s) \Bigr]\,.
\ee
The effective interaction Hamiltonian appearing in \pref{LindbladHFmom} is
\bea \label{VeffDefFmom}
  V_{\rm eff}(t) &=&  \int \exd^3x \; a^3(t) \, \ol\cO(t) \, \sigma(t,\bfx)   + \int \exd^3x \int \exd^3y \; a^6(t) \, h(t, \bfx-\bfy)\,  \sigma(t,\bfx) \sigma(t,\bfy) \nn\\
  &=&  (2\pi)^{3/2}  a^3(t) \, \ol\cO(t) \, \sigma_0(t)  + \int \exd^3k \; a^6(t) \, h_\bfk(t)\,  \sigma_\bfk(t) \sigma_{-\bfk}(t)  \,, 
\eea 
where $\ol\cO(t) = \llangle \,\cO(t,\bfx) \, \rrangle$ while $\sigma_0 := \sigma_{\bfk=0}$ and $h_\bfk$ is given by
\be \label{hkdef}
    h_\bfk(t) = \int_{t_0}^t \exd s \; \hbox{Im} \, \Bigl[\cW_\bfk(t,s)  \Bigr] \,.
\ee

\subsection{Late-time solutions}

Having the right-hand sides of \pref{LindbladHFmom} and \pref{VeffDefFmom} be quadratic in $\sigma_\bfk$ ensures two things. First, it ensures there is no mode mixing so the state for each mode $\bfk$ remains uncorrelated as time evolves provided this is also true of the initial conditions. For discretely normalized momentum states this means $\varrho(t) = \prod_\bfk \otimes \varrho_\bfk(t)$ and so \pref{LindbladHFmom} can be written as a separate evolution equation for each mode's density matrix, with 
\bea \label{LindbladHFmomfactor}
  \partial_t \varrho_\bfk  (t) &=&   - i   \Bigl[ V_{\rm eff}(t)  \, , \varrho_\bfk (t)  \Bigr] - a^6(t)   \kappa_\bfk(t) \left[  \Bigl\{ \sigma_\bfk(t) \sigma_{-\bfk}(t), \varrho_\bfk(t) \Bigr\} \phantom{\frac12} \right. \\
  && \qquad\qquad\qquad\qquad\qquad\qquad\qquad\qquad  \left. \phantom{\frac12} - \sigma_\bfk(t) \, \varrho_\bfk (t) \, \sigma_{-\bfk}(t) - \sigma_{-\bfk}(t) \, \varrho_\bfk (t) \, \sigma_{\bfk}(t) \right]\,, \nn
\eea 
where it is understood that the creation and annihilation operators appearing in $\sigma_\bfk$ are for discretely normalized momentum states. 

The second implication of having the right-hand sides of \pref{LindbladHFmom} and \pref{VeffDefFmom} be quadratic in $\sigma_\bfk$ is that it ensures an initially Gaussian system remains Gaussian. This allows the evolution \pref{LindbladHFmomfactor} to be solved using a Gaussian ansatz
\be\label{GaussianAnsatz}
  \langle \sigma | \varrho_\bfk | \tilde \sigma \rangle := \cZ_\bfk \, \exp\Bigl[ - \cA_\bfk(t) \, \sigma^* \sigma - \cA_\bfk^*(t) \, \tilde \sigma^* \tilde \sigma + \cB_\bfk (t) \, \sigma^* \tilde \sigma + \cB_\bfk^*(t) \, \sigma \tilde \sigma^* \Bigr] \,,
\ee 
as can be seen by taking the matrix elements of \pref{LindbladHFmomfactor} using \pref{GaussianAnsatz} and explicitly differentiating and equating the powers of $\sigma$ and $\tilde \sigma$ on both sides. The evolution equation \pref{LindbladHFmomfactor} is equivalent to the following evolution equation for $\cA_\bfk(t)$ and $\cB_\bfk(t)$ 
\be \label{Aevo}
   \partial_t \cA_\bfk = - \frac{i}{a^3} \Bigl( \cA_\bfk^2 - | \cB_\bfk|^2 \Bigr) + a^3 \left[ i  \left(  m^2 + \frac{k^2}{a^2}  +a^3 h_\bfk \right) + a^3 \kappa_\bfk \right]   \,,
\ee
and
\be \label{Bevo}
   \partial_t \cB_\bfk = - \frac{i}{a^3} \Bigl( \cA_\bfk -  \cA^*_\bfk \Bigr) \cB^*_\bfk + a^6   \kappa_\bfk  \,,
\ee
subject to the initial conditions $ \cA_\bfk(t_0) = \cA_{\bfk0}$ and $\cB_\bfk(t_0) = \cB_{\bfk0}$. 

If the coefficient functions $\kappa_\bfk(t)$ and $h_\bfk(t)$ do not also depend on $t_0$ the arguments of  \cite{Burgess:2022rdo, Burgess:2020tbq} then allow the solutions obtained by integrating \pref{Aevo} and \pref{Bevo} to be trusted at times beyond where perturbative methods would usually apply. Once obtained, these solutions completely determine the evolution of equal-time correlation functions and the purity, with the equal-time correlation function given by:
\be \label{WvscalA}
  W_\bfk(t,t) =  \Tr \, \Bigl[ \varrho_\bfk(t) \,\sigma_\bfk^*(t) \, \sigma_\bfk(t) \Bigr]  = \int \exd \sigma \, \exd \sigma^* \,( \sigma^*  \sigma) \, \langle \sigma | \varrho_\bfk(t) | \sigma \rangle  = \frac{1}{\cA_\bfk + \cA_\bfk^* - \cB_\bfk - \cB_\bfk^*} \,, 
\ee
while the purity similarly becomes
\be \label{purityvsAB}
   \gamma_\bfk(t) := \hbox{Tr}\left[  \varrho_\bfk^2(t)  \right]  = \int \exd \sigma \, \exd \sigma^* \,  \langle \sigma | \varrho^2_\bfk(t) | \sigma \rangle 
   =  \frac{1 - \cR_\bfk}{1 + \cR_\bfk}
   \qquad \hbox{where} \qquad
   \cR_\bfk := \frac{\cB_\bfk + \cB_\bfk^*}{\cA_\bfk+\cA_\bfk^*} \,.
\ee
Notice in particular that $\gamma_\bfk \to 1$ if $\mathrm{Re}[\cB_\bfk] \to 0$. Notice also that if $\cB_\bfk(t_0)$ is real -- as is true in particular for an initially pure state with $\cB_\bfk(t_0) = 0$ -- then the reality of $\kappa_\bfk$ in \pref{Bevo} ensures $\cB_\bfk(t)$ remains real for all later times. 

\subsubsection{Perturbation theory}

These solutions represent a resummation of perturbation theory at late time in the following sense. When perturbing about an initially pure state we have
\be \label{AkIntegralPure}
  \cA_\bfk = \cA_{\bfk0} + \delta \cA_\bfk \quad \hbox{and} \quad
   \cB_\bfk = \delta \cB_\bfk \,,
\ee
where the unperturbed solution satisfies the form of \pref{Aevo} obtained by dropping $h_\bfk$, $\kappa_\bfk$ and $\cB_\bfk$:
\be \label{AevoPT}
   \partial_t \cA_{\bfk0} = - \frac{i}{a^3} \, \cA_{\bfk0}^2  + i a^3  \left(  m^2 + \frac{k^2}{a^2} \right)   \,.
\ee
This can be solved through the change of variables $\cA_{\bfk0} = -i a^3 \partial_t \ln u_\bfk$, since substituting this into \pref{AevoPT} implies $u_\bfk$ satisfies
\be \label{BDevo}
   \ddot u_\bfk + 3 H \dot u_\bfk + \left(  m^2 + \frac{k^2}{a^2} \right) u_\bfk = 0 \,.
\ee
This last equation is recognized as the usual mode-function equation, whose solutions for de Sitter applications are the standard Bunch-Davies modes given in \pref{GendSMode}. 

For modes satisfying \pref{BDevo} the Wronskian $a^3 (u^*_\bfk \dot u_\bfk - u_\bfk \dot u^*_\bfk)$ is time independent and is conventionally used to impose the normalization condition 
\be
     a^3 \Bigl(u^*_\bfk \dot u_\bfk - u_\bfk \dot u^*_\bfk \Bigr)= i \,.
\ee
With this choice we have
\be
   \cA_{\bfk0} + \cA_{\bfk0}^* = -i a^3 \left( \frac{u^*_\bfk \dot u_\bfk - u_\bfk \dot u^*_\bfk}{|u_\bfk|^2} \right) = \frac{1}{|u_\bfk|^2} \,,
\ee
and so the unperturbed equal-time correlator becomes
\be
  W_{\bfk0} =  \frac{1}{ \cA_{\bfk0} + \cA_{\bfk0}^*} = |u_\bfk|^2   \,,
\ee
as expected.

Assuming $\cB_{\bfk0} = 0$ the leading order equation for $\delta \cB_\bfk = a^3 \beta_\bfk$ is obtained by linearizing \pref{Bevo}, giving
\be \label{BevoPT1}
   \dot \beta_\bfk + 3H \beta_\bfk + \frac{i}{a^3} \Bigl( \cA_{\bfk0} -  \cA^*_{\bfk0} \Bigr) \beta_\bfk   = \dot \beta_\bfk + 3H \beta_\bfk +  \frac{ \beta_\bfk \partial_t  |u_\bfk|^2}{|u_\bfk|^2} = a^3 \kappa_\bfk  \,,
\ee
which uses
\be
\label{eq:Ak0:Im}
   \cA_{\bfk0} - \cA_{\bfk0}^* = -i a^3 \left( \frac{u^*_\bfk \dot u_\bfk + u_\bfk \dot u^*_\bfk}{|u_\bfk|^2} \right) = -ia^3 \partial_t \ln \Bigl(|u_\bfk|^2 \Bigr) \,.
\ee
Eq.~\pref{BevoPT1} integrates  -- assuming $\beta_\bfk(t_0) = 0$ -- to give the real solution
\be \label{deltaBkSoln}
  \delta \cB_\bfk = a^3(t)  \beta_\bfk(t) = \frac{1}{ |u_\bfk(t)|^2} \int_{t_0}^t \exd s \; a^6(s) |u_\bfk(s)|^2 \, \kappa_\bfk(s) \,.
\ee
Because this solution implies $\cB_\bfk \sim \cO(\kappa_\bfk)$ it also shows that the $\cB_\bfk^2$ term can be dropped in the equation for $\delta \cA_\bfk$ obtained by linearizing \pref{Aevo} to leading order in the coupling parameters $\mu$ or $g$. Writing $\delta \cA_\bfk = a^3 \alpha_\bfk$ one finds
\be \label{AevoPT1}
    \dot \alpha_\bfk + 3H \alpha_\bfk + \frac{2\dot u_\bfk}{u_\bfk} \,    \alpha_\bfk  =a^3 \Bigl( i  h_\bfk +  \kappa_\bfk  \Bigr) \,,
\ee
which implies $\partial_t(a^3 u^2_\bfk \alpha_\bfk) = a^6 u^2_\bfk ( i h_\bfk + \kappa_\bfk)$ and so integrates to give
\be \label{deltaAkSoln}
u^2_\bfk(t) \, \delta \cA_\bfk(t) 
= u^2_{\bfk}(t_0) \delta \cA_{\bfk}(t_0) + \int_{t_0}^t \exd s \, a^6(s) \, u^2_\bfk(s)\Bigl[ i h_\bfk(s) + \kappa_\bfk(s) \Bigr] \,.
\ee 

The leading perturbative corrections to \pref{WvscalA} and \pref{purityvsAB} then become 
\be \label{PertWgamma}
   W_\bfk(t,t) \simeq \frac{1}{\cA_{\bfk0} + \cA_{\bfk0}^*} \left[	1 - \frac{ \delta \cA_\bfk + \delta \cA_\bfk^* - \delta \cB_\bfk - \delta \cB_\bfk^*}{\cA_{\bfk0}+ \cA_{\bfk0}^*}   \right] = |u_\bfk|^2 \left[	1 - |u_\bfk|^2 \Bigl( \delta \cA_\bfk + \delta \cA_\bfk^* - 2\delta \cB_\bfk  \Bigr)   \right] \,,
\ee
and $\gamma_\bfk(t) \simeq 1 - 2 \cR_\bfk(t)$ with
\be \label{PertWgamma2}
   \cR_\bfk(t)  \simeq  \frac{\delta \cB_\bfk + \delta \cB_\bfk^*}{\cA_{\bfk0}+ \cA_{\bfk0}^*}  =  2 |u_\bfk|^2 \delta \cB_\bfk  
   = 2 \int_{t_0}^t \exd s \; a^6(s) |u_\bfk(s)|^2 \, \kappa_\bfk(s)\,.
\ee

The main point is that expressions like \pref{WvscalA} and \pref{purityvsAB} have a broader domain of validity than perturbative expressions like \pref{PertWgamma} and \pref{PertWgamma2}, and remain true even if the perturbative parts $\delta \cA_\bfk$ and $\delta \cB_\bfk$ are not particularly small.  This broader domain of validity follows because these expressions rely only on the validity of the Gaussianity assumption \pref{GaussianAnsatz}, which in turn only relies on the evolution equation \pref{LindbladHFmomfactor} being quadratic in $\sigma_\bfk$. 

One might worry, however, that \pref{LindbladHFmomfactor} is itself only quadratic at leading order in perturbation theory because it is derived to lowest nontrivial order in the small quantities $\mu/H$ or $g/H$ (depending on whether the relevant system/environment interaction is $V_{\rm mix}$ of \pref{int=mix} or $V_c$ of \pref{int=c}). It is tempting therefore to conclude that it is inconsistent to use the full evolution equations \pref{Aevo} and \pref{Bevo} when they differ from their perturbative limits \pref{BevoPT1} and \pref{AevoPT1}, but the arguments of \cite{Burgess:2022rdo, Burgess:2020tbq} show why this conclusion can be wrong. 

It doesn't matter if perturbative predictions like \pref{deltaBkSoln} or \pref{deltaAkSoln} break down at late times due to the secular growth in time of the coefficients of perturbative parameters like $g/H$. So long as the differential evolution equations like \pref{Aevo} and \pref{Bevo} are themselves independent of the initial time $t_0$ their {\it solutions} can be trusted for later times, and thereby resum the late-time behaviour. (As argued in \cite{Burgess:2022rdo, Burgess:2020tbq}, this is much the same way that the exponential decay law $n\propto e^{-\Gamma t}$ predicted by the evolution $\exd n/\exd t = - \Gamma n$ can be trusted for times $t \gg 1/\Gamma$ even when $\Gamma$ is only known perturbatively.) 

We shall find examples of this in the next section, which applies the above expressions to decoherence in de Sitter and flat space. When so doing we find in some circumstances that approximate solutions for $\delta \cA_\bfk$ and $\delta \cB_\bfk$ grow without bound at late times (and so display the kind of secular growth that is usually dangerous for perturbative methods). But we nevertheless are still justified in using eqs.~\pref{WvscalA} and \pref{purityvsAB} even for times for which $\delta \cA_\bfk$ and $\delta \cB_\bfk$ are no longer small, provided we use the full evolution equations \pref{Aevo} and \pref{Bevo} in their full glory when predicting the time evolution. 

\section{Applications to decoherence} 
\label{sec:DecoApp}

The perturbative expressions for decoherence obtained in \S\ref{sec:MinkDec} and \S\ref{sec:dSdec} apply at early enough times but eventually break down once $t - t_0$ is sufficiently large. When Lindblad evolution provides a good approximation it gives the late-time evolution beyond where straight-up perturbation theory fails. If there is a domain where both perturbative and Lindblad methods apply then matching these two solutions to one another allows the perturbative evolution to be used to transfer the initial conditions into the Lindblad regime.  

We describe how this matching occurs between perturbative and Lindblad solutions for systems prepared in both the flat-space Minkowski vacuum or the Bunch-Davies vacuum in de Sitter space, and then return to explore the domain of validity of the Lindblad approximation itself. 

\subsection{Matching to the perturbative limit}

Earlier sections compute $\partial_t \gamma_\bfk(t)$ purely within perturbation theory in terms of correlation functions defined in the initial system state $\varrho_0 = \varrho(t_0)$. To make contact with this we next compute $\partial_t \gamma_\bfk$ within the Lindblad evolution.  Differentiating \pref{purityvsAB} gives 
\be \label{purityvsAB2}
    \partial_t \gamma_\bfk  =  2 \left[ \frac{(\cB_\bfk  + \cB_\bfk ^*) \partial_t (\cA_\bfk  + \cA_\bfk ^*) -  (\cA_\bfk  + \cA_\bfk ^*) \partial_t (\cB_\bfk  + \cB_\bfk ^*) }{(\cA_\bfk  + \cA_\bfk ^* + \cB_\bfk  + \cB_\bfk ^*)^2} \right] 
 = - \, \frac{4 a^6 \,  \kappa_\bfk \, \gamma_\bfk }{\cA_\bfk  + \cA_\bfk ^* + \cB_\bfk  + \cB_\bfk ^*}   \,,
\ee
where the second equality eliminates $\partial_t \cA_\bfk$ and $\partial_t \cB_\bfk$ using the evolution equations \pref{Aevo} and \pref{Bevo}. 

In the strictly perturbative limit this last expression further simplifies because its right-hand side can be evaluated using the unperturbed expression to an error that is subdominant in the system/environment interaction. If the unperturbed state is pure then $\cA_\bfk(t) \simeq \cA_{\bfk0}(t)$, $\gamma_\bfk(t) \simeq \gamma_{\bfk0}(t) = 1$ and $\cB_{\bfk}(t) \simeq \cB_{\bfk0}(t)  = 0$, leading to
\be \label{purityvsABMatch}
    \partial_t \gamma_\bfk(t) \simeq   -  \frac{4 \,a^6(t) \,  \kappa_\bfk(t)}{\cA_{\bfk0}(t)  + \cA_{\bfk0}^*(t)} = - 4 \, a^6(t)  \, W_{\bfk0}(t,t)  \int_{t_0}^t \exd s  \; \hbox{Re } \cW_\bfk(t,s) \quad \hbox{(perturbative Lindblad)} \,,
\ee
where \pref{WvscalA} is used, specialized to the initial pure state, with $W_{\bfk0}(t,s)$ being the momentum-space correlator in this initial state and eqs.~\pref{gammahdef} and \pref{CDefEqF} are used to express $\kappa_\bfk(t)$ in terms of the environmental correlator $\cW_\bfk(t,s)$.

Eq.~\pref{purityvsABMatch} differs from the cosmic-time version of the perturbative expression \pref{PurityChangeMasterEqFflatdS} used in earlier sections, repeated here for convenience of reference:
\be \label{purityPTMatch}
  \partial_t \gamma_\bfk  =   - 4 a^3(t)  \int_{t_0}^t  \exd t'\, a^3(t')  \; \hbox{Re}\Bigl[ W_{\bfk0}(t,t') \, \cW_{-\bfk}(t,t') \Bigr] \qquad \hbox{(perturbative)} \,,
\ee
differing by the replacement of $W_{\bfk0}(t,t')\,  a^3(t')$ with its real equal-time limit $W_{\bfk0}(t,t) \, a^3(t)$. This replacement is what would follow if the system variables vary much more slowly than the environment so that the difference between $t$ and $t'$ in $W_{\bfk0}(t,t') \, a^3(t')$ can be neglected in the integration regime for which $\cW_\bfk(t,t')$ has significant support. When this replacement is justified within the perturbative regime then both \pref{purityvsABMatch} and \pref{purityPTMatch} share a common domain of validity that can be used to match the initial perturbative evolution onto the later Lindblad evolution.  

We next ask when these two types of evolution agree using the explicit form for the correlation functions evaluated earlier for the Minkowski and Bunch-Davies vacua in flat and de Sitter space.

\subsubsection*{Minkowski vacuum in flat space}

We start with the Minkowski vacuum in flat space, starting with the Gaussian case where environment and system interact through the mixing term \pref{int=mix}. In this case both system and environmental correlation functions are given by \pref{FlatWk}
\be \label{FlatWkv2}
 W_{\bfk0}(t,t') =  \frac{1}{2 \varepsilon_{\rm sys}} \, e^{-i\varepsilon_{\rm sys}(t-t')} \qquad \hbox{and} \qquad
 \cW_{\bfk}(t,t') =  \frac{\mu^4}{2 \varepsilon_{\rm env}} \, e^{-i\varepsilon_{\rm env}(t-t')}
\,,
\ee
for which the perturbative purity evolution \pref{purityPTMatch} evaluates to \pref{PurityChangeMasterEqFflat}, repeated here:
\be  
  \partial_t \gamma_\bfk  
   =    \frac{\mu^4}{\varepsilon_{\rm sys} \varepsilon_{\rm env}(\varepsilon_{\rm sys} +\varepsilon_{\rm env})} \, \hbox{Im}\Bigl[ e^{-i(\varepsilon_{\rm sys}+\varepsilon_{\rm env})(t-t_0)} \Bigr] \qquad \hbox{(perturbative)} \,.
\ee 
This is to be compared with the perturbative Lindblad result \pref{purityvsABMatch}, which in this case evaluates to
\be \label{purityvsABMatch22}
    \partial_t \gamma_\bfk(t) \simeq    \frac{\mu^4}{\varepsilon_{\rm sys} \varepsilon_{\rm env}^2} \, \hbox{Im}\Bigl[ e^{-i\varepsilon_{\rm env}(t-t_0)} \Bigr] \qquad \hbox{(perturbative Lindblad)} \,,
\ee
showing that they agree up to $\varepsilon_{\rm sys}/\varepsilon_{\rm env}$ corrections when $\varepsilon_{\rm env}\gg \varepsilon_{\rm sys}$ (as is the case in particular when $m_{\rm env} \gg m_{\rm sys}, k$). This is intuitive; the system correlator oscillates much more slowly than the environment in this limit and this produces the hierarchy of scales underlying the Lindblad approximation. In this limit perturbative evolution matches smoothly onto the Lindblad limit, but this doesn't buy us much because in this case the perturbative evolution also remains under control at late times.

Similar considerations apply for flat-space decoherence mediated by the cubic interaction \pref{int=c}, for which the environmental correlator is given by \pref{FlatMassivecWk}. In this case again the Lindblad and perturbative evolutions agree with one another when $m_{\rm env} \gg k, m_{\rm sys}$ because 
\bea \label{FlatMassivecWk22}
W_{\bfk0}(t,t') \, \cW_{\bf{k}}(t,t') & = &  \frac{ig^2}{4(2\pi)^2 \varepsilon_{\rm sys} k (t-t')} \int_0^\infty \frac{p \, \exd p}{\varepsilon_{\rm env}}\, e^{-i(\varepsilon_{\rm sys}+  \varepsilon_{\rm env})  (t-s)} \left[ e^{-i \varepsilon_+(t-t')} - e^{-i \varepsilon_-(t-t')} \right] \nn\\
 & \simeq &  \frac{ig^2}{4(2\pi)^2 \varepsilon_{\rm sys} k (t-t')} \int_0^\infty \frac{p \, \exd p}{\varepsilon_{\rm env}}\, e^{-i \varepsilon_{\rm env} (t-s)} \left[ e^{-i \varepsilon_+(t-t')} - e^{-i \varepsilon_-(t-t')} \right] \\
 &=&  W_{\bfk0}(t,t) \, \cW_{\bf{k}}(t,t')\,,\nn
\eea
where the approximate equality neglects $\varepsilon_{\rm sys}$ relative to $\varepsilon_{\rm env}$ up to $m_{\rm sys}/m_{\rm env}$ or $k/m_{\rm env}$ corrections. Again the Lindblad and perturbative evolutions can have overlapping domains of validity, allowing early-time perturbative evolution to be matched onto later Lindblad evolution. Again this does not buy that much because the perturbative evolution in this case does not break down at the times of interest for decoherence. 

\subsubsection*{Bunch-Davies vacuum in de Sitter space}

Consider next de Sitter space with the scalar fields prepared in the remote past in the Bunch-Davies vacuum. This is the case of practical interest because we saw in \S\ref{sec:dSdec} that perturbative decoherence grows strongly with time and so Lindblad methods are needed to make late-time predictions for the purity evolution. 

We start for concreteness' sake assuming that the environment consists of a conformally coupled scalar, though the conclusions can be extended to apply more generally along the lines explored in \S\ref{sssec:FourierGeneral}. As we've seen above, overlap between the early-time perturbative evolution and the Lindblad evolution requires expressions like \pref{PurityChangeSingPart2} for the perturbative purity evolution,
\be  \label{PurityChangeSingPart3}
  \partial_\eta \gamma_\bfk  =  - \frac{g^2}{8\pi H}   \int_{-\infty}^\eta \frac{ \exd \eta' }{(H^2 \eta\eta')^{1/2}} \;  \hbox{Im} \left\{ \frac{e^{-ik(\eta - \eta')}}{\eta - \eta'} \;  H^{(1)}_{\nu_{\rm sys}}(-k\eta)  \left[ H^{(1)}_{\nu_{\rm sys}}(-k\eta') \right]^*  \right\}  
  \quad\text{(perturbative)}
   \,,  
\ee
to be well-approximated by the perturbative Lindblad form
\bea  
  \partial_\eta \gamma^{\rm LB}_\bfk  &= & - 4  \int_{-\infty}^\eta \frac{\exd \eta'}{ (H^2\eta\eta')^4} \hbox{Re}\left[ W_\bfk(\eta,\eta) \, \cW_{-\bfk}(\eta,\eta') \right] \nonumber \\
 & = &   \frac{g^2\eta}{8\pi H^2}  \left|  H^{(1)}_{\nu_{\rm sys}}(-k\eta)  \right|^2    \int_{-\infty}^\eta \frac{ \exd \eta' }{{\eta'}^{2}} \;  \hbox{Im} \left\{ \frac{e^{-ik(\eta - \eta')}}{\eta - \eta'} \; \right\}  \quad\text{(perturbative Lindblad)} \,.
 \label{PurityChangeSingPartLB}
\eea
These expressions do agree\footnote{This agreement can be made explicit for the special case of a MCMS system and CCMS environment, in which case \pref{PurityChangeSingPart3} is evaluated explicitly in \S\ref{App:MCMSsystemCCMSenvironmentExplicit} while \pref{PurityChangeSingPartLB} is evaluated in \pref{kappakBDx2} below.} -- but only for the super-Hubble regime to leading order in $k\eta$ -- because for small $\eta$ the integral is dominated by the $i \pi \delta(\eta-\eta')$ that controls the imaginary part of $1/(\eta-\eta')$ given that $\eta - \eta'$ carries a small negative imaginary part -- see eq.~\pref{0+Identity}. Perturbative evolution can therefore be used to evolve $\gamma_\bfk(t)$ into the super-Hubble regime where it starts to grow significantly. Tossing the torch of evolution to the Lindblad equation before perturbative methods fail then allows the purity to be evolved beyond the strictly perturbative domain using formulae like \pref{purityvsAB}.

\subsection{Lindblad evolution}

We pause here to record what these late-time expressions look like in the special case where the environmental correlator is computed within the Bunch-Davies vacuum in de Sitter space. It is intuitive that the Lindblad approximation should apply for super-Hubble timescales when using the Bunch-Davies de Sitter correlator because this is in many ways similar to a thermal correlator. 

\begin{figure}[h]
\begin{center}
\includegraphics[height=55mm]{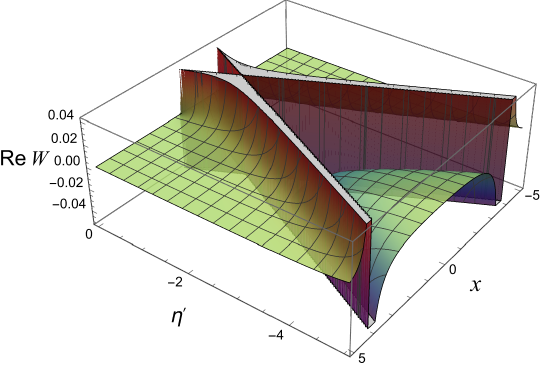}
\caption{\small Plot of $\mathrm{Re}\left[ \mathcal{W}(\eta,\bfx; \eta',0) \right]$ from Eq.~(\ref{WkFTdefdSm0}) as a function of $-5 < \eta' < 0$ and $-5 < x < 5$. All other parameters are held fixed, with $\eta = -1$ and $H = 1$ and $\zeta_{\text{env}} = 2$. Notice the strong peaking along the light cone.} \label{Fig:3DWPlot}
\end{center}
\end{figure}

Fig.~\ref{Fig:3DWPlot} shows a plot of $\text{Re}\left[ \mathcal{W}(\eta,\bfx; \eta',0)\right]$ as a function of $\eta'$ and spatial separation $x =|\mathbf{x}|$, with $\eta$ and $\zeta_{\mathrm{env}}$ held fixed. (In the figure the environment mass and nonminimal couplings are chosen such that $\zeta_{\rm env}= 2$ and so $\nu_{\rm env} = 1/2$). This plot reveals a strong and narrow ridge that runs along the light cone for which $\eta-\eta' = \pm x$, indicating a single narrow peak in time for each choice of $x$. This ridge is also sharply peaked in space at fixed time.

Converting to conformal time \pref{LindbladHFmom} becomes
\bea \label{LindbladHFmomCT}
  \partial_\eta \varrho  (\eta)    &=&  - i   \Bigl[ V_{\rm eff}(\eta)  \, , \varrho (\eta)  \Bigr] - a^7(\eta) \int \exd^3k  \; \kappa_\bfk(\eta) \left[  \Bigl\{ \sigma_\bfk(\eta) \sigma_{-\bfk}(\eta), \varrho(\eta) \Bigr\} \phantom{\frac12} \right. \\
  && \qquad\qquad\qquad\qquad\qquad\qquad\qquad \left. \phantom{\frac12} -  \sigma_\bfk(\eta) \, \varrho (\eta) \,\sigma_{-\bfk}(\eta) - \sigma_{-\bfk}(\eta) \, \varrho (\eta) \,\sigma_{\bfk}(\eta) \right]  \,, \nn
\eea 
with \pref{kappakdef} and \pref{hkdef} implying
\be \label{kappakdefCT}
    \kappa_\bfk(\eta) =  \int_{\eta_0}^\eta \exd \eta' \;a(\eta') \,\text{Re }\cW_\bfk(\eta,\eta')   
\quad\hbox{and} \quad
    h_\bfk(\eta) =   \int_{\eta_0}^\eta \exd \eta' \; a(\eta') \, \text{Im } \cW_\bfk(\eta,\eta') \,,
\ee
with $\eta_0$ chosen to lie within the domain of overlap between perturbative and Lindblad evolution. Of these only $\kappa_\bfk(\eta)$ is relevant for decoherence calculations. 

Because Lindblad methods only agree with perturbative evolution in the super-Hubble limit we can assume both $k\eta$ and $k \eta_0$ are small when evaluating eqs.~\pref{kappakdefCT}. For a conformal environment we can use \pref{cWkdScubicc}, so eqs.~\pref{kappakdefCT} become 
\bea \label{kappakBDx}
    \kappa_\bfk(\eta)  &\simeq&  -  \frac{g^2 H^3 \eta^2}{8\pi^2} \int_{\eta_0}^\eta \exd \eta' \;\text{Im} \left[\frac{\eta'}{ (\tilde\eta - \eta')} \, e^{-ik(\tilde\eta-\eta')} \right]  \nn\\
    \hbox{and} \quad
    h_\bfk(\eta) &\simeq&   \frac{g^2 H^3 \eta^2}{8\pi^2} \int_{\eta_0}^\eta \exd \eta' \;\text{Re} \left[\frac{\eta'}{ (\tilde\eta - \eta')} \, e^{-ik(\tilde\eta-\eta')} \right] \,,
\eea
where $\tilde \eta = \eta - i \epsilon$ with $\epsilon \to 0$ at the end, with the regularization used since $h_\bfk(\eta)$ is formally divergent. The required integral is
\bea \label{kappakBDx2}
    \int_{\eta_0}^\eta \exd \eta' \;  \left[\frac{\eta'}{ (\tilde\eta - \eta')} \, e^{-ik(\tilde\eta-\eta')} \right]  &=&   \left\{ \frac{i e^{-i k (\tilde \eta - \eta')}}{k}- \tilde\eta\, \text{Ei}[-i k (\tilde\eta -\eta')]  \right\}_{\eta'=\eta_0}^{\eta'=\eta} \nn\\ 
    &=& \eta \ln \left[ \frac{\eta - \eta_0}{\epsilon} \right]  + \frac{i\pi \eta}{2}  - (\eta-\eta_0) + \cdots  \,,
\eea
which evaluates the small-argument limits of Ei$(-2k\epsilon)$ and Ei$[-2ik(\eta-\eta_0)]$ using \pref{AppEizsmall} and drops all terms involving positive powers of $k\eta$ or $k\eta_0$. 

Only the imaginary part is relevant for decoherence and this remains finite as $\epsilon \to 0$. Notice also the agreement between the leading term in the imaginary part and the delta-function contribution contribution from \pref{0+Identity} (keeping in mind the factor of 2 coming because the delta-function has support at the edge of the integration region, see \pref{eq:Dirac:prop}). Combining everything gives the following leading super-Hubble expression 
\be  \label{kappakBDx3}
    \kappa_\bfk(\eta)  \simeq   - \frac{g^2 H^3\eta^3}{16\pi}  = \frac{g^2}{16\pi a^3(\eta)}  
    \qquad \hbox{and} \qquad
 h_\bfk(\eta)  \simeq   - \frac{g^2 H^3 \eta^3}{8\pi^2} \ln \epsilon = \frac{g^2}{16\pi a^3(\eta)} \,  \ln \epsilon 
\,.
\ee
Notice that the combinations $a^3 \kappa_\bfk \simeq g^2/(16\pi)$ and $a^3 h_\bfk \simeq g^2/(16\pi)$ are time independent in the super-Hubble limit, and although $h_\bfk$ diverges as $\epsilon \to 0$ the time-independence and $k$- and $\eta$-independence of the divergent part of $a^3 h_\bfk$ are precisely what are required for the divergence to be absorbed into the particle mass $m$, such as in eq.~\pref{Aevo}. 

\subsubsection{Perturbative evolution (again)}

We may now evaluate the perturbative expressions \pref{deltaAkSoln} and \pref{deltaBkSoln} for $\delta \cA_\bfk$ and $\delta \cB_\bfk$ respectively. Doing so also requires the super-Hubble limit of the Bunch-Davies mode function \pref{GendSMode}, which using the asymptotic form \pref{AppHankelSmall} becomes
\be 
    u_\bfk(\eta) \simeq \frac{iH}{\sqrt{4\pi k^3}} \, 2^{\nu} \Gamma(\nu) (k \eta)^{\frac32-\nu} \,,
\ee
in the limit $|k\eta| \ll 1$. Using this in \pref{deltaBkSoln} then gives (after switching to conformal time)
\bea \label{deltaBkSolndS}
  \delta \cB_\bfk(\eta) &=&   \frac{g^2}{16\pi |u_\bfk(\eta)|^2} \int_{\eta_0}^\eta \exd \eta' a^4(\eta') \; |u_\bfk(\eta')|^2 
  \simeq  \frac{g^2}{16\pi H^4 |\eta|^{3-2\nu_{\rm sys}}} \int_{\eta_0}^\eta  \exd \eta'  \; |\eta'|^{-1-2\nu_{\rm sys}} \nn\\
  &\simeq& -  \frac{g^2}{32\pi H^4 \nu_{\rm sys} \eta^3}  \left[1 - \left( \frac{\eta}{\eta_0} \right)^{2\nu_{\rm sys}} \right]   \qquad \hbox{(for real $\nu_{\rm sys}$)} \,. 
\eea
where we use the late-time limit $\eta_0 /\eta \gg 1$. When $\nu_{\rm sys} \to 0$ \pref{deltaBkSolndS} instead predicts
\be
    \delta \cB_\bfk \to - \frac{g^2}{16\pi H^4 \eta^3 } \ln \left( \frac{\eta_0}{\eta} \right) = \frac{g^2(t-t_0)}{16\pi}\, e^{3Ht} \qquad\qquad \hbox{(MMCS)}
\ee
for a minimally coupled massless scalar. These expressions grow without bound in the far future (for which $\eta \to 0^-$ or $t - t_0 \to \infty$), and so provide explicit examples of the kind of secular growth that can imperil late-time perturbative methods.

The leading correction to $\delta \cA_\bfk$ is given by expression \pref{deltaAkSoln}, and so has the super-Hubble form
\be \label{deltaAkSoln1}
    \delta \cA_\bfk(\eta)  =  \left[ \frac{u^2_\bfk(\eta_0)}{u^2_\bfk(\eta)} \right] \delta \cA_{\bfk}(\eta_0) +  \int_{\eta_0}^\eta \exd \eta' \, a^7(\eta') \, \left[ \frac{u^2_\bfk(\eta')}{u^2_\bfk(\eta)} \right] \Bigl[ i h_\bfk(\eta') + \kappa_\bfk(\eta') \Bigr] \,,
\ee
where $\delta \cA_\bfk(\eta_0)$ is in principle found by matching to the perturbative evolution shortly after the modes in question pass to the super-Hubble limit. In what follows we neglect $\delta \cA_\bfk(\eta_0)$ whenever $\eta_0/\eta \gg 1$, and so neglect the relatively small effects accrued before and during horizon exit relative to the larger contributions that secular growth generates at later times. 

Notice that $h_\bfk(\eta')$ drops out of the combination $\delta \cA_\bfk + \delta \cA^*_\bfk$ relevant to the purity $\gamma_\bfk$ and to the corrections to the equal-time correlator $W_\bfk(\eta)$. In the super-Hubble limit we have
\be  \label{deltaAkSoln2}
    \delta \cA_\bfk(\eta) + \delta \cA^*_\bfk(\eta)   
    \simeq  -  \frac{g^2}{16\pi H^4 \nu_{\rm sys} \eta^3}  \left[1 - \left( \frac{\eta}{\eta_0} \right)^{2\nu_{\rm sys}} \right]   \qquad \hbox{(for real $\nu_{\rm sys}$)} \,,
\ee 
which indeed dominates $\delta \cA_\bfk(\eta_0) + \delta \cA^*_\bfk(\eta_0)$ at late times (when $\eta_0/\eta \gg 1$). For comparison purposes recall that the unperturbed equal-time correlator in the super-Hubble limit is given by
\be \label{dSWkAsymp2}
  \cA_{\bfk0} + \cA^*_{\bfk0}  
  = \frac{1}{|u_\bfk(\eta)|^2}  \simeq \frac{4\pi k^3}{H^2} \left| \frac{1}{2^{\nu_{\rm sys}} \Gamma(\nu_{\rm sys}) }\right|^2 (-k \eta)^{2\nu_{\rm sys} - 3}  \,.
\ee

Using \pref{kappakBDx3}, the super-Hubble expression for the perturbative purity \pref{PertWgamma2} similarly becomes
\be \label{PertRgamma}
    \gamma_\bfk(\eta) \simeq 1 - 2 \cR_\bfk(\eta)  
\ee
with
\be \label{PertRgamma2}
    \cR_\bfk(\eta)  \simeq  2 \int_{\eta_0}^\eta \exd \eta' \; a^7(\eta') |u_\bfk(\eta')|^2 \, \kappa_\bfk(\eta') 
    \simeq    \frac{g^2}{64\pi^2 H^2 \nu_{\rm sys}} \Bigl| 2^{\nu_{\rm sys}} \Gamma(\nu_{\rm sys}) \Bigr|^2 (-k\eta)^{-2\nu_{\rm sys}}  \left[1 - \left( \frac{\eta}{\eta_0} \right)^{2\nu_{\rm sys}} \right]  \,,
\ee
in precise agreement with the perturbative result \pref{PurityChangeSingPart2}.

\subsubsection{Beyond perturbative evolution}

The secular growth of the perturbative expressions like \pref{PertRgamma2} soon takes the prediction beyond the domain of perturbation theory, such as by violating the assumption that $\delta \cA_\bfk$ must be small relative to $\cA_{\bfk0}$ that was used to linearize the evolution equation \pref{Aevo} to derive the time dependence in \pref{deltaAkSoln2}. In this section we use Open EFT techniques -- {\it i.e.}~we exploit a hierarchy of scales $k\ll aH$ to derive Lindblad evolution -- to extract reliable late-time evolution even once straight-up perturbation theory fails. This can be done by relying exclusively on expressions like \pref{WvscalA} and \pref{purityvsAB} and inferring the time evolution only using eqs.~\pref{Aevo} and \pref{Bevo}. 

\begin{figure}[tbp]
	\centering
	\includegraphics[width=0.8\textwidth]{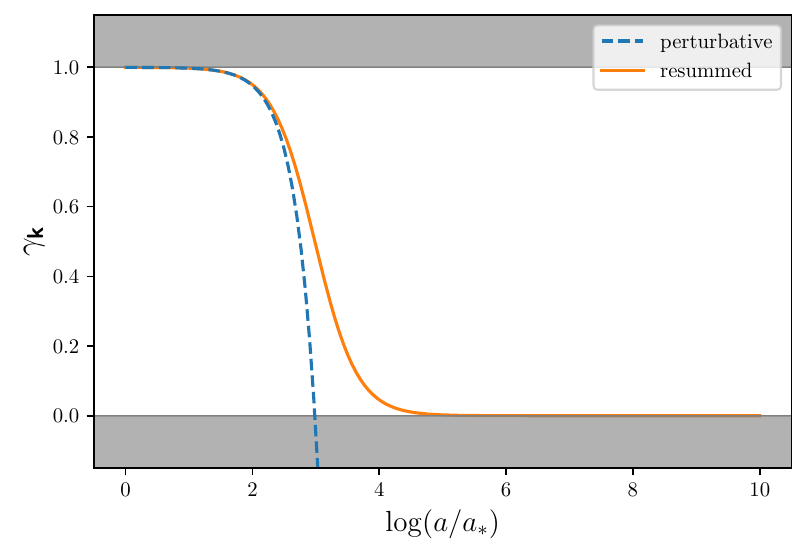}
	\caption{Purity $\gamma_\bfk$ as a function of the number of efolds after Hubble crossing $a_* = k/H$. We considered $m = 0.1H$ and $g = 0.1H$. The perturbative result (in \textit{blue}) is obtained from Eq.~\pref{purityvsABMatch} and the non-perturbative solution (in \textit{orange}) from Eq.~\pref{eq:fin}. The overlap of the solutions at early time is manifest. Then, secular effects become dominant and drive the perturbative result to unphysical values (in \textit{grey}). The Lindblad equation, when solved non-perturbatively, provides a reliable resummation of the perturbative result which remains valid at late times.}
	\label{fig:resum}
\end{figure}

A convenient expression for the purity that relies only on \pref{purityvsAB} first defines the quantity $\cX_\bfk$ by
\be
   \frac{1}{\cR_\bfk} := 1 + \frac{1}{\cX_\bfk} \qquad \hbox{so that} \qquad \cR_\bfk = \frac{ \cX_\bfk}{1 + \cX_\bfk} \qquad \hbox{and} \qquad
   \gamma_\bfk = \frac{1 - \cR_\bfk}{1+\cR_\bfk} = \frac{1}{1 + 2 \cX_\bfk} \,.
\ee
Clearly $0 \leq \gamma_\bfk \leq 1$ for all $\cX_\bfk \geq 0$. At leading order in perturbation theory we have $\cR_\bfk \simeq \cX_\bfk$ but this is not true beyond leading order. For $\cX_\bfk \gg 1$ we have instead $\cR_\bfk \simeq 1- \cX_\bfk^{-1}$. This definition is motivated by comparing the perturbative expressions \pref{deltaBkSolndS} and \pref{deltaAkSoln2}, which imply $\delta \cA_\bfk + \delta\cA_\bfk^* = 2 \delta \cB_\bfk$ in the super-Hubble regime, since using this in the definition 
\be
  \cR_\bfk = \frac{\cB_\bfk + \cB_\bfk^*}{\cA_\bfk + \cA_\bfk^*} = \frac{2\delta \cB_\bfk}{\cA_{\bfk0} + \cA_{\bfk0}^* + 2 \delta \cB_\bfk}
\ee
suggests that using the corrected versions of both $\cA_\bfk$ and $\cB_\bfk$ in the exact expression for $\cR_\bfk$ is equivalent to the choice 
\be
  \cX_\bfk = \frac{2\delta \cB_\bfk}{\cA_{\bfk0} + \cA_{\bfk0}^*}  \simeq    \frac{g^2}{64\pi^2 H^2 \nu_{\rm sys}} \Bigl| 2^{\nu_{\rm sys}} \Gamma(\nu_{\rm sys}) \Bigr|^2 (-k\eta)^{-2\nu_{\rm sys}} \,,
\ee
corresponding to the resummed purity prediction
\be\label{eq:fin}
  \gamma_\bfk(\eta) \simeq \left[ 1 +  \frac{g^2}{32\pi^2 H^2 \nu_{\rm sys}} \Bigl| 2^{\nu_{\rm sys}} \Gamma(\nu_{\rm sys}) \Bigr|^2 (-k\eta)^{-2\nu_{\rm sys}} \right]^{-1} \,.
\ee
The proposal is that Lindblad evolution guarantees that this resummed formula also holds in the late-time limit where $\eta \to 0$ and so $\cX_\bfk \gg 1$ and so $\gamma_\bfk \to 0$ -- see Fig.~\ref{fig:resum} for a comparison with the direct perturbative expression. (See also Appendix \ref{app:transport} for a derivation of \pref{eq:fin} starting directly with the correlation functions rather than the density matrix.) 

But the explicit expressions that suggest $\cA_\bfk + \cA_\bfk^* \to \cB_\bfk + \cB_\bfk^*$ (and so also $\cR_\bfk \to 1$) in the large-time limit are derived in perturbation theory and so break down once the corrections are large. Why believe these quantities should approach one another even beyond the perturbative regime? The key to understanding this lies in solving the full evolution equations \pref{Aevo} and \pref{Bevo}, which in particular imply
\bea \label{AevoR}
   \partial_t \Bigl(\cA_\bfk + \cA^*_\bfk \Bigr) &=& -\frac{i}{a^3} \Bigl( \cA_\bfk - \cA^*_\bfk \Bigr)  \Bigl( \cA_\bfk +  \cA^*_\bfk \Bigr) + 2a^6 \kappa_\bfk    \nn\\
     \partial_t \Bigl(\cB_\bfk + \cB^*_\bfk \Bigr)&=&- \frac{i}{a^3} \Bigl( \cA_\bfk - \cA^*_\bfk \Bigr)  \Bigl( \cB_\bfk +  \cB^*_\bfk \Bigr) +2 a^6 \kappa_\bfk   \,,
\eea
and so
\be \label{BevoR}
  \partial_t \Bigl(\cA_\bfk +  \cA^*_\bfk - \cB_\bfk - \cB^*_\bfk \Bigr) =- \frac{i}{a^3} \Bigl( \cA_\bfk - \cA^*_\bfk \Bigr)  \Bigl(\cA_\bfk +  \cA^*_\bfk - \cB_\bfk - \cB^*_\bfk \Bigr)   \,,
\ee
showing that $\cA_\bfk + \cA_\bfk^* - \cB_\bfk - \cB_\bfk^*$ is a fixed point of the exact evolution. Furthermore \pref{eq:Ak0:Im} shows that $\cA_{\bfk0}$ has a negative imaginary part, at least in models where $\vert u_\bfk \vert $ grows (such as during slow-roll inflation), in which case this is an attractor solution of the dynamics. 

In general one solves \pref{Aevo} and \pref{Bevo}, perhaps numerically, and uses this to evolve $\cX_\bfk$ into the far future. Convenient variables for such a solution are found by writing $\cB_\bfk = a^3 \, \beta_\bfk$ for real $\beta_\bfk$ and $\cA_\bfk = a^3 \, \alpha_\bfk$ with $\alpha_\bfk = -i \dot w_\bfk/w_\bfk$ with $w_\bfk$ and $\beta_\bfk$ to be determined using the evolution equations \pref{Aevo} and \pref{Bevo}, which in these variables become the coupled equations
\be
    \ddot w_\bfk + 3H \dot w_\bfk + \left[  m^2 + \frac{k^2}{a^2}  + \beta_\bfk^2 +a^3( h_\bfk -i \kappa_\bfk) \right] w_\bfk = 0 \,.
\ee
and 
\be \label{NewBevo}
    \dot \beta_\bfk + \Bigl[ 3H  + \partial_t \ln \Bigl(|w_\bfk|^2\Bigr) \Bigr] \beta_\bfk   = a^3\kappa_\bfk   \,.
\ee
Notice that the state being Gaussian, this is strictly equivalent to solving the transport equations controlling the dynamical evolution of the system two-point functions as done in Appendix \ref{app:transport}.

\Eq{NewBevo} can be formally integrated to give
\be
 a^3 |w_\bfk|^2 \beta_\bfk(t) = a^3 |w_{\bfk}|^2 \beta_{\bfk}(t_0) +  \int_{t_0}^t \exd t' \; a^6 |w_\bfk|^2 \kappa_\bfk (t') \,.
\ee
Notice that the presence of nonzero $\kappa_\bfk$ implies the Wronskian for the `mode' functions $w_\bfk$ is no longer time-independent:
\be
 \partial_t \Bigl[ a^3(w^*_\bfk \dot w_\bfk - w_\bfk  \dot w^*_\bfk ) \Bigr] = a^3 \Bigl[ w^*_\bfk(\ddot w_\bfk + 3H \dot w_\bfk) - w_\bfk(\ddot w^*_\bfk + 3H \dot w^*_\bfk) \Bigr] = 2i a^6 \kappa_\bfk |w_\bfk|^2 \,.
\ee

\section{Conclusions}
\label{sec:Conclusions}
How inflationary perturbations become classical has been a pressing issue since the realization that quantum fluctuations during the accelerated expansion phase of the universe could be the seeds of structure formation. We here focus our work on how decoherence occurs in these fluctuations, and more generally how reliable calculations of the rate of decoherence can be made. To this end, the tools of Open EFTs play a crucial role in resumming the secular growth of the purity diagnostic we use to see how decoherence unfolds once an environment is traced out. Because we work with spectator scalars the results can be calculated fairly explicitly, allowing us to test several features of these types of calculations.

We enumerate our findings as follows: 
\begin{itemize}
\item Integrating out the environmental field generically causes the state of the system field to decohere, though on flat space the total decoherence remains fixed over time (after the passage of initial transients) while on de Sitter space the decoherence grows with time into the late time regime (in a special case of the phenomenon of `secular growth' in perturbation theory). This means the perturbative semiclassical calculation of the evolution of super-Hubble modes eventually fails for super-Hubble modes on de Sitter space at sufficiently late times. 
\item We explore how decoherence depends on the properties of the system and environment fields, and in particular the decoupling limit where the massive field is much more massive than the system field. In all cases decoherence becomes suppressed as the heavy field decouples, falling to zero with its mass. The large-mass decoupling limit is somewhat subtle because it does not commute with taking the small imaginary $i \epsilon$ in the time difference of any Wightman function to zero. These limits raise a number of interesting questions about how heavy fields decouple, since decoupling normally implies that effects depending on inverse powers of heavy masses can be captured by effective interactions within an effective Hamiltonian, and any type of Hamiltonian evolution cannot in itself generate decoherence. We intend to expand on these conceptual questions in a later paper.
\item Within perturbation theory the decoherence rate remains small over time-scales characteristic of the environment (such as the temperature for a thermal environment) because it is explicitly suppressed by powers of the small perturbative couplings. This ensures that perturbative decoherence rates remain long compared with the Hubble time for quantum fields on de Sitter space.
\item For super-Hubble modes perturbative purity evolution is amplified by secular growth and this eventually pushes the discussion of late-time behaviour beyond to domain of perturbation theory. We show how Open EFT techniques can be used to resum this late-time secular growth and so determine the evolution into the late-time regime. This resummation is performed by showing how the evolution is well-described by a Lindblad equation, whose solutions can be trusted over longer times than are the direct perturbative calculations. Because the domain of validity of perturbative evolution overlaps with the domain of validity of Lindblad evolution (which only applies for super-Hubble modes) the information about the initial Bunch-Davies configuration of the field in the remote past can be transferred to the Lindblad evolution, thereby determining its subsequent evolution. It appears that super-Hubble Markovian dynamics, as also encountered in \cite{Burgess:2022nwu}, emerge once again in this work, suggesting that this is a very general phenomenon worth understanding in more generality in future work.
\item Although we do not compute here the evolution of the metric or inflaton fluctuations themselves in this paper, our calculation does shed light on some of the puzzles associated with such a calculation \cite{Burgess:2022nwu}. In particular, our calculation here shows that for de Sitter evolution secular growth implies the environmental modes that are the most important for decohering super-Hubble system modes are those with wavelengths not too much shorter than the system modes (and so are themselves also super-Hubble). Such long wavelengths and late times are important because small coupling ensures the perturbative decoherence rate is very slow compared with typical Hubble rates. This observation is also why it was sufficient to focus on super-Hubble environmental modes in \cite{Burgess:2022nwu} (which is useful because it allows the neglect of effective interactions involving time derivatives).  
\end{itemize}

We believe our results provide a formalism that allows \emph{reliable} and explicit calculations of how primordial fluctuations decohere. In principle, this formalism should be useful to understand proposals aimed at distinguishing observationally whether cosmological structure formation is seeded by classical or quantum precursors. 

Although the decoherence rate we find here agrees with \cite{Burgess:2022nwu} in being very rapid, it is worth keeping in mind that the precise amount of decoherence required to erase a particular quantum feature can vary. For instance, in \cite{Martin:2021znx}, it was shown that decohered states in a de Sitter universe still carry a large quantum discord if decoherence is sufficiently slow (with $\gamma_{\bm{k}}\propto a^{-p}$ with $p<4$). For comparison the calculation of \cite{Burgess:2022nwu} predicts $p=3$ for the decoherence of perturbation by gravitational self-interactions, suggesting that although decoherence is very efficient, the erasure of quantum discord is not. Small scales that spend too few $e$-folds beyond the horizon might also be expected to retain some quantum features. 

The proposal that large-scale structure is seeded by primordial quantum perturbations has been so successful that it has become the modern paradigm and like any core tenet of cosmology it therefore cries out for observational testing. Calculations of cosmic decoherence are not meant to discourage searching for such signals (even if decoherence is efficient, as it seems to be). The goal of such calculations is to determine how decoherence depends on model parameters so that we can learn from whatever is ultimately observed. Indeed, from this point of view finding direct evidence for quantum coherence amongst primordial fluctuations would be both surprising and arguably the most attractive option: since it would likely teach us the most.

\section*{Acknowledgements}

We thank Tim Cohen and Jerome Martin for many useful discussions. CB's research was partially supported by funds from the Natural Sciences and Engineering Research Council (NSERC) of Canada. Research at the Perimeter Institute is supported in part by the Government of Canada through NSERC and by the Province of Ontario through MRI. This work has been supported by STFC consolidated grant ST/X001113/1, ST/T000791/1, ST/T000694/1 and ST/X000664/1.


\appendix

\section{Flat space correlators}

In the text, we encounter several correlators of the form:
\bea \label{AppCorrDef}
  \cW_\cO(t,\bfx, s,\bfy)  &=& \llangle \, \cO(t,\bfx) \, \cO(s,\bfy) \,\rrangle - \llangle \, \cO(t,\bfx) \, \rrangle \; \llangle \, \cO(s,\bfy) \, \rrangle \nn\\
  &=& \Tr \Bigl[ \cO(t,\bfx) \, \cO(s,\bfy) \, \Xi_{\rm env}\Bigr] - \Tr \Bigl[ \cO(t,\bfx)  \, \Xi_{\rm env}\Bigr]\Tr \Bigl[ \cO(s,\bfy) \, \Xi_{\rm env}\Bigr] 
\eea
where $\cO = \sigma$ for the system and the trace is over the system states and $\cO = \mu^2 \phi$ or $\cO = g \phi^2$ when the trace is over the environmental states. This appendix evaluates these expressions in several simple examples in flat space.

\subsection{Flat space free correlator}

When $\cO =\phi$ or $\cO = \sigma$ and the state is prepared in the Minkowski vacuum $\varrho_0 = | \vac \rangle \, \langle \vac |$ then the correlation function $W_{\bf{k}}(t,s)$ is
\be
  W_{\bf{k}}(t,s) = \int \exd^3 q \; u_{q}(t)  u^{\ast}_{q}(s) \, \delta^{3}(\bfq - \bfk) =    u_{k}(t)  u^{\ast}_{k}(s)  \,,
\ee
which in flat space reduces to
\be \label{AppFlatWk}
W_{\bf{k}}(t,s) =  \frac{1}{2 \varepsilon_k} \, e^{-i\varepsilon_k(t-s)}
\,,
\ee
In position space this becomes (using $k^0 = \varepsilon_k$ and $x^0 = t$ and $y^0 = s$)
\bea \label{FinalDelta+}
   W(x-y) &=&  \int \frac{ \exd^3k }{(2\pi)^32\varepsilon_k} \; e^{ik \cdot (x-y)} \nn\\
   &=& \frac{m^2}{4\pi^2z} \; K_1(z) \qquad\qquad\qquad\qquad\quad\quad\;\; \hbox{for }\;\; (x-y)^2 > 0 \\ 
&=& \frac{m^2}{8\pi z} \; \left[ Y_1(z) + \hbox{sign}\left( z^0 \right) \; iJ_1(z) \right] \qquad \hbox{for } \;\; (x-y)^2 < 0 \,, \nn
\eea
in which the variables $z$ and $z^0$ are defined by $z^0 \equiv x^0 - y^0 = t-s$ and $z = m \sqrt{|(x-y)^2|}$. Here $J_1$ and $Y_1$ are the standard Bessel functions and $K_1$ is a modified Bessel function, and we use the integral representations
\bea
Y_1(z) &=& \frac{2z}{\pi} \int_0^\infty \frac{u^2 \; \exd u}{\sqrt{u^2+1}}  \;
\cos\left( z\sqrt{u^2+1}\right) \nn\\
J_1(z) &=& -\frac{2z}{\pi} \int_0^\infty
\frac{u^2 \; \exd u}{\sqrt{u^2+1}} \; \sin\left( z\sqrt{u^2+1}\right)
\eea
and
\be
K_1(z) = \int_0^\infty \frac{ u \; \exd u}{\sqrt{u^2 + 1}} \; \sin( uz).
\ee
The integral in \pref{FinalDelta+} converges absolutely if the time interval has a small imaginary part, so $t-s \to t-s -i\delta$. Using the definitions $H^{(1)}_1(u) = J_1(u) + i Y_1(u)$ and $H^{(2)}_1(u) = J_1(u) - i Y_1(u)$ and 
\bea
   K_1(u) &=& - \frac{\pi}{2} \, H^{(1)}_1(iu) \qquad \hbox{if} \qquad - \pi < \hbox{arg}\, u \leq \frac{\pi}{2} \nn\\
   \hbox{and} \qquad  K_1(u) &=&   - \frac{\pi}{2} \, H^{(2)}_1(-iu) \quad\; \hbox{if} \qquad - \frac{\pi}2 < \hbox{arg}\, u \leq \pi  \,,
\eea
shows that $W$ can be more compactly written
\be \label{AppFinalDeltaComplex}
   W(x-y) =  \frac{im^2}{8\pi w} \;  H^{(2)}_1(w)  \qquad \hbox{where } w := m \sqrt{-(x-y)^2} \,,
\ee
when the time difference has a negative imaginary part, provided the proper phases are kept when moving from the future to the past light cone.\footnote{For instance with Mathematica conventions we must take $w = -i z$ for spacelike separations.}

\subsubsection*{Massless limit}

In the massless limit this evaluates to 
\be
  W(t-s,\bfx-\bfy) = \langle \vac | \sigma(t,\bfx) \sigma(t',\bfx') | \vac \rangle = \frac{1}{4\pi^2 \big[ - (t - s - i \epsilon)^2 + |\bfx - \bfy|^2 \big]} \,,
\ee
This can be rewritten as a partial fraction, so
\be
  W(t-s,\bfx-\bfy) = \frac{1}{8\pi^2|\bfx-\bfy|} \left[ \frac{1}{|\bfx-\bfy| - (t - s - i \epsilon)} + \frac{1}{|\bfx-\bfy| + (t-s-i\epsilon)} \right] \,,
\ee
where $\epsilon\to 0^+$ at the end of the calculation. Together with the identity
\be \label{App0+Identity}
  \frac{1}{x+i \epsilon} = \frac{x}{x^2 + \epsilon^2} - \frac{i\epsilon}{x^2 + \epsilon^2} \to P \left( \frac{1}{x} \right) - i \pi \delta(x) \,,
\ee
this allows $W$ to be written in terms of explicitly real coordinates as
\bea
   W(t-s,\bfx-\bfy) &=& \frac{1}{8\pi^2|\bfx-\bfy|} \left\{ P \left[\frac{1}{|\bfx-\bfy| - (t - s)} \right] + P \left[ \frac{1}{|\bfx-\bfy| + (t-s)} \right] \right. \nn\\
   && \qquad\qquad \left. \phantom{\frac12} -i \pi \delta \Bigl[|\bfx-\bfy| - (t - s) \Bigr] + i \pi \delta \Bigl[|\bfx-\bfy| + (t - s) \Bigr] \right\} \,.    
\eea

\subsection{Flat space composite correlator}

The correlator of interest are those of \pref{AppCorrDef} with $\cO = g \phi^2$, in which we use the field expansion from Eq.~\pref{FieldExpansion} in the main text. Assuming the state satisfies $\mfc_{\mathbf{k}} |0 \rangle = 0$, this leads to
\be 
    \langle 0 | \phi^2(t,\bfx) | 0 \rangle = \int \frac{\exd^3k \, \exd^3 l}{(2\pi)^3} \; u_k(t) \, u^*_l(t) \, e^{i (\bfk+\bfl)\cdot \bfx} \langle 0 | \mfc_\bfk \, \mfc^*_{-\bfl} | 0 \rangle = \int \frac{\exd^3k}{(2\pi)^3} \;  u_k(t) u^*_k(t) \,,
\ee
and so 
\bea \label{correlatorIntermed1}
  \langle 0 | \phi^2(t,\bfx) \, \phi^2(s,\bfy) | 0 \rangle &=&  \int \frac{\exd^3k \, \exd^3 l \, \exd^3p \, \exd^3q}{(2\pi)^6} \; \Bigl[ u_k(t) u_l(t) u^*_p(s) u^*_q(s) \langle 0 | \mfc_\bfk \mfc_\bfl \mfc^*_{-\bfp} \mfc^*_{-\bfq} | 0 \rangle \\
  && \qquad + u_k(t) u^*_l(t) u_p(s) u^*_q(s) \langle 0 | \mfc_\bfk \mfc^*_{-\bfl} \mfc_\bfp \mfc^*_{-\bfq} | 0 \rangle \Bigr]  \, e^{i (\bfk+\bfl)\cdot \bfx + (\bfp+\bfq) \cdot \bfy} 
\,. \nn
\eea
Using the matrix elements
\bea
   \langle 0 |  \mfc_\bfk \mfc_\bfl \mfc^*_{-\bfp} \mfc^*_{-\bfq} | 0 \rangle &=&
\delta^3(\bfp+\bfk) \delta^3(\bfq+\bfl) +  \delta^3(\bfp+\bfl) \delta^3(\bfq+\bfk) \nn\\
   \hbox{and} \quad 
  \langle 0 |  \mfc_\bfk \mfc^*_{-\bfl} \mfc_\bfp \mfc^*_{-\bfq} | 0 \rangle &=& 
  \delta^3(\bfk+\bfl) \delta^3(\bfp+\bfq) \,, 
\eea
we see that the final line of \pref{correlatorIntermed1} is $\langle \phi^2(x) \rangle \, \langle \phi^2(y) \rangle$ and so \pref{AppCorrDef} becomes
\bea \label{AppCorrDefphisq}
  \cW_{\phi^2}(t,\bfx; s, \bfy)  &=& 
   2g^2 \int \frac{\exd^3p \, \exd^3 q}{(2\pi)^3} \; \, e^{i (\bfp+\bfq)\cdot (\bfx -\bfy)} u_p(t) u_q(t) u^*_p(s) u^*_q(s) \,.
\eea
Comparing \pref{AppCorrDefphisq} with \pref{cWtocWk} then shows $\cW_{\bf{k}}(t,s)$ takes the form
\be \label{AppWkCompo}
\cW_{\bf{k}}(t,s)  =  2g^2 \int \frac{\exd^3p \, \exd^3 q}{(2\pi)^3}\; u_{q}(t) u_{p}(t) u^{\ast}_{q}(s) u^{\ast}_{p}(s) \; \delta^{3}(\bf{q} + \bf{p} - \bf{k}) \,.
\ee
Once evaluated using the Minkowski vacuum and free mode functions \pref{AppMinkModes} this becomes
\be
\cW_{\bf{k}}(t,s)  =  \frac{g^2}{2(2\pi)^3} \int \exd^3 q \int \exd^3p \; \frac{1}{\varepsilon_p \, \varepsilon_q} \, e^{-i(\varepsilon_p + \varepsilon_q)(t-s)} \,   \delta^{3}(\bf{q} + \bf{p} - \bf{k}) \,.
\ee
Using rotational invariance in $\bfk$ and integrating over the delta function to perform the $\exd^3q$ integral implies $\bfq = \bfk - \bfp$ and so $q^2 = k^2 + p^2 - 2pk \cos \theta$ where $\theta$ is the angle between $\bfk$ and $\bfp$. Performing the integration using polar coordinates for $\bfp$ then leads to
\be
\cW_{\bf{k}}(t,s)  =  \frac{g^2}{2(2\pi)^2} \int_0^\infty \exd p \left( \frac{p^2}{\varepsilon_p} \right)\, e^{-i \varepsilon_p(t-s)} \int_{-1}^1 \exd u \; \frac{1}{\varepsilon_q} \, e^{-i\varepsilon_q(t-s)} \, \,,
\ee
where 
\be
   \varepsilon_p = \sqrt{p^2 + m^2} \quad \hbox{and} \quad
   \varepsilon_q = \sqrt{p^2 + k^2 + m^2 - 2kpu} \,.
\ee

Since $\exd \varepsilon_q/\exd u = kp/\varepsilon_q$ the $u$ integration can be traded for an integral over $\varepsilon_q$ with $p$ and $k$ fixed. This implies
\be
   \int_{-1}^1 \exd u \; \frac{1}{\varepsilon_q} \, e^{-i\varepsilon_q(t-s)} =\frac{1}{kp}  \int_{\varepsilon_-}^{\varepsilon_+} \exd \varepsilon \, e^{-i\varepsilon(t-s)} = \frac{i}{kp(t-s)} \left[ e^{-i \varepsilon_+(t-s)} - e^{-i \varepsilon_-(t-s)} \right]\,,
\ee
where $\varepsilon_\pm := \sqrt{(p \pm k)^2 + m^2}$. Therefore 
\bea
\cW_{-\bf{k}}(t,s)  = \cW_\bfk(t,s) &=& 
 \frac{ig^2}{2(2\pi)^2k(t-s)} \int_0^\infty \exd \varepsilon_p \, e^{-i \varepsilon_p  (t-s)} \left[ e^{-i \varepsilon_+(t-s)} - e^{-i \varepsilon_-(t-s)} \right]  \,,
\eea
where the last line uses $\varepsilon_p = \sqrt{p^2+m^2}$ and so $\varepsilon_p \exd \varepsilon_p = p \,\exd p$. 

\subsubsection*{Massless environment}

Although the remaining integral is messy when $m \neq 0$ it is simple when $m = 0$, since then $\varepsilon_p = p$ and $\varepsilon_\pm = \sqrt{(p\pm k)^2} = |p\pm k|$. We require
\bea
 I &:=& \int_0^\infty \exd p \, e^{-ip(t-s)} \Bigl[ e^{-i (p+k)(t-s)} - e^{-i|p-k|(t-s)} \Bigr] \nn\\ 
 &=& \int_0^\infty \exd p \,  e^{-i (2p+k)(t-s)} - \int_0^k \exd p \, e^{-ik(t-s)}  -  \int_k^\infty \exd p \,  e^{-i(2p-k)(t-s)}  \nn\\
 &=& \frac{1}{2i(t-s)}\,  e^{-ik(t-s)} - k e^{-ik(t-s)} -   \frac{1}{2i(t-s)}\,  e^{-ik(t-s)} 
 =   - k e^{-ik(t-s)}  \,.
\eea
This uses that the time interval $t-s$ has a small negative imaginary part, which makes the integral converge absolutely at $p \to \infty$. The final expression for the massless environmental correlator then is
\be \label{AppPhiSqMasslessWk}
\cW_{\bf{k}}(t,s)  =\cW_{-\bf{k}}(t,s)  = -  \frac{ig^2}{2(2\pi)^2(t-s)} \,  e^{-ik(t-s)} \,.
\ee

\section{de Sitter space correlators}

This appendix evaluates correlators of the form of \Eq{AppCorrDef} in several simple examples in de Sitter. It also contains the details of the evaluation of the purity for a minimally coupled system with a conformal environment in App. \ref{App:MCMSsystemCCMSenvironmentExplicit}.
 
\subsection{de Sitter free correlator}
\label{AppdSFreeCorr}

For the case of mixed fields in de Sitter space we can use the mode functions $u_k(\eta) \, e^{i \bfk \cdot \bfx}$ where we switch to conformal time and have (for massless scalars) the time-dependence  
\begin{eqnarray} \label{AppdSModes}
u_{k}(\eta) = ( - H \eta ) \;  \frac{e^{- i k \eta}}{\sqrt{ 2 k}} \left(1 - \frac{i}{k\eta} \right)  \qquad \qquad (\mathrm{de\ Sitter})
\end{eqnarray}
using the Bunch-Davies mode functions. This is conformal to the flat space result -- so $a u_k(\eta) \to u^{\rm flat}_k(\eta)$ -- in the remote past ($k\eta \to - \infty$) while $u_k(\eta) \to (iH/\sqrt{2k^3})$ in the far future ($k\eta \to 0$). For this choice we have
\begin{eqnarray}
  W(\eta,\bfx, \eta',\bfy) & = & \int \frac{ \exd^3k}{(2\pi)^3} \; W_{\bf{k}}(\eta,\eta') e^{+ i \bf{k} \cdot (\bf{x} - \bf{y})} \,,
\end{eqnarray}
with
\be \label{AppdSWk}
  W_{\bf{k}}(\eta,\eta') =    u_{k}(\eta)  u^{\ast}_{k}(\eta')  =  \frac{H^2\eta\eta'}{2 k} \, e^{-ik(\eta-\eta')} \left(1 - \frac{i}{k\eta} \right)  \left(1 + \frac{i}{k\eta'} \right) \,.
\ee

\subsubsection*{Arbitrary mass}

The massive mode function on de Sitter is given by
\be
   u_k(\eta) = C_k \, (-k\eta)^{3/2} H_\nu^{(1)}(-k\eta) \,,
\ee
where $\nu^2 = \frac94 - \frac{m^2}{H^2} - 12 \xi = \frac94 - \zeta^2$ in conventions where the nonminimal coupling is $\cL_{\rm nm} = +\frac12 \xi \sqrt{-g} \; \phi^2 R$. To fix $C_k$ expand $au_k$ the above for $- k \eta \gg 1$ in the remote past giving
\bea
a(\eta) u_k(\eta) \simeq C_{k} \frac{k}{H} \sqrt{\frac{2}{\pi}}\; e^{- i k \eta -\tfrac{i \pi \nu}{2} - \tfrac{ i \pi }{4}} \left\lbrace   1 + \mathcal{O}\left[ (- k \eta)^{-1}\right] \right\rbrace
\eea
and so matching to the flat space result $u_k^{\mathrm{flat}} \simeq e^{- i k \eta} / \sqrt{2k}$ sets
\bea
C_{k} = \frac{\sqrt{\pi} H}{2 k^{3/2}}\; e^{\tfrac{i \pi \nu}{2} + \tfrac{ i \pi }{4}} \quad \mathrm{and\ so} \quad  u_k(\eta) = \frac{\sqrt{\pi} H}{2}  e^{\tfrac{i \pi \nu}{2} + \tfrac{ i \pi }{4}} (-\eta)^{3/2} H_\nu^{(1)}(-k\eta) \ .
\eea
This shows that the massive correlator becomes
\be \label{AppWkmassivedS}
   W_k(\eta,\eta') =   \frac{\pi \, H^2 }{4} \, (\eta\eta)^{\tfrac{3}{2}} H_\nu^{(2)}(k\eta) \, \Bigl[  H_\nu^{(2)}(k\eta') \Bigr]^* \,.
\ee
\subsubsection*{Conformally coupled scalar}

For a conformally coupled scalar this gives the position-space result
\begin{eqnarray} \label{AppWkFTdefdSm0}
  W(\eta,\bfx; \eta',\bfy) & = & \int \frac{ \exd^3k}{(2\pi)^3} \; W_{\bf{k}}(\eta,\eta') e^{+ i \bf{k} \cdot (\bf{x} - \bf{y})} \nn\\
  &=& \frac{H^2 \eta\eta'}{4\pi^2 [-(\eta - \eta')^2-(\bfx-\bfy)^2]} \qquad \hbox{(conformally coupled)}  \,,
\end{eqnarray}
which (as expected) is conformal to the flat result. This mode sum can be performed more generally and for arbitrary mass and nonminimal coupling gives 
\cite{Bunch:1978yq}
\be \label{AppWkFTdefdSmass}
  W(\eta,\bfx; \eta',\bfy) = \frac{H^2}{16\pi} \left(\tfrac14 - \nu^2 \right) \sec(\pi \nu)\;  {}_2F_1\left[\tfrac32 + \nu, \tfrac32 - \nu; 2; z(\eta,\bfx; \eta', \bfy) \right]    \,, 
\ee
where
\be \label{Appzetadef}
   z(\eta,\bfx;\eta', \bfy) := 1 - \frac{1}{4\eta\eta'} \Bigl[-(\eta - \eta')^2 + (\bfx - \bfy)^2 \Bigr] \,,
\ee
and ${}_2F_1[a,b;c;z]$ is the usual hypergeometric function. The prefactor $(\frac14 - \nu^2) \sec(\pi \nu)$ can be usefully re-expressed using $\Gamma(1+z)=z\Gamma(z)$ and  the reflection formula $\Gamma(1-z)\Gamma(z) = \pi \csc(\pi z)$ to write
\be \label{AppDiffSecID}
 \Gamma(\tfrac32+\nu) \Gamma(\tfrac32-\nu) = (\tfrac14-\nu^2) \Gamma(\tfrac12+\nu) \Gamma(\tfrac12-\nu) = (\tfrac14-\nu^2) \pi \csc[\pi (\tfrac12 -\nu)]  = (\tfrac14-\nu^2) \pi \sec(\pi \nu) \,.
\ee

Notice the following useful transformation formulae that relate the solutions expanded about each of the three singular points ($z=0$, $z=1$ and $z=\infty$) 
\bea \label{Ftransformations}
  {}_2F_1[a,b;c;z] &=& (1-z)^{c-a-b} {}_2F_1[c-a,c-b;c;z] \nn\\
   {}_2F_1[a,b;c;z] &=& (1-z)^{-a} {}_2F_1\left[a,c-b;c;\frac{z}{z-1} \right]= (1-z)^{-b} {}_2F_1\left[c-a,b;c;\frac{z}{z-1} \right] \nn\\
  {}_2F_1[a,b;c;z]  &=& \frac{\Gamma(c)\Gamma(c - a - b)}{ \Gamma(c - a)\Gamma(c - b)}\,  F (a, b; a + b - c + 1;1 - z) \nn\\
 && \qquad +(1 - z)^{c-a-b }\frac{\Gamma(c)\Gamma(a + b - c)}{\Gamma(a)\Gamma(b)} \, F (c - a, c - b; c - a - b + 1;1 - z)  \nn\\
      {}_2F_1[a,b;c;z] &=& \frac{\Gamma(c) \Gamma(b-a)}{\Gamma(b) \Gamma(c-a)} (1-z)^{-a} {}_2F_1\left[a,c-b;a+1-b; \frac{1}{1-z} \right] \nn\\
  && \qquad\qquad + \frac{\Gamma(c) \Gamma(a-b)}{\Gamma(a) \Gamma(c-b)} (1-z)^{-b} {}_2F_1\left[b,c-a;b+1-a; \frac{1}{1-z} \right] \\
   {}_2F_1[a,b;c;z] &=& \frac{\Gamma(c) \Gamma(b-a)}{\Gamma(b) \Gamma(c-a)} (-z)^{-a} {}_2F_1\left[a,a+1-c;a+1-b; \frac{1}{z} \right] \nn\\
   && \qquad\qquad
   + \frac{\Gamma(c) \Gamma(a-b)}{\Gamma(a) \Gamma(c-b)} (-z)^{-b} {}_2F_1\left[b,b+1-c;b+1-a; \frac{1}{z} \right]    \,,\nn
\eea
where the last of these assumes $|\hbox{arg}\,z|<\pi$ and that $a-b$ is not an integer. The first of these identities implies $W$ has an equivalent representation
\be \label{AppWkFTdefdSmass2}
  W(\eta,\bfx; \eta',\bfy) = \frac{H^2}{16\pi} \left(\tfrac14 - \nu^2 \right) \sec(\pi \nu) \frac{{}_2F_1\left[\tfrac12 - \nu, \tfrac12 + \nu; 2; z(\eta,\bfx; \eta', \bfy) \right]  }{1-z}   \,.
\ee

A check on the above explores the coincident limit, $\bfy \to \bfx$ and $\eta' \to \eta$, in which case 
\be \label{Appzetaexpn}
   1-z(\eta,\bfx;\eta', \bfy) = \frac{1}{4\eta\eta'} \Bigl[-(\eta - \eta')^2 + (\bfx - \bfy)^2 \Bigr] \to 0 \,.
\ee
In this limit the third line of \pref{Ftransformations} implies the dominant part of the hypergeometric function is
\be
  {}_2F_1\left[\tfrac32 + \nu, \tfrac32 - \nu; 2; z \right]   \simeq   \frac{1}{\Gamma(\tfrac32 + \nu)\Gamma(\tfrac32 - \nu)} \, (1 - z)^{-1}    + \cdots = \frac{\cos(\pi \nu)}{\pi(\tfrac14-\nu^2)}  \, (1-z)^{-1} + \cdots \,, 
\ee 
which uses \pref{AppDiffSecID}. Eq.~\pref{AppWkFTdefdSmass} then becomes in this limit
\be 
  W(\eta,\bfx; \eta',\bfy) \simeq \frac{H^2}{16\pi^2(1-z)} + \cdots =   \left( \frac{H^2\eta\eta'}{4\pi^2} \right) \frac{1}{-(\eta - \eta')^2 + (\bfx - \bfy)^2} + \cdots  \,, 
\ee
which has the required small-separation Hadamard form \cite{Hadamard:1923, DeWitt:1960fc, Fulling:1978ht} $1/(4\pi^2 s)$ where 
\bea
  s(x,x') &:=& a(\eta) a(\eta')[-(\eta - \eta')^2 + (\bfx - \bfy)^2] \nn\\
  &=&  -\frac{1}{ H^2}  \left[ e^{H(t-t')} + e^{-H(t-t')} - 2 \right]^2 + e^{H(t+t')} (\bfx - \bfy)^2] \\
  &\simeq& - (t-t')^2 + e^{2Ht} (\bfx - \bfy)^2 + \cdots \,.\nn
\eea 

Conformally coupled scalars have\footnote{Notice we use the right metric and Weinberg curvature conventions while Birrel and Davies use the wrong metric and MTW curvature conventions.} $m = 0 $ and $\xi = + \frac16$ and so $\nu = \frac12$. In the limit $\nu \to  \frac12$ we have
\be
   \left( \tfrac14 - \nu \right)^2 \sec(\pi \nu) =  \frac{\left(\nu - \tfrac12 \right) \left(\nu +  \tfrac12 \right) }{\sin[\pi(\nu-\tfrac12)]} \to  \frac{1}{\pi} 
\ee
and 
\be
  {}_2F_1\left[\tfrac32 + \nu, \tfrac32 - \nu; 2; z] \right] \to  {}_2F_1\left[2, 1; 2; z \right] = \sum_{n=0}^\infty \frac{\Gamma(n+1)}{n!} z^n = \sum_{n=0}^\infty z^n = \frac{1}{1-z} \,,
\ee
and so \pref{AppWkFTdefdSmass} reproduces \pref{AppWkFTdefdSm0} in this limit.

\subsubsection*{Minimally coupled massless}

Minimally coupled massless fields (like the axion), by contrast, have $m = \xi = 0$ and so $\nu = \frac32$. In this case 
\be
   \sec(\pi \nu) = \frac{1}{\sin[\pi(\nu - \frac32)]} \to \frac{1}{\pi(\nu-\frac32)}  + \frac{\pi}{6} \left( \nu - \tfrac32 \right) + \cO \left[(\nu-\tfrac32)^3 \right] \,,
\ee
while 
\be
  {}_2F_1\left[\tfrac32 + \nu, \tfrac32 - \nu; 2; z \right] \to  {}_2F_1\left[3, 0; 2; z \right] = 1\,,
\ee
and so $W(\eta,\bfx;\eta',\bfy)$ famously diverges in this limit (due to strong IR fluctuations). This divergence comes purely as an additive constant, as can be seen by using \pref{AppDiffSecID} to write
\bea
  W(x,y) &=& \frac{H^2}{16\pi} (\tfrac14- \nu^2) \sec(\pi\nu) F[\tfrac32 + \nu, \tfrac32 - \nu; 2 ; z(x,y)] \nn\\
  &=& \frac{H^2}{16\pi} \Gamma(\tfrac32+\nu) \Gamma(\tfrac32 - \nu) F[\tfrac32+\nu, \tfrac32 - \nu; 2; z(x,y)] \nn\\
  &=&  \frac{H^2}{16\pi} \sum_{n=0}^\infty \frac{ \Gamma(\tfrac32 + \nu + n) \, \Gamma(\tfrac32 - \nu + n)}{\Gamma(2+n) \, n!} \; [z(x,y)]^n \,,
\eea
for which only the first term in the sum diverges in the limit $\nu \to \tfrac32$, leaving
\bea
 W(x,y) &\to&  \frac{H^2}{16\pi} (\tfrac14- \nu^2) \sec(\pi\nu) F[\tfrac32 + \nu, \tfrac32 - \nu; 2 ; z(x,y)] \nn\\
 &=& \Gamma(3) \, \Gamma(\tfrac32-\nu) + \sum_{n=1}^\infty \frac{ \Gamma(3 + n) \, \Gamma(n)}{\Gamma(2+n) \, n!} \; [z(x,y)]^n \nn\\
  &=& \frac{2}{\tfrac32 - \nu} + \sum_{n=1}^\infty \frac{(n+2)}{n} \, [z(x,y)]^n \\
  &=& \frac{2}{\tfrac32-\nu} + \frac{z(x,y)}{1-z(x,y)} - 2\log[1-z(x,y)]  \,.\nn
\eea
Finally, using \pref{Appzetadef} the massless minimally-coupled correlator becomes
\bea
   W(x,y) &=& C + \frac{1}{1-z} - 2 \log(1-z) \\
   &=& C +  \frac{4\eta\eta'}{-(\eta - \eta')^2 + (\bfx - \bfy)^2} - 2 \log \left[  \frac{-(\eta - \eta')^2 + (\bfx - \bfy)^2}{4\eta\eta'}\right] \,.\nn
\eea

The asymptotic approach to this singular form can be had by taking $\zeta^2:=(m/H)^2 + 12 \xi$ small but nonzero and so 
\be
   \nu = \sqrt{\frac94-\zeta^2} \simeq \frac32 - \frac{\zeta^2}3   + \cO(\zeta^4) \qquad \hbox{(for $\zeta \ll 1$)}\,. 
\ee
Using
\bea
  {}_2F_1[a,\epsilon;c;z] &=& 1 + \frac{a\epsilon}{c} \, z + \frac{a(a+1)\epsilon(\epsilon+1)}{c(c+1) } \left( \frac{z^2}{2!} \right) + \cdots \nn\\
  &=& 1 + \frac{a\epsilon z}{c}\left[1 +  \frac{(a+1)(\epsilon+1)}{c+1} \left( \frac{z}{2!} \right)+\cdots \right]
\eea
we can see that the expansion of $a=3+\cO(\epsilon)$ first contributes at order $\epsilon^2$ and so 
\be
  {}_2F_1[3+\cO(\epsilon),\epsilon;2;z] =  {}_2F_1[3,\epsilon;2;z]  + \cO(\epsilon^2) 
  = 1 - \frac{\epsilon}{2}\left[ \frac{z}{z-1}+2 \log (1-z) \right] + \cO(\epsilon^2) \,.
\ee
Therefore the minimally coupled small-mass limit of the propagator \pref{AppWkFTdefdSmass} becomes
\begin{eqnarray} \label{AppWkFTdefdSsmallmass}
  W(\eta,\bfx; \eta',\bfy) & = &  \frac{H^2}{16\pi} \left(\tfrac14 - \nu^2 \right) \sec(\pi \nu)\;  {}_2F_1\left[\tfrac32 + \nu, \tfrac32 - \nu; 2; z(\eta,\bfx;\eta',\bfy) \right] \nn\\
  &=&   -  \frac{H^2}{16\pi^2} \left\{  \frac{2}{ (\nu - \frac32)} + 3  +  \left[ \frac{z}{z-1}+2 \log (1-z) \right] _{z=z(\eta,\bfx;\eta',\bfy)} + \cO\left[ (\nu-\tfrac32 ) \right] \right\} \\
  &=&    \frac{H^2}{8\pi^2} \left\{  \frac{3}{ \zeta^2} - 2  +  \frac{2\eta\eta'}{-(\eta-\eta')^2+(\bfx-\bfy)^2} + \log \left[ \frac{4\eta\eta'}{-(\eta-\eta')^2+(\bfx-\bfy)^2} \right]  + \cO\left( \zeta^2 \right) \right\} \,. \nn
\end{eqnarray}

Notice that once the $3/\zeta^2$ is factored out then the coefficient inside the bracket of $\log|\bfx-\bfy|$ for large spatial separation is $-2\zeta^2/3$, as would be expected when a factor like $|\bfx-\bfy|^{-2\zeta^2/3}$ is expanded in powers of $\zeta^2$. This is the power found above on general grounds for small $\zeta$ (but without requiring that $\zeta^2 \log|\bfx-\bfy|$ also be small. To see why, notice that the correlation function falls off at spatial infinity (with $\eta$ and $\eta'$ fixed) and a rate that can be read off by recognizing that $|\bfx-\bfy| \to \infty$ corresponds to $-z \simeq  (\bfx-\bfy)^2/(4\eta\eta') \to \infty$. The last two of the identities \pref{Ftransformations} are useful because they allow the asymptotic forms in this limit to be read off once used together with $F[a,b;c;0]=1$. When used in \pref{WkFTdefdSmass}, together with the duplication formula $\Gamma(z) \Gamma(\tfrac12+z) = 2^{1-2z}\sqrt\pi \, \Gamma(2z)$, this leads to 
\bea \label{AppWkFTdefdSmass5}
  W(\eta,\bfx; \eta',\bfy) 
  &\simeq&  \frac{H^2}{16\pi} \left(\tfrac14 - \nu^2 \right) \sec(\pi \nu)\; \left\{ \frac{ \Gamma(-2\nu)}{\Gamma(\frac32-\nu) \Gamma(\frac12-\nu)} \left[ \frac{4\eta\eta'}{(\bfx-\bfy)^2} \right]^{\frac32+\nu} \left(1+ \cdots\right)   \right. \nn\\
  && \qquad \qquad \left. +  \frac{ \Gamma(2\nu)}{\Gamma(\frac32+\nu) \Gamma(\frac12+\nu)}  \left[ \frac{4\eta\eta'}{(\bfx-\bfy)^2 }\right]^{\frac32-\nu} \left(1+ \cdots\right)  \right\} \nn\\
  &=&   \frac{H^2}{16\pi^2} \left\{ \frac{\Gamma(\frac32+ \nu) \Gamma(-2\nu)}{ \Gamma(\frac12-\nu)} \left[ \frac{4\eta\eta'}{(\bfx-\bfy)^2} \right]^{\frac32+\nu}\left(1+ \cdots\right)   \right. \\
  && \qquad \qquad \left. +  \frac{\Gamma(\frac32-\nu) \Gamma(2\nu)}{ \Gamma(\frac12+\nu)}  \left[ \frac{4\eta\eta'}{(\bfx-\bfy)^2 }\right]^{\frac32-\nu} \left(1+ \cdots\right)  \right\}\,, \nn 
\eea
as $|\bfx-\bfy| \to \infty$. Both terms fall like $|\bfx-\bfy|^{-3}$ when $\nu$ is imaginary ({\it i.e.}~when $m \gg H$). It would predict $|\bfx-\bfy|^{-2}$ as found above when $\nu = \frac12$  (a conformal scalar as given in \pref{WkFTdefdSm0}) and no falloff at all if $\nu = \frac32$ (minimally coupled massless scalar --  see below). If $\nu = \frac32 - \frac13 \zeta^2$ for small nonzero $\zeta^2$ (such as for a light minimally coupled scalar) then the falloff power is $|\bfx-\bfy|^{-2\zeta^2/3}$. 

\subsection{Momentum-space de Sitter composite correlator}
\label{AppdSCompCorr}

The de Sitter correlator of interest again uses $\cO = g \phi^2$, which leads to the same formula \pref{AppWkCompo}. We here evaluate \pref{AppWkCompo} using the Bunch-Davies mode functions \pref{CCMSmodes} for conformally coupled massless fields. This leads to the simpler answer
\be
\cW_{-\bf{k}}(\eta, \eta')  =  \frac{g^2(H^2\eta\eta')^2}{2(2\pi)^3} \int \exd^3 q \int \exd^3p \; \frac{1}{p \,q} \, e^{-i(p+q)(\eta-\eta')}   \delta^{3}(\bf{q} + \bf{p} + \bf{k}) \,.
\ee
Using the delta function to perform the $\exd^3q$ integral implies $\bfq =-( \bfk + \bfp)$ and so $q^2 = k^2 + p^2 + 2pk \cos \theta$ where $\theta$ is the angle between $\bfk$ and $\bfp$. Performing the integration using polar coordinates for $\bfp$ then leads to
\be
\cW_{-\bf{k}}(\eta,\eta') =  \frac{g^2(H^2\eta\eta')^2}{2(2\pi)^2} \int_0^\infty \exd p \, p \, e^{-i p(\eta-\eta')}    \int_{-1}^1 \exd u \; \frac{1}{q} \, e^{-iq(\eta - \eta')}    \, \,,
\ee
where $q(u) = \sqrt{p^2 + k^2 + 2kpu}$.  Since $\exd q/\exd u = kp/q$ the $u$ integration can be traded for an integral over $q$ with $p$ and $k$ fixed, giving
\be
 \int_{-1}^1 \exd u \; \frac{1}{q} \, e^{-iq(\eta-\eta')} = \frac{1}{pk}  \int_{|p-k|}^{p+k} \exd q \, e^{-iq(\eta - \eta')}= \frac{i}{ pk (\eta-\eta')} \Bigl[e^{-i (p+k) (\eta-\eta')}  -   e^{-i |p-k| (\eta-\eta')}  \Bigr]  \,,  
\ee
and so
\be \label{AppcubicWkdSc}
\cW_{-\bf{k}}(\eta,\eta')  = \frac{ig^2( H^2 \eta \eta')^2}{(2\pi)^2 2k  (\eta-\eta')} \int_0^\infty \exd p \,  e^{-i p(\eta-\eta')}  \Bigl[e^{-i (p+k) (\eta-\eta')}  -   e^{-i |p-k| (\eta-\eta')}  \Bigr] \,. 
\ee
To simplify the absolute value, break the integration region into the interval $0 < p < k$ and $k < p$ so
\bea
\cW_{-\bf{k}}(\eta,\eta') & = & \frac{ig^2( H^2 \eta \eta')^2}{(2\pi)^2 2k  (\eta-\eta')} \bigg\{ \int_0^k \exd p \, e^{-i p(\eta-\eta')}  \Bigl[e^{-i (p+k) (\eta-\eta')}  -   e^{-i (k-p) (\eta-\eta')}  \Bigr] \\
&& + \int_k^\infty \exd p   \Bigl[e^{-i (2p+k) (\eta-\eta')}  -   e^{-i (2p-k) (\eta-\eta')}  \Bigr] \bigg\} \, .\notag
\eea
Evaluating the above gives rise to 
\be \label{AppcWkdScubicc}
 \cW_{-\bfk}(\eta, \eta') = -\frac{ig^2( H^2 \eta \eta')^2}{2(2\pi)^2(\eta - \eta')} \, e^{-ik(\eta - \eta')} \,.
\ee
as quoted in Eq.~(\ref{cWkdScubicc}) in the main text.

\subsection{Purity for minimally coupled system with conformal environment}
\label{App:MCMSsystemCCMSenvironmentExplicit}

All integrals can be done very explicitly in the case of cubic coupling with the environment given by a conformally coupled massless scalar ($\nu_{\rm env}=\frac12$) and the system given by a minimally coupled massless scalar. In this case \pref{dSWkMM} gives the system correlator 
\be \label{AppdSWkMM2}
  W_{\bf{k}}(\eta,\eta')  =  \frac{H^2\eta\eta'}{2 k} \, e^{-ik(\eta-\eta')} \left(1 - \frac{i}{k\eta} \right)  \left(1 + \frac{i}{k\eta'} \right) \,,
\ee
and \pref{cWkdScubicc} gives the environmental correlator
\be 
 \cW_{-\bfk}(\eta, \eta')  = -\frac{ig^2( H^2 \eta \eta')^2}{2(2\pi)^2(\eta - \eta')} \, e^{-ik(\eta - \eta')}  \,.
\ee

With these choices \pref{PurityChangeMasterEqFflatdS2} evaluates to
\be   \label{AppPurityChangeMasterEqFflatdS2w}
  \partial_\eta \gamma_\bfk   =   - 4  \int_{-\infty}^\eta \frac{\exd \eta'}{ (H^2\eta\eta')^4} \hbox{Re}\Bigl[ W_\bfk(\eta,\eta') \, \cW_{-\bfk}(\eta,\eta') \Bigr] =  - \frac{g^2 }{4\pi^2k} \; \cI \,,
\ee
with integral
\bea  \label{AppPurityChangeMasterEqFflatdS2wy}
  \cI &:=&   \int_{-\infty}^\eta \frac{\exd \eta'}{  H^2\eta\eta' }\; \hbox{Im}\left[ \left(1 - \frac{i}{k\eta} \right)  \left(1 + \frac{i}{k\eta'} \right)  \frac{ e^{-2ik(\eta - \eta')}}{\eta - \eta'} \right] \nn\\
  &=&  - \frac{1 }{ H^2k^2\eta^4}  \text{Im} \left\{  (1+ ik \eta) \left[ -(1+ i k \eta) \text{Ei}(2 i k \eta' ) \,  e^{-2 i k \eta} \phantom{\frac12} \right.\right. \\
  && \qquad\qquad\qquad\qquad \left.\left. + (1-i k \eta) \text{Ei}[-2 i k (\eta -\eta')]+ \frac{\eta}{\eta'} \, e^{-2 i k(\eta- \eta') }\right]  \right\}_{\eta'=-\infty}^{\eta'=\eta}   \,. \nn
\eea
The limits must be evaluated with care, keeping in mind that $\eta$ has a small negative imaginary part though $\eta'$ does not (because we require $\eta - \eta'$ to have the imaginary part in order for subsequent $k$ integrations to converge for large $k$). Writing the complex-valued time coordinate as $\tilde \eta := \eta - i \epsilon$ for $\epsilon > 0$ taken to zero at the end, the precise integral to be evaluated is  
\bea  \label{AppPurityChangeMasterEqFflatdS2wd}
  \cI  &=&  - \frac{1 }{ H^2k^2\eta^4} \text{Im} \left\{  (1+ ik \tilde\eta) \left[ -(1+ i k \tilde\eta) \text{Ei}(2 i k \eta' ) \,  e^{-2 i k \tilde\eta} \phantom{\frac12} \right.\right. \nn\\
  && \qquad\qquad\qquad\qquad \left.\left. + (1-i k \tilde\eta) \text{Ei}[-2 i k (\tilde\eta -\eta')]+ \frac{\tilde\eta}{\eta'} \, e^{-2 i k(\tilde\eta- \eta') }\right]  \right\}_{\eta'=-\infty}^{\eta'=\eta} \\
  &=&   - \frac{1 }{ H^2k^2\eta^4} \text{Im} \left\{  (1+ ik \tilde\eta) \left[ -(1+ i k \tilde\eta) \Bigl( \text{Ei}(2 i k \eta ) + i \pi \Bigr) \,  e^{-2 i k \tilde\eta} \phantom{\frac12} \right.\right. \nn\\
  && \qquad\qquad\qquad\qquad \left.\left. + (1-i k \tilde\eta)\Bigl( \text{Ei}[-2 i k (\tilde\eta -\eta)] + i \pi \Bigr) + \frac{\tilde\eta}{\eta} \, e^{-2 i k(\tilde\eta- \eta) }\right]  \right\}    \,,  \nn
\eea
which uses \pref{AppEizImag} to evaluate the asymptotic form Ei$(-i\infty) = - i \pi$ for large argument. 

Now comes the main subtlety. The difference between $\tilde \eta = \eta - i \epsilon$ and $\eta$ only really matters when evaluating the singular limit Ei$[-2ik(\tilde \eta-\eta)] = \text{Ei}(-2k\epsilon)$. For this we use \pref{AppEizsmall} to get
\be\label{EiSmallRealz}
   \text{Ei}(-2k\epsilon) = \log(-2ik\epsilon) + \mfC - \frac{i\pi}{2} + \cO(\epsilon)  = \log(2k\epsilon) + \mfC - i\pi + \cO(\epsilon) \,.
\ee
Using this together with Im$\,\{(1-ik \tilde \eta)(1+ik\tilde \eta)[ \log(2k\epsilon) + \mfC]\} \to 0$ as $\epsilon \to 0$ leads in this limit to
\be   \label{AppPurityChangeMasterEqFflatdS2wcd}
  \cI   \to   - \frac{1 }{ H^2k^2\eta^4} \text{Im} \left\{  (1+ ik \eta) \left[ -(1+ i k \eta) \Bigl( \text{Ei}(2 i k \eta ) + i \pi \Bigr) \,  e^{-2 i k \eta}   + 1\right]  \right\}    \,.
\ee

We can now evaluate the small-$\eta$ limit of this result -- again using \pref{AppEizsmall} -- to capture the super-Hubble evolution, giving 
\bea  \label{AppPurityChangeMasterEqFflatdS2cw2}
  \cI  &=&   - \frac{1 }{ H^2k^2\eta^4} \text{Im} \Bigl[  1+ ik \eta  -(1+ i k \eta)^2 \Bigl( \text{Ei}(2 i k \eta ) + i \pi \Bigr) \,  e^{-2 i k \eta}   \Bigr]  \nn\\
   &\simeq&   - \frac{1 }{ H^2k^2\eta^4} \text{Im} \left\{  1+ ik \eta  - (1+ i k \eta)^2 \left[ \log(-2k\epsilon) + \mfC + \frac{i\pi}{2} + 2ik\eta - (k\eta)^2 + \cdots \right]   e^{-2 i k \eta}   \right\} \nn\\
  &=& \frac{1 }{ H^2k^2\eta^4} \Bigl\{ \frac{\pi}{2}\left[1 + (k\eta)^2 \right] + k \eta  + \cO[(k\eta)^3] \Bigr\}  \,,  
\eea
and so \pref{AppPurityChangeMasterEqFflatdS2w} becomes
\be   \label{AppPurityChangeMasterEqFflatdS2w22}
  \partial_\eta \gamma_\bfk  =  - \frac{g^2 }{4\pi^2k} \; \cI = - \frac{g^2 }{4\pi^2 H^2k^3\eta^4} \Bigl\{ \frac{\pi}{2} + k \eta  + \cO[(k\eta)^2] \Bigr\} \,,
\ee
Eq.~\pref{AppPurityChangeMasterEqFflatdS2cw2} can be compared with the contribution $\frac12 i\pi$ coming from the $i \pi\delta(\eta-\eta')$ in the imaginary part of $(\eta - \eta')^{-1}$ which if used directly in \pref{AppPurityChangeMasterEqFflatdS2wy} would have given 
\be  \label{AppPurityChangeMasterEqFflatdS2wy2}
  \cI =   \int_{-\infty}^\eta \frac{\exd \eta'}{  H^2\eta\eta' }\; \hbox{Im}\left[ \left(1 - \frac{i}{k\eta} \right)  \left(1 + \frac{i}{k\eta'} \right)  \frac{ e^{-2ik(\eta - \eta')}}{\eta - \eta'} \right]  \simeq   \frac{\pi}{2  H^2 k^2\eta^4 }\Bigl[1 + (k \eta)^2 \Bigr]  \,,  
\ee 
which agrees on the leading term.

\section{Special functions and asymptotic forms}
	\label{AppAsymptotic}

This appendix collects for completeness several useful asymptotic forms used in the main text.

The main text uses the exponential integral function, defined by 
\be \label{AppEizDef}
   \hbox{Ei}(z) := - \int_{-z}^\infty \exd t \;  \frac{e^{-t}}{t}   \,,
\ee
in which the principal part is taken if the integration passes through $t = 0$. This has the asymptotic form
\be \label{AppEizsmall}
  \hbox{Ei}(-ix) \simeq  \log (x) + \mfC - \frac{i \pi}{2} -i x-\frac{x^2}{4}+\frac{i x^3}{18}+\frac{x^4}{96}-\frac{i x^5}{600}+\cO\left(x^6\right)\qquad (\hbox{if } 0 < x \ll 1) \,,
\ee
where $\mfC \simeq 0.577216...$ is Euler's constant, and 
\bea \label{AppEizImag}
  \hbox{Ei}(-ix) &\simeq& -i \pi + \left(\frac{i}{x}-\frac{1}{x^2}-\frac{2 i}{x^3}+\frac{6}{x^4}+\frac{24 i}{x^5} \right) e^{-i x} +\cO\left(x^{-6}\right) \qquad (\hbox{if } x \gg 1 )\,.
\eea

Also used is the cosine integral function
\be \label{AppCizDef}
   \hbox{Ci}(z) := - \int_{-z}^\infty \exd t \;  \frac{\cos t}{t}   \,,
\ee
in which the principal part is taken if the integration passes through $t = 0$. This has the asymptotic form
\be \label{AppCizsmall}
  \hbox{Ci}(x) \simeq  \log (x) + \mfC -\frac{x^2}{4}+\frac{ x^4}{96}+\cO\left(x^6\right)\qquad (\hbox{if } 0 < x \ll 1) \,,
\ee
and 
\bea \label{AppCizImag}
  \hbox{Ci}(x) &\simeq& \cos x \left[- \frac{1}{x^2} + \frac{6}{x^4} + \cdots\right]
  + \sin x \left[ \frac{1}{x} - \frac{2}{x^3} + \frac{24}{x^5} + \cdots \right] + \cO(x^{-6}) \,,
\eea
for large arguments with $x \gg 1$. Similarly
\be
  \text{Si}(z) := \int_0^z \exd t \; \frac{\sin t}{t} = - \text{Si}(-z) \,,
\ee
has the small-argument expansion 
\be
   \text{Si}(z) = \sum_{n=0}^\infty (-)^n \frac{z^{2n+1}}{(2n+1)(2n+1)!} = z - \frac{z^3}{18} + \frac{z^5}{600} + \cdots \,.
\ee 
For large arguments one instead finds
\be
  \text{Si}(z) = \frac{\pi}{2} - \frac{\cos z}{z} + \cO[z^{-2}] \,.
\ee

The generalized hypergeometric functions are defined for small arguments by the series
\be
   {}_1F_2[a; b,c; z] = \frac{\Gamma(b) \, \Gamma(c)}{\Gamma(a)} \sum_{n=0}^\infty \frac{\Gamma(n+a)}{\Gamma(n+b) \Gamma(n+c)} \, \frac{z^n}{n!}
\ee
and so ${}_1F_2[a;b,c;0] = 1$. They have the following large-argument limits
\bea \label{App1F2Asymptotics}
   {}_1F_2\left[ \tfrac14; \tfrac54, \tfrac32; -x^2\right] &=& \frac{1}{(4x)^2} \Bigl( e^{-2ix+i\pi} + e^{2ix-i\pi} \Bigr) + \cO(x^{-3}) \nn\\
   {}_1F_2\left[ -\tfrac14; \tfrac12, \tfrac34; -x^2\right] &=& - \frac{1}{8x} \Bigl( e^{-2ix+i\frac{\pi}{2}} + e^{2ix-i\frac{\pi}2} \Bigr) + \cO(x^{-2})   \,,
\eea

The Hankel function asymptotic forms are given for small $z$ by
\be\label{AppHankelSmall}
  H^{(1)}(z) \simeq \left( \frac{z}{2} \right)^{-\nu} \bigg[ -\frac{i \Gamma (\nu ) }{\pi } + \mathcal{O}(z^2) \bigg] + \left( \frac{z}{2} \right)^{\nu} \left[ \frac{ 1+i \cot (\pi  \nu )}{\Gamma (\nu +1)} + \mathcal{O}(z^2) \right]
\,.
\ee

The Struve functions $\pmb{H}_\nu(z)$ similarly have the asymptotic forms
\be
  \pmb{H}_\nu(z) = \frac{1}{\sqrt\pi \Gamma(\frac32+\nu)} \left( \frac{z}{2} \right)^{\nu+1} \left[ 1 - \frac{z^2}{3(3+2\nu)} + \cO(z^4) \right] 
\ee
for small argument and
\be
 \pmb{H}_\nu(z) \simeq \frac{1}{\sqrt\pi \Gamma(\frac12+\nu)} \left( \frac{2}{z} \right)^{1-\nu} - \sqrt{\frac{2}{\pi z}}  \sin \left[z + \frac{\pi}{4} (1+2\nu)\right]   + \cdots  
\ee
for large argument.

\section{Non-perturbative purity from transport equation}\label{app:transport}

In this Appendix, we derive an alternative non-perturbative expression for the purity of the system from the dynamical equations of the two-point functions known as transport equations \cite{ehrenfest_bemerkung_1927}. Indeed, Gaussianity allows us to directly extract the purity from the determinant of the covariance matrix of the system \cite{Serafini:2003ke}. Hence, following \cite{Colas:2022hlq}, we adopt the following strategy which consists in deriving the transport equations, solving them non-perturbatively from which we deduce the purity. 

Starting from the Lindblad equation of Eq.~\pref{LindbladHFmom}, we first derive the adjoint master equation \cite{breuerTheoryOpenQuantum2002} which controls the evolution of the expectation value of an operator $\langle {\mathcal{O}} \rangle (t) = \mathrm{Tr}[\mathcal{O}  \varrho  (t)]$
\be\label{eq:adjoint}
	\partial_t \langle {\mathcal{O}} \rangle (t) = i \langle \left[{H}_0(t) + {V}_{\mathrm{eff}}(t), {\mathcal{O}}\right]\rangle(t) - a^6(t)   \int \exd^3k  \kappa_\bfk(t)  \langle \left[ {\sigma}_{\bfk}, \left[ {\sigma}_{-\bfk}, {\mathcal{O}}\right]\right]\rangle (t)
\ee
where $H_0(t)$ is the unperturbed Hamiltonian. ${V}_{\mathrm{eff}}(t)$ is defined in Eq.~\pref{VeffDefFmom} and encodes the renormalization of the mass of the $\sigma$ field due to its interactions with $\phi$. $\kappa_\bfk(t)$ is given in Eq.~\pref{kappakdef} and controls quantum diffusion due to the fluctuations of $\phi$ which acts as a source term for quadratic observables due to the double commutator structure.

From the adjoint master equation \pref{eq:adjoint}, one can derive the dynamical equation of the covariance matrix of a mode $\bfk$ defined as 
\be
	\bm{\Sigma}_{\bfk,ij}(t) \equiv \frac{1}{2}\mathrm{Tr} \left[\left( {\bm{z}}^i_{\bfk} {\bm{z}}^j_{-\bfk} + {\bm{z}}^j_{\bfk} {\bm{z}}^i_{-\bfk}\right) \varrho  (t)\right] 
\ee
where we defined the phase-space operator $ {\bm{z}}_{\bfk}  \equiv \left( {\sigma}_{\bfk}  , \pi_{\bfk} \right)^{\mathrm{T}}$, $\pi_{\bfk}$ being conjugate momentum of ${\sigma}_{\bfk}$. Making use of the canonical commutation relations, we obtain from Eq.~\pref{eq:adjoint}
\be
	\partial_t  \bm{\Sigma}_\bfk = \bm{\omega} \bm{H}_\bfk \bm{\Sigma}_\bfk - \bm{\Sigma}_\bfk \bm{H}_\bfk  \bm{\omega} + \bm{D}_\bfk
\ee
where we defined $\bm{\omega} \equiv \begin{pmatrix}
	0 & 1 \\
	-1 & 0
\end{pmatrix}$, the Hamiltonian matrix and the diffusion matrix being given by
\be
{H}_0(t) + {V}_{\mathrm{eff}}(t) \equiv \frac{1}{2} \int \exd^3k \bm{{z}}_{\bfk}^{\dag} \bm{H}_\bfk \bm{{z}}_{\bfk} \qquad \mathrm{and} \qquad	\bm{D}_\bfk \equiv \begin{pmatrix}
		0 & 0 \\
		0 &  a^6(t)  \kappa_\bfk(t)
	\end{pmatrix}
\ee
respectively. A non-perturbative solution to this equation is given in terms of the sum of a homogeneous and inhomogeneous part 
\be\label{eq:non-pert}
	\bm{\Sigma}_\bfk(t) = \bm{\Sigma}^{(h)}_\bfk(t) + \int_{t_0}^t \dd t' \bm{G}_\bfk(t,t').\bm{D}_\bfk(t').\bm{G}^{\mathrm{T}}_\bfk(t,t')
\ee
where the retarded Green's matrix is obtained from the modified mode functions $u_k(t)$ and $\pi_k(t)$
\be
	\bm{G}_\bfk(t,t') = 2 \begin{pmatrix}
		\Ima\left[u_k(t) \pi_k^*(t')\right] & -  \Ima\left[u_k(t) u_k^*(t')\right]\\
		\Ima\left[\pi_k(t) \pi_k^*(t')\right] & -\Ima\left[\pi_k(t) u_k^*(t')\right]
	\end{pmatrix}
\ee
which are obtained by updating the equations of motion of the field variables by including the mass renormalization $m^2 \rightarrow m^2 + 2 a^3(t) h_{\bm{q}}(t)$. The homogenous covariance matrix captures the unitary corrections to the power spectra 
\be
	\bm{\Sigma}^{(h)}_\bfk(t) = \begin{pmatrix} \left|u_k(t) \right|^2 &  \Rea{\left[u_k(t)\pi_k^*(t)\right]} \\
		\Rea{\left[u_k(t)\pi_k^*(t)\right]}        & \left|\pi_k(t) \right|^2
	\end{pmatrix}. 
\ee

For Gaussian states, the purity directly relates to the determinant of the covariance matrix through \cite{Serafini:2003ke, Colas:2021llj}
\be\label{detcov}
	\gamma_\bfk(t) = \frac{1}{4 \mathrm{det}\left[ \bm{\Sigma}_\bfk(t) \right]}.
\ee
We can now compute $\mathrm{det}\left[ \bm{\Sigma}_\bfk(t)\right]$ directly from \Eq{eq:non-pert}. Using the fact that $ \mathrm{det}[ \bm{\Sigma}^{(h)}_\bfk(t)]= 1/4$ from the unitarity of the Hamiltonian evolution and the Wronskian condition $u_k(t) \pi_k^*(t) - u_k^*(t) \pi_k(t)  = i$, we obtain the non-perturbative expression of the purity 
\be\label{eq:purnonpert}
	\gamma_\bfk(t) = \left[1 + 4 \int_{t_0}^t \dd t' a^6(t') \kappa_\bfk(t) \left|u_k(t') \right|^2 \right]^{-1}.
\ee
Notice that this expression matches Eq.~\pref{purityvsABMatch} in its perturbative limit, as it should. Injecting the (mass dressed) mode functions solutions together the late-time limit of $\kappa_\bfk$ given in Eq.~\pref{kappakBDx3}, we recover the main text result
\be
\gamma_\bfk(\eta) \simeq \left[ 1 +  \frac{g^2}{32\pi^2 H^2 \nu_{\rm sys}} \Bigl| 2^{\nu_{\rm sys}} \Gamma(\nu_{\rm sys}) \Bigr|^2 (-k\eta)^{-2\nu_{\rm sys}} \right]^{-1}.
\ee
The perturbative and resummed results are compared in \Fig{fig:resum}. 

\section{Environmental mass dependence in perturbation theory}
\label{App:PT_mass}

This section supplements the results from \S\ref{sssec:FourierGeneral} in the main text. In particular, this Appendix fleshes out the details on how the purity in perturbation theory depends on an environment with arbitrary mass, parametrized by (see Eq.~(\ref{nu_xi_def}))
\be
\nu_{\rm env} = \sqrt{ \tfrac{9}{4} - \zeta_{\rm env}^2 } \qquad\ \mathrm{with}\ \zeta_{\rm env}^2 = \frac{m_{\rm env}^2}{H^2} + 12 \xi_{\rm env}\ .
\ee
Assuming $\eta_0 \to - \infty$ in de Sitter space gives the purity in perturbation theory
\be
  \gamma_\bfk = 1 - \frac{g^2}{128 H^2} \, \cF(-k\eta) \,,
\ee
with the dimensionless function $\cF$ defined by
\bea
   \cF(-k\eta) &:=& \int \exd^3 p \, \exd^3q \; \Bigl| \widehat \cN_{pqk}(\eta)  \Bigr|^2 \delta^3(\bfp+\bfq-\bfk) \\
   \hbox{with} \qquad 
  \widehat \cN_{pqk}(\eta)  &:=&  \int_{-\infty}^\eta \exd s \;  \sqrt{ - s } \; H_{\nu_{\rm sys}}^{(1)}(- k s) H_{\nu_{\rm env}}^{(1)}(- p s) H_{\nu_{\rm env}}^{(1)}(- q s)  \nn
\eea
as given in Eq.~(\ref{PurityMasterEqFNestedMixCosmo321}). Ultimately our goal is to understand the scaling of $1 - \gamma_\bfk \propto \mathcal{F}$ in the super-Hubble limit $- k \eta \ll 1$. As discussed in \S\ref{sssec:FourierGeneral} we provide numerical evidence as well as analytic formulas in this limit.

\subsection{Numerical Evaluation}

The function $\cF$ is usefully written for analytic manipulation after integrating over the angles $(\theta_p,\varphi_p)$ and $(\theta_{q}, \varphi_{q})$ in the $\delta^3$-function (as done in \cite{Burgess:2022nwu}), which writes the above as
\be
   \cF(-k\eta) = \frac{2\pi}{k} \iint_{U_k} \exd p\; \exd q \; p q \; \Bigl| \widehat \cN_{pqk}(\eta)  \Bigr|^2
\ee
which integrates $(p,q)$ over the tilted rectangular region $U_k = \left\{ - 2 p k < p^2 + k^2 - q^2 < 2 p k \right\}$ (arising from a $\delta$-function integration over the polar angle $\theta_p$). It is then convenient to rotate the variables $(p,k)$ and define the integration variables 
\be
a = \frac{p - q}{k} \ , \qquad b = \frac{p+q}{k} \ , \qquad z' =  - k s
\ee
which simplifies the above expression to 
\be \label{Ffunction_ab}
   \cF(z)  =  \frac{\pi}{2} \int_{0}^{1} \exd a \int_{1}^\infty \exd b\;   (b^2 -a^2) \bigg| \int_{z}^\infty \exd z' \; \sqrt{ z' } \; H_{\nu_{\rm sys}}^{(1)}(z') H_{\nu_{\rm env}}^{(1)}( \tfrac{b-a}{2} z')  H_{\nu_{\rm env}}^{(1)}( \tfrac{b+a}{2} z')  \bigg|^2
\ee
after also using symmetry of the $a$-integrand under reflections and using the shorthand
\be
z := - k \eta > 0 
\ee
which is positive and dimensionless. We emphasize that the form (\ref{Ffunction_ab}) in terms of $a$ and $b$ is useful for numerical and analytical investigation since the $a$-integral has a finite range. It turns out that integrating over all four integrals in (\ref{Ffunction_ab}) is numerically very exhaustive. However, instead writing $\cF(z)$ in (\ref{Ffunction_ab}) as
\be
   \cF(z)  = \int_{0}^{1} \exd a\; g(z,a)
\ee
with the definition
\be \label{gnu_def}
 g(z,a) := \frac{\pi}{2} \int_{1}^\infty \exd b\;   (b^2 -a^2) \bigg| \int_{z}^\infty \exd z' \; \sqrt{ z' } \; H_{\nu_{\rm sys}}^{(1)}(z') H_{\nu_{\rm env}}^{(1)}( \tfrac{b-a}{2} z')  H_{\nu_{\rm env}}^{(1)}( \tfrac{b+a}{2} z')  \bigg|^2 \ ,
\ee
it turns out that numerical integration of $g$ turns out to be reasonably efficient, which is the reason why we study $g$ in Figure \ref{Fig:NumericalEnvironment}. Since $\mathcal{F}$ integrates $g$ over a {\it finite} range of $a$, one can conclude that the late-time scaling of $g(z,a)$ is the same as that of $\cF(z)$ for $z \ll 1$, which is why we plot $g$ in Figure \ref{Fig:NumericalEnvironment} in the main text. 

\subsection{Super-Hubble scaling when $\frac{3}{4} < \nu_{\rm env} < \frac{3}{2}$}
\label{App:34nu32}

The super-Hubble scaling in this case turns out to be of the form $\cF(z) \propto z^{3 - 2 \nu_{\rm sys}  - 4 \nu_{\rm env}}$. The easiest way to see this is to take expression (\ref{Ffunction_ab}) and rescale the integration variable
\be
y = z' / z
\ee
which turns (\ref{Ffunction_ab}) into
\be \label{Ffunction_yy}
   \cF(z) = \frac{\pi z^3}{2} \int_{0}^{1} \exd a \int_{1}^\infty \exd b\; (b^2 -a^2) \bigg| \int_{1}^\infty \exd y \; \sqrt{ y } \; H_{\nu_{\rm sys}}^{(1)}(z y) H_{\nu_{\rm env}}^{(1)}( \tfrac{b-a}{2} z y )  H_{\nu_{\rm env}}^{(1)}( \tfrac{b+a}{2} z y )  \bigg|^2 \ .
\ee
The simplest thing one can do at this stage is expand the above integrand for $z \ll 1$. Since $H_{\nu}(x) \propto x^{ - \mathrm{Re}[\nu]}$ for $x\ll 1$, means that $ \mathcal{F} \propto z^{3 - 2 \mathrm{Re}[\nu_{\rm sys}]  - 4 \mathrm{Re}[\nu_{\rm env}]}$, however the remaining integrals are not always convergent in this limit which provides restrictions on $\nu_{\rm sys}$ and $\nu_{\rm env}$. One can see in this same limit that the integrand $y$-integrand scales as $y^{{1}/{2} - \mathrm{Re}[\nu_{\rm sys}] + 2 \mathrm{Re}[\nu_{\rm env}] }$, and the $a$- and $b$-integrand scales as $(b^2 - a^2)^{1 - 2 \nu_{\rm env}}$. All of these integrals converge only when
\be \label{B_region}
\mathrm{Re}[\nu_{\rm sys}] \geq 0 \qquad \mathrm{and} \qquad \tfrac{3}{4} < \nu_{\rm env}< \tfrac{3}{2} 
\ee
which means there is no restriction on $\nu_{\rm sys}$. Assuming this holds, and furthermore that $\nu_{\rm sys} > 0$ for simplicity, one finds the leading-order behaviour
\be
   \cF(z) \simeq \; B(\nu_{\rm sys},\nu_{\rm env}) \; z^{3 - 2 \nu_{\rm sys}  - 4 \nu_{\rm env}} + \ldots \qquad (\mathrm{when} \ \nu_{\rm sys} > 0)
\ee
where the coefficient $B$ is given by
\be \label{B_coeff_def}
B(\nu_{\rm sys},\nu_{\rm sys})  :=  \frac{ 2^{8 \nu _{\rm env}+2 \nu _{\rm sys}-1} \Gamma^2(\nu _{\rm sys}) \Gamma^4(\nu _{\rm env})}{\pi ^5} \int_{0}^{1} \exd a \int_{1}^\infty \exd b\; (b^2 -a^2)^{ 1 - 2 \nu_{\rm env}} \bigg| \int_{1}^\infty \exd y \; y^{\tfrac{1}{2} - \nu_{\rm sys} -2 \nu_{\rm env}}   \bigg|^2 \ .
\ee
which comes from the leading-order behaviour $H_{\nu}(x) \simeq - \frac{i}{\pi} \Gamma(\nu) \left( \frac{x}{2} \right)^{-\nu} + \ldots $ for $\nu>0$. In the region \pref{B_region} this is convergent and explicitly evaluates to 
\be \label{B_coeff_result}
B(\nu_{\rm sys},\nu_{\rm sys}) = \frac{ 4^{\nu _{\rm sys}+ 4 \nu _{\rm env}-1} \Gamma^2(\nu _{\rm sys}) \Gamma^4(\nu _{\rm env}) \tan \left(\pi  \nu _{\rm env}\right) \sec \left(2 \pi  \nu _{\rm env}\right)  }{\pi ^{7/2} (3 - 2 \nu_{\rm sys} - 4 \nu_{\rm env})^2 \left(\nu _{\rm env}-1\right) \Gamma \left(\frac{5}{2}-2 \nu _{\rm env}\right) \Gamma \left(2 \nu _{\rm env}-1\right) }
\ee
In the case that $\mathrm{Re}[\nu_{\rm sys}] = 0$ ({\it ie.} when $\zeta_{\rm sys} > 9/4$) we write $\nu_{\rm sys} = i \mu_{\rm sys}$, and it turns out the leading-order behaviour of 
\be
   \cF(z) \simeq B_{-} \;  z^{3 - 2 i \mu_{\rm sys}  - 4 \nu_{\rm env}} + B_{+} \; z^{3 + 2 i \mu_{\rm sys}  - 4 \nu_{\rm env}} + \ldots \qquad (\mathrm{when} \ \mathrm{Re}[\nu_{\rm sys}] =0\ \mathrm{with\ } \nu_{\rm sys} = i \mu_{\rm sys})
\ee
where the coefficients $B_{\pm}$ are calculable by the same method as above but we do not compute here. 

\subsection{Super-Hubble scaling when $\mathrm{Re}(\nu_{\rm env}) < \frac{3}{4}$}

For the opposing case of $\mathrm{Re}(\nu_{\rm env}) \leq \frac{3}{4}$, one must take a different approach since the coefficients in the series computed in Appendix \ref{App:34nu32} above diverge here. In Appendix \ref{App:34nu32}, the restriction that $\nu_{\rm env}> \frac{3}{4}$ is what allows the $b$-integral to converge in the UV ({\it ie.}~for large momenta $ b \to \infty$) --- to make the $b$-integral better convergent in the UV (in the required limit of $z\ll 1$) it suffices to rescale the integration variable such that
\be
x = b z
\ee
but will change the $z$-scaling of the overall function. This turns (\ref{Ffunction_yy}) into
\be
   \cF(z) = \frac{\pi}{2} \int_{0}^{1} \exd a \int_{z}^\infty \exd x\;   (x^2 -a^2 z^2 ) \bigg| \int_{1}^\infty \exd y \; \sqrt{ y } \; H_{\nu_{\rm sys}}^{(1)}( z y ) H_{\nu_{\rm env}}^{(1)}( \tfrac{x-az}{2} y)  H_{\nu_{\rm env}}^{(1)}( \tfrac{x+az}{2} y)  \bigg|^2 \ .
\ee
This form gives the leading-order behaviour of $\cF(z)$ when expanded for $z \ll 1$, such that
\be \label{Fz_asymp}
   \cF(z) \simeq  A(\nu_{\rm sys},\nu_{\rm env}) \; z^{- 2 \nu_{\rm sys} } + \ldots
\ee
with the definition
\be \label{A_coeff_def}
  A(\nu_{\rm sys},\nu_{\rm env}) := \frac{2^{2 \nu_{\rm sys}}\Gamma^2(\nu_{\rm sys})}{2\pi} \int_{0}^\infty \exd x\; x^2 \; \bigg| \int_{1}^\infty \exd y \; y^{\tfrac{1}{2} - \nu_{\rm sys}} \left[ H_{\nu_{\rm env}}^{(1)}( \tfrac{x}{2} y) \right]^2  \bigg|^2 
\ee
where we've used the leading-order form of $H_{\nu_{\rm sys}}^{(1)}( z y )$ for $z \ll 1$, again assuming that $\nu_{\rm sys} > 0 $ for simplicity (and have trivially performed the $a$-integral in this particular limit). Note that the $x$-integrand of $A(\nu_{\rm sys},\nu_{\rm env})$ scales as $\propto x^{2 - 4 \nu_{\rm env}}$ for $x \ll 1$ which means that the integral converges in the IR only in the case that
\be
\mathrm{Re}(\nu_{\rm env}) < \tfrac{3}{4} 
\ee
and otherwise the integrals converge for any $\nu_{\rm sys} > 0$. We are unable to find a closed form solution for $A(\nu_{\rm sys},\nu_{\rm env})$, however it is possible to perform the $y$-integral for generic $\nu_{\rm sys}$ --- for simplicity, consider the special case of $\nu_{\rm sys} = \frac{3}{2}$ where one is able to express (\ref{A_coeff_def}) as
\be \label{A32_App}
A(\tfrac{3}{2}, \nu_{\rm env}) = \int_{0}^\infty \exd x \; x^2 \; \left| \; \frac{x}{4 \nu_{\rm env}} \left[ H_{\nu_{\rm env}}^{(1)}\big(\tfrac{x}{2} \big) \frac{\partial H_{\nu_{\rm env}+1}^{(1)}\big(\tfrac{x}{2}\big) }{\partial \nu_{\rm env}} - \frac{\partial H_{\nu_{\rm env}}^{(1)}\big(\tfrac{x}{2} \big) }{\partial \nu_{\rm env}} H_{\nu_{\rm env}+1}^{(1)}\big(\tfrac{x}{2} \big) \right] - \frac{\big[ H_{\nu_{\rm env}}^{(1)}\big(\tfrac{x}{2} \big) \big]^2}{2\nu_{\rm env}} \; \right|^2
\ee
which uses Eq.~(5.11.14) from \cite{watson1922treatise}. The precise scaling of the above integrand is $\simeq \frac{4\Gamma^4(\nu_{\rm env})}{\pi^4\nu^2} \left( \frac{x}{4} \right)^{2- 4 \nu_{\rm env}}$ for $x\ll 1$ and $\simeq \frac{16}{\pi^2 x^2}$ for $x \gg 1$. 

Of particular interest in the above formula is the heavy mass limit where $\nu_{\rm env}= i \mu_{\rm env}$ (when $\zeta_{\rm env}> 9/4$), which ends up making the index on the Hankel function imaginary and large such that $\mu_{\rm env}\gg 1$. One may use the expansion \cite{dunster1990bessel}
\be
 H_{i \mu}^{(1)}(\mu_{\rm env}w) \simeq e^{+ \tfrac{\pi}{2} \mu - \tfrac{i \pi}{4} }  \sqrt{ \frac{2}{\pi\mu_{\rm env}} } \; (1 + w^2)^{-\tfrac{1}{4}} \; e^{i \mu_{\rm env}\left[ \sqrt{1 + w^2} - \mathrm{csch}^{-1}(w) \right] } \; \bigg[ 1 -  \frac{i \left(3 w^2-2\right)}{24 (w^2+1)^{3/2} \mu_{\rm env}} + \mathcal{O}(\mu_{\rm env}^{-2}) \bigg] 
\ee
for fixed $w>0$ to eventually derive in (\ref{A32_App})
\be
A(\tfrac{3}{2}, i\mu_{\rm env})  \simeq  \int_{0}^\infty \exd w \;  \frac{8 w^2}{\pi ^2 \mu_{\rm env} \left(w^2+1\right)^2} \bigg[1 + \frac{8 w^2-5 w^4}{2 \mu_{\rm env}^2 \left(w^2+1\right)^3} + \mathcal{O}(\mu_{\rm env}^{-4}) \bigg] \qquad \mathrm{for} \ \mu_{\rm env}\gg 1 
\ee
which makes the variable change $w = \frac{x}{2\mu_{\rm env}}$ for simplicity. The individual terms can be safely integrated where one finds that
\be \label{Aint_asympt_mu}
A(\tfrac{3}{2}, i\mu_{\rm env}) \simeq  \frac{2}{\pi  \mu_{\rm env}} \bigg[ 1-\frac{1}{128 \mu_{\rm env}^2} + \mathcal{O}(\mu_{\rm env}^{-4} ) \bigg] \qquad \mathrm{for} \ \mu_{\rm env}\gg 1 \ .
\ee

\bibliographystyle{JHEP}
\bibliography{PurityRate}

\end{document}